%% file: JFluidMech_01.tex
\documentclass{jfm}
\usepackage{amssymb}
\usepackage{amsfonts}
\usepackage{amsmath}
\usepackage{euscript}
\usepackage{color}
\usepackage{graphicx}
\usepackage{rotating}
\usepackage{bm}
\usepackage{curves}
\usepackage[svgnames,dvipsnames]{pstricks}
\usepackage{lettrine}
\usepackage{mathrsfs}
\usepackage{booktabs}
\usepackage{scalefnt}
\usepackage[normalem]{ulem}
\usepackage{dcolumn}
\usepackage{natbib}
\usepackage{tikz}
\DeclareGraphicsExtensions{{},.pdf}
\providecommand{\tsn}[1]{{\text{\scalefont{0.80}#1}}}

\providecommand{\tabref}[1]{{\textup{(Tab.~\ref{#1})}}}
\providecommand{\figref}[1]{{\textup{(Fig.~\ref{#1})}}}

\providecommand{\tabrefnp}[1]{{\textup{Tab.~\ref{#1}}}}
\providecommand{\figrefnp}[1]{{\textup{Fig.~\ref{#1}}}}

\providecommand{\eqrefsatob} [2]{\textup{(\ref{#1}--\ref{#2})}}
\providecommand{\eqrefsab}   [2]{\textup{(\ref{#1}, \ref{#2})}}
\providecommand{\eqrefsabc}  [3]{\textup{(\ref{#1}, \ref{#2}, \ref{#3})}}

\providecommand{\figrefsab}   [2]{{\textup{(Figs.~\ref{#1}, \ref{#2})}}}
\providecommand{\figrefsabc}  [3]{{\textup{(Figs.~\ref{#1}, \ref{#2}, \ref{#3})}}}
\providecommand{\figrefsatob} [2]{{\textup{(Figs.~\ref{#1}--\ref{#2})}}}
\providecommand{\figrefsatobnp} [2]{{\textup{Figs.~\ref{#1}--\ref{#2}}}}
\providecommand{\parref}[1]{{\textup{(\S\ref{#1})}}}

\providecommand{\parrefsatob}[2]{\textup{(\S\ref{#1}--\S\ref{#2})}}
\providecommand{\parrefnp}[1]{{\textup{\S\ref{#1}}}}
\providecommand{\parrefsatob}[2]{{\textup{(\S\ref{#1}--\S\ref{#2})}}}
\providecommand{\const}{{\rm const}}

\providecommand{\ie}{{\em ie~}}
\providecommand{\eg}{{\em eg~}}
\providecommand{\viz}{{\em viz~}}

\newcommand{\abs}[1]{\left\lvert#1\right\rvert}

\title[Compressible wall-turbulence]{Pressure, density, temperature and entropy fluctuations in compressible turbulent plane channel flow}
\author[G.A. Gerolymos and I. Vallet]
{G.\ns A.\ns G\ls E\ls R\ls O\ls L\ls Y\ls M\ls O\ls S,\ns
I.\ns V\ls A\ls L\ls L\ls E\ls T\footnote{Email address for correspondence: isabelle.vallet@upmc.fr}}
\affiliation{\scalefont{0.94}{Faculty of Engineering, Universit\'e Pierre-et-Marie-Curie (UPMC), 4 place Jussieu, 75005 Paris, France}}
\date{\today}

\begin{document}

\maketitle

\begin{abstract}
We investigate the fluctuations of thermodynamic state-variables in compressible aerodynamic wall-turbulence,
using results of direct numerical simulation (\tsn{DNS}) of compressible turbulent plane channel flow.
The basic transport equations governing the behaviour of thermodynamic variables (density, pressure, temperature and entropy)
are reviewed and used to derive the exact transport equations for the variances and fluxes (transport by the fluctuating velocity field)
of the thermodynamic fluctuations. The scaling with Reynolds and Mach number of compressible turbulent plane channel flow is discussed.
Correlation coefficients and higher-order statistics of the thermodynamic fluctuations are examined.
Finally, detailed budgets of the transport equations for the variances and fluxes of the thermodynamic variables from a well-resolved \tsn{DNS}
are analysed. Implications of these results both to the understanding of the thermodynamic interactions in compressible wall-turbulence and
to possible improvements in statistical modelling are assessed. Finally, the required extension of existing \tsn{DNS} data to fully characterise this canonical flow is discussed.
\end{abstract}

\begin{keywords}
compressible turbulence, wall-turbulence, \tsn{DNS}, turbulent transport budgets, density fluctuations, turbulent mass-fluxes, turbulent heat-fluxes
\end{keywords}

%
%
%
%
%
%
%
%
%
\section{Introduction}\label{PDTEFCTPCF_s_I}
%
%
%
%
%
%
%
%
%

Analysis of compressible turbulence is invariably related to the dynamics of thermodynamic fluctuations, {\em viz} $\rho'$, $p'$, $T'$ and $s'$
(where $\rho$ is the density, $p$ is the pressure, $T$ is the temperature, and $s$ is the entropy).
This is exemplified by the early work of \citet{Kovasznay_1953a} on the modal decomposition of compressible turbulence.
From this analysis \citep{Kovasznay_1953a} it follows that vortical, acoustic and entropic fluctuation-modes are only coupled one with another by nonlinear (high fluctuation amplitude) effects \citep{Chu_Kovasznay_1958a}.
Building upon this modal description, \citet{Morkovin_1962a} formulated his hypothesis that in supersonic boundary-layers (external flow Mach number $\bar M_e\lessapprox5$) 
the vorticity-entropy coupling, which is expected to be the dominant "{\em compressible feedback mechanism}"
is not sufficiently strong to "{\em affect the essential dynamics of the boundary-layer but rather modulates them through (stratified) mean values of $\bar\rho(y)$ and $\bar T(y)$}" (where $y$ is the wall-normal coordinate).
The main corollary of Morkovin's hypothesis is \citep{Bradshaw_1977a} that turbulent structure (correlation coefficients and spectra) does not change substantially compared to
constant-density flow. The correlation coefficient between streamwise-velocity and temperature fluctuations $c_{u'T'}$ is central in the analysis of \citet{Morkovin_1962a}.
\citet{Bradshaw_1977a} uses the relative value of density-rms \smash{$[\rho']_\mathrm{rms}:=[\overline{\rho'^2}]^\frac{1}{2}$} (where rms stands for root-mean-square, $\overline{(\cdot)}$ denotes Reynolds-averaging,
and $(\cdot)'$ are Reynolds-fluctuations) with respect to mean density $\bar\rho$ as the main criterion of validity of Morkovin's hypothesis \citep{Morkovin_1962a},
setting the limit at $[\rho']_\mathrm{rms}\lessapprox\tfrac{1}{10}\bar\rho$.

Although measurement techniques for turbulent fluctuations of pressure $p'$ \citep*{Tsuji_Fransson_Alfredsson_Johansson_2007a},
density $\rho'$ \citep{Panda_Seasholtz_2002a} and temperture $T'$ \citep{Teitel_Antonia_1993a}
are available, there are complex cross-correlations, especially in the near-wall region, that are still beyond experimental reach.
Therefore, in line with incompressible flow practice \citep{Kim_2012a}, direct numerical simulation (\tsn{DNS})
"{\em its limitation to low-$Re$ notwithstanding, $\cdots$ has played an essential role by providing accurate and detailed
information, which are not readily available from experimental results}" (where $Re$ denotes an appropriate Reynolds number).

Recently, \citet{Donzis_Jagannathan_2013a} studied the statistics of $\{p',\rho',T'\}$ obtained from \tsn{DNS} of sustained compressible homogeneous isotropic turbulence (\tsn{HIT}), for Taylor-microscale Reynolds numbers
$Re_\lambda:=[\tfrac{2}{3}\mathrm{k}]^\frac{1}{2}\lambda_g\breve\nu^{-1}\in[30,430]$ and turbulent Mach numbers $\breve M_\tsn{T}:=\breve a^{-1}[2\mathrm{k}]^\frac{1}{2}\in[0.1,0.6]$
(where $\mathrm{k}$ is the turbulent kinetic energy, $\lambda_g$ is the Taylor microscale, and $\breve\nu$ and $\breve a$ are the average kinematic viscosity and the sound-speed at mean conditions, respectively),
corresponding to relative density-fluctuation rms-levels (coefficients of variation of $\rho$) $[\bar\rho^{-1}\rho']_\mathrm{rms}\in[\tfrac{3}{10}\%,16\%]$. We prefer using the coefficient of variation of density
as a measure of compressibility when transposing the \tsn{HIT} results to wall-bounded flows, because the turbulent Mach number $\breve M_\tsn{T}$ is not an appropriate scaling for shear flows \citep{Sarkar_1995a,
                                                                                                                                                                                                       Pantano_Sarkar_2002a}.
In sustained compressible \tsn{HIT} the coefficients of variation $\{[\bar\rho^{-1}\rho']_\mathrm{rms},[\bar T^{-1}T']_\mathrm{rms},[\bar p^{-1}p']_\mathrm{rms}\}$ scale reasonably well 
with $\breve M^2_\tsn{T}$ \citep[Fig. 2, p. 227]{Donzis_Jagannathan_2013a}, and the fluctuations follow on the average a polytropic exponent $n_\mathrm{p}\approxeq1.2$,
slighlty increasing with increasing $[\bar\rho^{-1}\rho']_\mathrm{rms}$ \citep[Fig. 7, p. 227]{Donzis_Jagannathan_2013a}.
However, wall-turbulence includes several phenomena that are absent in \tsn{HIT}, such as mean shear inducing rapid pressure fluctuations \citep{Chou_1945a},
mean density and temperature gradients producing fluctuations $\rho'$ and $T'$ \citep{Taulbee_vanOsdol_1991a,
                                                                                    Tamano_Morinishi_2006a},
and the presence of solid walls introducing wall-echo effects on $p'$ \citep*{Gerolymos_Senechal_Vallet_2013a},
and requires therefore specific study.

The detailed study of canonical incompressible wall-bounded flows, \viz flat-plate zero-pressure-gradient (\tsn{ZPG}) boundary-layer and internal
fully developed (streamwise-inva\-riant in the mean) flows (plane channel and pipe), has substantially advanced our understanding of wall turbulence \citep*{Smits_McKeon_Marusic_2011a},
although there is still debate on the differences in behaviour with increasing Reynolds number between boundary-layers and internal flows \citep*{Monty_Hutchins_Ng_Marusic_Chong_2009a}, 
especially concerning the scaling of the streamwise-velocity fluctuation $\overline{u'^2}$ near-wall peak \citep*{Hultmark_Bailey_Smits_2010a}
and of wall-pressure variance $\overline{p_w'^2}$ \citep*{Tsuji_Imayama_Schlatter_Alfredsson_Johansson_Marusic_Hutchins_Monty_2012a}.
In regard with incompressible \tsn{DNS} studies, \citet{Schlatter_Orlu_2010a} show that boundary-layer results are sensitive to inflow and boundary conditions used by different authors,
contrary to fully developed internal flows which apply unambiguous streamwise-periodicity conditions.
This also applies to the compressible boundary-layer, with the added difficulty of prescribing inflow conditions for the thermodynamic quantities as well \citep{Xu_Martin_2004a}.
For this reason, always keeping in mind possible differences in behaviour at the high-$Re$ limit \citep{Monty_Hutchins_Ng_Marusic_Chong_2009a},
the study of compressible turbulent plane channel flow can provide insight to the physics of "{\em near-wall boundary-layer flow as it appears in high-speed flight}" \citep{Friedrich_2007a}.

\citet*{Coleman_Kim_Moser_1995a} established the canonical case of compressible turbulent plane channel flow between two isothermal walls, made streamwise-invariant in the mean by the introduction of a body-acceleration term,
adjusted to enforce the target massflow. Their results included distributions of $\{[p']_\mathrm{rms},[\rho']_\mathrm{rms},[T']_\mathrm{rms}\}$ \citep[Figs. 10 and 14, pp. 170 and 172]{Coleman_Kim_Moser_1995a},
of 2-point correlations in the homogeneous streamwise ($\{R_{p'p'}^{(x)},R_{T'T'}^{(x)}\}$) and spanwise (\smash{$\{R_{p'p'}^{(z)},R_{T'T'}^{(z)}\}$}) directions \citep[Figs. 3--4, pp. 165--166]{Coleman_Kim_Moser_1995a},
and joint-pdfs (probability-density functions) of $\{(T',\rho'),(T',u')\}$ \citep[Fig. 11, p. 171]{Coleman_Kim_Moser_1995a}.
The data obtained by \citet{Coleman_Kim_Moser_1995a} were further analysed by \citet*{Huang_Coleman_Bradshaw_1995a}, who found that for the cases studied (centerline Mach-number $\bar M_\tsn{CL}\in\{1.5,2.2\}$)
the influence of density fluctuations $\rho'$ on the Reynolds shear-stress $\overline{\rho u'' v''}$ and on the wall-normal turbulent temperature transport $\overline{\rho v'' T''}$ were quite small
(where $(\cdot)''$ are Favre fluctuations, $u$ is the streamwise velocity-component,  and $v$ is the wall-normal velocity-component),
that the strong Reynolds analogy (\tsn{SRA}) which assumes negligible total temperature fluctuations \citep{Morkovin_1962a} is unsatisfactory for these flows, and that
the Reynolds averages of Favre fluctuations $\{\overline{u''},\overline{v''},\overline{T''}\}$ could be reasonably well modelled by assuming an isobaric polytropic process for the thermodynamic fluctuations.
\citet*{Lechner_Sesterhenn_Friedrich_2001a} studied scatter plots of $\{(T',\rho'),(T',u'),(T',v'),(\rho',p'),(s',\rho'),(s',p'),(s',T')\}$ \citep[Figs. 8--11, pp. 12--14]{Lechner_Sesterhenn_Friedrich_2001a},
briefly discussed, without plotting detailed budgets, the transport equations for the variances of density $\overline{\rho'^2}$ \citep[(32), p. 10]{Lechner_Sesterhenn_Friedrich_2001a}
and pressure $\overline{p'^2}$ \citep[(42), p. 13]{Lechner_Sesterhenn_Friedrich_2001a}, and studied the budgets of the transport equations for the Reynolds-stresses $\overline{\rho u_i''u_j''}$
and the turbulent kinetic energy $\tfrac{1}{2}\overline{\rho u_i''u_i''}$, this last equation having also been scrutinized by \citet{Huang_Coleman_Bradshaw_1995a}.
From the study of the scatter plots for the thermodynamic variables, \citet[Figs. 8--11, pp. 12--14]{Lechner_Sesterhenn_Friedrich_2001a}
concluded that near the wall ($y^+\approxeq9$ where $y^+$ is the distance from the wall in wall-units) considering isobaric fluctuations (neglecting $p'$) was a reasonable approximation.
However, examination of the scatter plots \citep[Figs. 8--11, pp. 12--14]{Lechner_Sesterhenn_Friedrich_2001a} at the centerline invalidates the isobaric assumption in the outer part of the flow.
\citet*{Morinishi_Tamano_Nakabayashi_2004a} investigated both the standard symmetric configuration with two isothermal walls at equal temperatures and the asymmetric case of a cold wall and a hot wall, either adiabatic
or isothermal at the adiabatic-wall temperature \citep{Tamano_Morinishi_2006a}, and studied the budgets of the transport equation for the temperature variance $\overline{\rho T''^2}$ \citep{Tamano_Morinishi_2006a}.
\citet{Senechal_2009a} used data from coarse-grid \tsn{DNS} computations to calculate the budgets of the transport equations for the variances of thermodynamic fluctuations
($\overline{\rho'^2}$, $\overline{\rho T''^2}$, $\overline{p'^2}$), and observed that the coefficient of variation of density $[\bar\rho^{-1}\rho']_\mathrm{rms}$ at constant
friction Reynolds number \eqref{Eq_PDTEFCTPCF_s_S_ss_CCF_001b} $Re_{\tau_w}\approxeq230$ and $\bar M_\tsn{CL}\in\{1.5,0.34\}$ , plotted against the outer-scaled wall-distance
$\delta^{-1}(y-y_w)$ ($\delta$ is the channel half-height) varied as $\bar M_\tsn{CL}^2$.
Finally, \citet{Wei_Pollard_2011a} examined the influence of Mach number ($\bar M_\tsn{CL}\in\{0.2,0.7,1.5\}$) at constant bulk Reynolds number $Re_{\tsn{B}_w}$ \eqref{Eq_PDTEFCTPCF_s_S_ss_CCF_001a} on the variance, skewness and flatness of
$\{u',v',w',p',\rho',T'\}$  (where $w$ is the spanwise velocity-component). They further analyzed these \tsn{DNS} results \citep{Wei_Pollard_2011b} to study the influence of Mach number on the correlation coefficient
$c_{\rho'p'}$ \citep[Fig. 1, p. 6]{Wei_Pollard_2011b} and on correlations between the fluctuating pressure-gradient and the fluctuating vorticity-gradients (and also vorticity fluxes).
They also presented budgets of the transport equation for $\overline{\rho'\omega_z'}$ (where $\vec\omega:=\mathrm{rot}\vec{V}$ is the vorticity and $z$ is the spanwise direction), and found
a surprisingly strong influence of the term $\overline{\rho\omega_z\Theta}$ (where $\Theta:=\mathrm{div}\vec{V}$ is the dilatation-rate) in this equation \citep[(3.7), p. 12]{Wei_Pollard_2011b} with decreasing Mach number \citep[Fig. 8, p. 13]{Wei_Pollard_2011b}.

\citet*{Foysi_Sarkar_Friedrich_2004a} focussed on the behaviour of pressure fluctuations $p'$ in compressible turbulent plane channel flow, using the compressible analogue \citep[(A 1e), p. 46]{Gerolymos_Senechal_Vallet_2013a}
of the incompressible flow Poisson equation for $\nabla^2 p'$ \citep{Chou_1945a}, and found that the observed reduction with increasing Mach number of the absolute magnitude of pressure-strain correlations could be satisfactorily accounted for by
mean-density stratification $\bar\rho(y)$, in line with Morkovin's hypothesis \citep{Morkovin_1962a}, the terms associated with $\rho'$ in \citep[(4.1), p. 213]{Foysi_Sarkar_Friedrich_2004a} having marginal influence.
This contrasts with the free shear-layer case, where acoustic propagation of $\rho'$ effects \citep[(4.7), p. 347]{Pantano_Sarkar_2002a} were found to be important \citep*{Mahle_Foysi_Sarkar_Friedrich_2007a}.
\citet*{Ghosh_Foysi_Friedrich_2010a} also studied the case of compressible turbulent pipe flow and found again that Morkovin's hypothesis \citep{Morkovin_1962a} was applicable.
The only potential issue in the analysis of compressible turbulent plane channel flow is that, as the Mach number increases, strong viscous heating \parref{PDTEFCTPCF_s_S_ss_CCF} substantially increases mean temperature in the
centerline (outer) region, and radiative heat transfer may need to be taken into account. Nonetheless, computations by \citet*{Ghosh_Friedrich_Pfitzner_Stemmer_Cuenot_ElHafi_2011a} indicate that, although
radiative heat transfer slightly reduces the effects of compressibility, its influence is not expected to be important for airflow at the Reynolds numbers investigated in the present work.

Several related studies concern the compressible turbulent \tsn{ZPG} boundary-layer developing on a flat plate \citep*{Guarini_Moser_Shariff_Wray_2000a,
                                                                                                                       Maeder_Adams_Kleiser_2001a,
                                                                                                                       Pirozzoli_Grasso_Gatski_2004a,
                                                                                                                       Martin_2007a,
                                                                                                                       Duan_Beekman_Martin_2010a,
                                                                                                                       Duan_Beekman_Martin_2011a,
                                                                                                                       Duan_Martin_2011a,
                                                                                                                       Lagha_Kim_Eldredge_Zhong_2011a,
                                                                                                                       Lagha_Kim_Eldredge_Zhong_2011b,
                                                                                                                       Bernardini_Pirozzoli_2011a,
                                                                                                                       Shahab_Lehnasch_Gatski_Comte_2011a},
covering, although in no systematic way, the range $\bar M_e\in[2.2,8]$ and $\bar T_w\in[\bar T_e,12\bar T_e]$ (where $(.)_e$ indicates the boundary-layer edge).
A few computations at higher Mach numbers \citep{Duan_Beekman_Martin_2010a,
                                                 Lagha_Kim_Eldredge_Zhong_2011a}
correspond to very high wall temperatures, for which more complex thermodynamics \citep{Duan_Martin_2011a} and/or radiative heat transfer \citep{Ghosh_Friedrich_Pfitzner_Stemmer_Cuenot_ElHafi_2011a} should be taken into account.
In the boundary-layer case the external and wall temperatures can be
chosen arbitrarily, independent of one another or of $\bar M_e$, contrary to the compressible plane channel case, where the difference $\bar T_\tsn{CL}-\bar T_w$ is the result of intense viscous heating \citep{Coleman_Kim_Moser_1995a},
and is therefore a unique function of the Mach and Reynolds numbers. More importantly, in the compressible plane channel case generated heat can only be evacuated across the walls,
implying that $\bar T_w<\bar T_\tsn{CL}$, contrary to the boundary-layer studies where $\bar T_w\geq\bar T_e$ was invariably chosen, with the exception
of the high-enthalpy cases studied by \citet[Tab. 2, p. 21]{Duan_Beekman_Martin_2011a}.
\citet{Guarini_Moser_Shariff_Wray_2000a} studied a $\bar M_e\approxeq2.5$ turbulent boundary-layer along an adiabatic wall and explained the success \citep[Fig. 16, p. 18]{Guarini_Moser_Shariff_Wray_2000a}
of the \tsn{HCB}-modified \citep[(4.10), p. 208]{Huang_Coleman_Bradshaw_1995a}
\tsn{SRA} by showing that this form implied the approximate equality of the correlation coefficients for wall-normal turbulent transport of velocity and temperature, \viz \smash{$c_{u'v'}\approxeq-c_{T'v'}$}
\citep[(4.20), p. 20]{Guarini_Moser_Shariff_Wray_2000a}, which was approximately satisfied over a large part of the boundary-layer \citep[Fig. 19, p. 21]{Guarini_Moser_Shariff_Wray_2000a}.
This conclusion on the validity of the \tsn{HCB}-modified \citep[(4.10), p. 208]{Huang_Coleman_Bradshaw_1995a} \tsn{SRA} has been confirmed in the other boundary-layer studies
(\citealp[Fig. 20, p. 208]{Maeder_Adams_Kleiser_2001a};
 \citealp[Fig. 16, p. 540]{Pirozzoli_Grasso_Gatski_2004a};
 \citealp[Fig. 10, p. 431]{Duan_Beekman_Martin_2010a};
 \citealp[Fig. 10, p. 255]{Duan_Beekman_Martin_2011a};
 \citealp[Fig. 20, p. 393]{Shahab_Lehnasch_Gatski_Comte_2011a}) as well.
\citet[Figs. 18--19, pp. 390--391]{Shahab_Lehnasch_Gatski_Comte_2011a} presented budgets of the transport equations for the fluxes $\overline{\rho u_i''T''}$,
and \citet[Figs. 18--21, pp. 11--12]{Lagha_Kim_Eldredge_Zhong_2011b} for the tranport equation for the spanwise vorticity variance $\overline{\omega_z'^2}$,
which was used to analyse the near-wall dynamics.

With the exception of the transport equations for the temperature-fluctuations variance \citep{Tamano_Morinishi_2006a} and fluxes \citep{Shahab_Lehnasch_Gatski_Comte_2011a},
the transport equations associated with the fluctuations of thermodynamic quantities have not been analysed in \tsn{DNS} studies of wall-turbulence. Nonetheless, these transport equations are necessary to understand the associated phenomena,
and are used in theoretical and modelling studies of compressible turbulence.
\citet{Taulbee_vanOsdol_1991a} studied the transport equations for the density variance $\overline{\rho'^2}$ \citep[(17), p. 4]{Taulbee_vanOsdol_1991a}
and for the turbulent mass-flux $\overline{\rho'u_i'}=-\bar\rho^{-1}\overline{u_i''}$ \citep[(23), p. 4]{Taulbee_vanOsdol_1991a}, proposed closures for the unknown terms in these equations, and applied their model to supersonic \tsn{ZPG} turbulent boundary-layers
($\bar M_e\in[1.7,3,4.7]$). Although the absence of \tsn{DNS} data of compressible wall-bounded turbulence at that time \citep{Taulbee_vanOsdol_1991a} made impossible the term-by-term analysis of the equations
and of the proposed closures, comparison with experimental data of $\overline{\rho'^2}$ and $\overline{u_i''}$ was encouraging \citep[Figs. 1--3, p. 8]{Taulbee_vanOsdol_1991a}.
\citet{Yoshizawa_1992a} developed a model for the density variance equation based on \tsn{TSDIA} (two-scale direct-interaction approximation), and remarked that closures obtained for various
compressible terms \citep[(B7--B13), p. 3303]{Yoshizawa_1992a} were proportional to $[\bar\rho^{-1}\rho']_\mathrm{rms}$, suggesting that this is presumably the most generally applicable scaling of compressibility effects.
\citet{Hamba_Blaisdell_1997a} further developed this type of \tsn{TSDIA}-based closure, which they calibrated using \tsn{DNS} data of compressible homogeneous turbulence.
\citet{Hamba_1999a} used approximate transport equations for $\overline{p'^2}$ \citep[(11), p. 1625]{Hamba_1999a} and $\overline{s'^2}$ \citep[(12), p. 1625]{Hamba_1999a},
in which all viscous terms, except for heat-conduction \citep[(13), p. 1625]{Hamba_1999a}, were dropped. \citet*{Yoshizawa_Matsuo_Mizobuchi_2013a} studied theoretically the transport equations for $\overline{\rho'^2}$,
$\overline{\rho'u_i'}$ and $\overline{\rho'p'}$, but viscous terms were neglected from the outset.
Finally, notice that the equation for the dilatation variance $\overline{\Theta'^2}$ used by \citet[(16), p. 3244]{Erlebacher_Sarkar_1993a} can also be interpreted, because of the continuity equation \citep[(A 1b), p. 45]{Gerolymos_Senechal_Vallet_2013a},
as a transport equation for $\overline{(\rho^{-1}D_t\rho)'^2}$ (where $D_t$ is the substantial derivative).
\begin{table}
\begin{center}
\input{JFluidMech_01_Tab_DNS_computations}
\caption{Parameters of the \tsn{DNS} computations
         [$L_x$, $L_y$, $L_z$ ($N_x$, $N_y$, $N_z$) are the dimensions (number of grid-points) of the computational domain ($x=$ homogeneous streamwise, $y=$ normal-to-the-wall, $z=$ homogeneous spanwise direction);
          $\delta$ is the channel halfheight;
          $\Delta x^+$, $\Delta y_w^+$, $\Delta y_\tsn{CL}^+$, $\Delta z^+$ are the mesh-sizes in wall-units \eqref{Eq_PDTEFCTPCF_s_S_ss_WUs_001a};
          $(\cdot)_w$ denotes wall and $(\cdot)_\tsn{CL}$ centerline values;
          $N_{y^+\leq10}$ is the number of grid points between the wall and $y^+=10$;
          $Re_{\tau^\star}:=\sqrt{\bar\rho(y)\;\bar\tau_w}\delta[\bar\mu(y)]^{-1}$ is the friction Reynolds number in \tsn{HCB}-scaling \eqref{Eq_PDTEFCTPCF_s_S_ss_CCF_001c};
          $\bar M_\tsn{CL}$ is the centerline Mach number;
          $\Delta t^+$ is the computational time-step in wall-units \eqref{Eq_PDTEFCTPCF_s_S_ss_WUs_001a};
          $t_\tsn{OBS}^+$ is the observation period in wall units \eqref{Eq_PDTEFCTPCF_s_S_ss_WUs_001a} over which single-point statistics were computed;
          $\Delta t_s^+$ is the sampling time-step for the single-point statistics in wall-units \eqref{Eq_PDTEFCTPCF_s_S_ss_WUs_001a}].}
\label{Tab_PDTEFCTPCF_s_BEqsDNSC_ss_DNSCs_001}
\end{center}
\end{table}

In the present work we study the behaviour of the fluctuations of thermodynamic variables in compressible turbulent plane channel flow.
In \parrefnp{PDTEFCTPCF_s_BEqsDNSC} we summarize the flow model \parref{PDTEFCTPCF_s_BEqsDNSC_ss_FM} and the \tsn{DNS} computations \parref{PDTEFCTPCF_s_BEqsDNSC_ss_DNSCs}, and recall the transport equations for
the thermodynamic variables \parref{PDTEFCTPCF_s_BEqsDNSC_ss_TTV} which are implied by the system of the Navier-Stokes equations \parref{PDTEFCTPCF_s_BEqsDNSC_ss_FM}.
In \parrefnp{PDTEFCTPCF_s_S} we discuss the basic scalings of compressible turbulent plane channel flow and study the effect of Mach and Reynolds numbers on turbulence structure (correlation coefficients).
In \parrefnp{PDTEFCTPCF_s_B} we work out the transport equations for the variances and fluxes of $\{p',\rho',T'',s''\}$ and study their budgets obtained from a well-resolved \tsn{DNS}.
Finally, in \parrefnp{PDTEFCTPCF_s_C} we summarize the basic results of the present analysis and discuss directions for further research.

%
%
%
%
%
%
%
%
%
\section{Basic equations and DNS computations}\label{PDTEFCTPCF_s_BEqsDNSC}
%
%
%
%
%
%
%
%
%

The compressible Navier-Stokes equations \parref{PDTEFCTPCF_s_BEqsDNSC_ss_FM} used in the present \tsn{DNS} computations \parref{PDTEFCTPCF_s_BEqsDNSC_ss_DNSCs} imply transport equations for the thermodynamic variables \parref{PDTEFCTPCF_s_BEqsDNSC_ss_TTV},
from which transport equations for their variances and fluxes can be obtained. Favre-averaging does not commute with differentiation, and care should be taken to avoid notational misuse of the operator $\widetilde{(\cdot)}$ \parref{PDTEFCTPCF_s_BEqsDNSC_ss_SA_sss_NAN}.

%
%
%
%
%
\subsection{Flow model}\label{PDTEFCTPCF_s_BEqsDNSC_ss_FM}
%
%
%
%
%

The \tsn{DNS} computations were performed using the solver developed in \citet{Gerolymos_Senechal_Vallet_2010a}.
The flow is modelled by the compressible Navier-Stokes equations \citep[(34--37), pp. 785--786]{Gerolymos_Senechal_Vallet_2010a}
\begin{subequations}
                                                                                                                                    \label{Eq_PDTEFCTPCF_s_BEqsDNSC_ss_FM_001}
\begin{alignat}{4}
\frac{\partial\rho}
     {\partial t  } + \frac{\partial       }
                           {\partial x_\ell}(\rho u_\ell)         =\;&0
                                                                                                                                    \label{Eq_PDTEFCTPCF_s_BEqsDNSC_ss_FM_001a}\\
\frac{\partial\rho u_i}
     {\partial t      } + \frac{\partial       }
                               {\partial x_\ell}(\rho u_i u_\ell) =\;&-\frac{\partial p  }
                                                                            {\partial x_i} + \frac{\partial\tau_{i\ell}}
                                                                                                  {\partial x_\ell     } + \rho f_{\tsn{V}_i}
                                                                                                                                    \label{Eq_PDTEFCTPCF_s_BEqsDNSC_ss_FM_001b}\\
\frac{\partial\rho e_t}
     {\partial t      } + \frac{\partial       }
                               {\partial x_\ell}(\rho h_t u_\ell) =\;& \frac{\partial       }
                                                                            {\partial x_\ell}( u_m \tau_{m\ell}-q_\ell) + \rho f_{\tsn{V}_m} u_m
                                                                                                                                    \label{Eq_PDTEFCTPCF_s_BEqsDNSC_ss_FM_001c}
\end{alignat}
with perfect-gas constant-$c_p$ thermodynamics
\begin{eqnarray}
p=\rho R_g T                               \;;\;
R_g ={\rm const}                           \;;\;
c_p=\frac{\gamma}{\gamma-1}R_g={\rm const} \;;\;
c_v=\frac{R_g   }{\gamma-1}   ={\rm const} \;;\;
a=\sqrt{\gamma\dfrac{p}{\rho}}\qquad\;
                                                                                                                                    \label{Eq_PDTEFCTPCF_s_BEqsDNSC_ss_FM_001d}
\end{eqnarray}
and linear constitutive relations
\begin{alignat}{4}
\tau_{ij}=& 2\mu\left(S_{ij}-\tfrac{1}
                                   {3}\Theta\delta_{ij}\right)+\mu_{\rm b}\Theta\delta_{ij}
\;;\;
S_{ij}:=  \tfrac{1}{2}\left(\frac{\partial u_i}
                                 {\partial x_j}+\frac{\partial u_j}
                                                     {\partial x_i}\right)
\;;\;
\Theta:=\frac{\partial u_\ell}
             {\partial x_\ell}=S_{\ell\ell}
                                                                                                                                    \label{Eq_PDTEFCTPCF_s_BEqsDNSC_ss_FM_001e}\\
q_i      =&-\lambda\frac{\partial T  }
                        {\partial x_i}
                                                                                                                                    \label{Eq_PDTEFCTPCF_s_BEqsDNSC_ss_FM_001f}
\end{alignat}
with
\begin{alignat}{4}
\mu_{\rm b}=0 &\quad;\quad&
\mu(T)=\mu_0\left[\frac{T        }
                       {T_{\mu_0}}\right]^{\frac{3}
                                                {2}}\frac{S_\mu+T_{\mu_0}}
                                                         {S_\mu+T        }&\quad;\quad&
\lambda(T)=\lambda_0\frac{\mu(T)}
                         {\mu_0 }[1+A_{\lambda}(T-T_{\mu_0})]
                                                                                                                                    \label{Eq_PDTEFCTPCF_s_BEqsDNSC_ss_FM_001g}
\end{alignat}
\end{subequations}
In \eqref{Eq_PDTEFCTPCF_s_BEqsDNSC_ss_FM_001},
$t$ is the time,
$x_i \in \{x,y,z\}$ are the Cartesian space coordinates,
$u_i \in \{u,v,w\}$ are the velocity components,
$h$ is the enthalpy,
$h_t:=h+\tfrac{1}{2}u_iu_i$ is the total enthalpy,
$e_t:=e+\tfrac{1}{2}u_iu_i=h_t-\rho^{-1}p$ is the total energy,
$e$ is the internal energy,
$\tau_{ij}$ is the viscous-stress tensor,
$q_i$ is the molecular heat-flux,
$f_{\tsn{V}_i}$ are body-acceleration terms,
$R_g$ is the gas-constant,
$\gamma$ is the isentropic exponent,
$c_p$ is the heat-capacity at constant pressure,
$c_v$ is the heat-capacity at constant volume (density),
$a$ is the sound-speed,
$S_{ij}$ is the strain-rate tensor,
$\Theta$ is the dilatation-rate,
$\mu$ is the dynamic viscosity,
$\mu_{\rm b}$ is the bulk viscosity, and
$\lambda$ is the heat-conductivity.
The present computations model airflow, for which the various coefficients and constants are \citep{Gerolymos_1990c,
                                                                                                    Gerolymos_Senechal_Vallet_2010a}
$R_g=287.04\;{\rm m}^{2}\;{\rm s}^{-2}\;{\rm K}^{-1}$, $\gamma=1.4$,
$\mu_0:=\mu(T_{\mu_0})=17.11\times10^{-6}\;{\rm Pa}\;{s}$, $T_{\mu_0}=273.15\;{\rm K}$, $S_\mu=110.4\;{\rm K}$,
$\lambda_0:=\lambda(T_{\mu_0})=0.0242\;{\rm W}\;{\rm m}^{-1}\;{\rm K}^{-1}$, $A_{\lambda}=0.00023\;{\rm K}^{-1}$.

%
%
%
%
%
\subsection{DNS computations}\label{PDTEFCTPCF_s_BEqsDNSC_ss_DNSCs}
%
%
%
%
%

\tsn{DNS} computations \tabref{Tab_PDTEFCTPCF_s_BEqsDNSC_ss_DNSCs_001} were run using an $O(\Delta x^{17})$ upwind-biased scheme \citep{Gerolymos_Senechal_Vallet_2009a} for the convective terms
and an $O(\Delta x^2$) conservative centered scheme \citep{Gerolymos_Senechal_Vallet_2010a} for the viscous terms, using
explicit dual-time-stepping time-integration \citep{icp_Gerolymos_Senechal_Vallet_2009a}.
Statistics were acquired using an onboard moving-averages technique \citep[\S4.4, p. 791]{Gerolymos_Senechal_Vallet_2010a}.
The numerical methodology has been validated \citep{Gerolymos_Senechal_Vallet_2010a,
                                                    Gerolymos_Senechal_Vallet_2013a}
by comparison with standard \tsn{DNS} data for incompressible \citep{Kim_Moin_Moser_1987a,
                                                                     Moser_Kim_Mansour_1999a,
                                                                     delAlamo_Jimenez_Zandonade_Moser_2004a}
and compressible \citep{Coleman_Kim_Moser_1995a,
                        Lechner_Sesterhenn_Friedrich_2001a}
fully developed turbulent channel flow,
and carefully assessed \parref{PDTEFCTPCF_s_B} for grid-resolution and statistical convergence of results.

Following standard practice \citep{Coleman_Kim_Moser_1995a}, a spatially constant body-accelera\-tion $f_{\tsn{V}_x}$ is applied in the streamwise direction
to counteract viscous friction and to apply a target bulk massflow $\dot m_\tsn{B}$ \citep[(46b, 48), p. 791]{Gerolymos_Senechal_Vallet_2010a},
bulk density $\rho_\tsn{B}$ is maintained constant at every subiteration \citep[(46a, 47), p. 791]{Gerolymos_Senechal_Vallet_2010a},
isothermal no-slip wall conditions are applied at $y\in\{0,2\delta\}$ \citep[(45), p. 790]{Gerolymos_Senechal_Vallet_2010a}, and periodic conditions
are applied in the homogeneous streamwise ($x$) and spanwise ($z$) directions \citep[p. 790]{Gerolymos_Senechal_Vallet_2010a}.

Notice that the algebraic correction of density \citep[(46a, 47), p. 791]{Gerolymos_Senechal_Vallet_2010a} applied at every numerical subiteration
to enforce constant $\rho_\tsn{B}$ is tantamount to including a source-term $Q_\rho$ in the continuity equation. Detailed examination of the time-evolution of
the computations \tabref{Tab_PDTEFCTPCF_s_BEqsDNSC_ss_DNSCs_001} reveals that this term is quite small (in wall-units) and its influence
($\bar Q_\rho$ and $Q'_\rho$ should appear in most turbulence transport equations) was neglected, in line with standard practice \citep{Coleman_Kim_Moser_1995a,
                                                                                                                                        Huang_Coleman_Bradshaw_1995a,
                                                                                                                                        Lechner_Sesterhenn_Friedrich_2001a,
                                                                                                                                        Morinishi_Tamano_Nakabayashi_2003a,
                                                                                                                                        Morinishi_Tamano_Nakabayashi_2004a,
                                                                                                                                        Foysi_Sarkar_Friedrich_2004a,
                                                                                                                                        Tamano_Morinishi_2006a,
                                                                                                                                        Kreuzinger_Friedrich_Gatski_2006a,
                                                                                                                                        Friedrich_2007a,
                                                                                                                                        Gerolymos_Senechal_Vallet_2010a,
                                                                                                                                        Gerolymos_Senechal_Vallet_2013a,
                                                                                                                                        Wei_Pollard_2011a,
                                                                                                                                        Wei_Pollard_2011b,
                                                                                                                                        Ghosh_Foysi_Friedrich_2010a,
                                                                                                                                        Ghosh_Friedrich_Pfitzner_Stemmer_Cuenot_ElHafi_2011a}.

In the present work we analyse \tsn{DNS} computations \tabref{Tab_PDTEFCTPCF_s_BEqsDNSC_ss_DNSCs_001} covering, although without a systematic variation
of both parameters, the range $Re_{\tau^\star}\in[64,344]$ and $\bar M_\tsn{CL}\in[0.34,2.47]$,
corresponding to $Re_{\tau_w}\in[99,525]$ and $M_{\tsn{B}_w}\in[0.3,3.83]$ \tabref{Tab_PDTEFCTPCF_s_S_ss_CCF_001}.
Available \tsn{DNS} data \citep{Coleman_Kim_Moser_1995a,
                                Friedrich_Foysi_Sesterhenn_2006a,
                                Tamano_Morinishi_2006a,
                                Gerolymos_Senechal_Vallet_2010a,
                                Wei_Pollard_2011b}
cover the range $Re_{\tau^\star}\in[150,300]$ and $\bar M_\tsn{CL}\in[0.35,2.25]$, again without systematic variation of both parameters.
The spatial resolution of the present large-box ($L_x\times L_y\times L_z=8\pi\delta\times2\delta\times4\pi\delta$) computations \tabref{Tab_PDTEFCTPCF_s_BEqsDNSC_ss_DNSCs_001}
in generally satisfactory, except for the $Re_{\tau^\star}\approxeq344$ simulation, for which computations with a finer grid are required to fully substantiate grid-convergence
of results (especially $\Delta z^+$ refinement; \tabrefnp{Tab_PDTEFCTPCF_s_BEqsDNSC_ss_DNSCs_001}).
Nonetheless, previous grid-convergence studies \citep[Figs. 5--6, pp. 794--795]{Gerolymos_Senechal_Vallet_2010a} indicate that results obtained on this grid are meaningfull, except
for a slight overestimation of $p'_\mathrm{rms}$, and they were therefore included to illustrate the influence of $Re_{\tau^\star}$ on the $\bar M_\tsn{CL}\approxeq1.5$ statistics.
Notice that the resolution for the $(Re_{\tau^\star},\bar M_\tsn{CL})=(344,1.51)$ large-box computations is similar to the one used in the $(Re_{\tau^\star},\bar M_\tsn{CL})=(350,2.25)$
simulations of \citet[Tab. 1, p. 208]{Foysi_Sarkar_Friedrich_2004a}.
For $(Re_{\tau^\star},\bar M_\tsn{CL})\in\{(150,1.50),(176,0.34)\}$, computations in a smaller box but with higher resolution were run \tabref{Tab_PDTEFCTPCF_s_BEqsDNSC_ss_DNSCs_001}, to verify the large-box computations, and,
in the case $(Re_{\tau^\star},\bar M_\tsn{CL})=(150,1.50)$, to obtain well-resolved statistics for the budgets of the transport equations \parref{PDTEFCTPCF_s_B}.

%
%
%
%
%
\subsection{Transport of thermodynamic variables}\label{PDTEFCTPCF_s_BEqsDNSC_ss_TTV}
%
%
%
%
%

Assuming bivariate thermodynamics \citep{Liepmann_Roshko_1957a,
                                         Kestin_1979a},
the knowledge of the transport equations for 2 thermodynamic variables suffices to obtain the transport equations for all of the others, using thermodynamic derivatives \citep{Bridgman_1961a,
                                                                                                                                                                               Kestin_1979a}.
The fundamental conservation equations \eqref{Eq_PDTEFCTPCF_s_BEqsDNSC_ss_FM_001} already contain the transport equation for the density $\rho$, which follows from the
continuity equation \eqref{Eq_PDTEFCTPCF_s_BEqsDNSC_ss_FM_001a}. Furthermore they can be combined, by substracting the momentum equation \eqref{Eq_PDTEFCTPCF_s_BEqsDNSC_ss_FM_001b} multiplied by $u_i$ (kinetic energy equation)
from the total energy equation \eqref{Eq_PDTEFCTPCF_s_BEqsDNSC_ss_FM_001c}, to obtain the transport equation for the entropy $s$ \citep{White_1974a,
                                                                                                                                        Sesterhenn_2001a}.
Using thermodynamic derivatives \citep{Bridgman_1961a,
                                       Kestin_1979a},
transport equations for $p$, $T$ and $h$ are readily obtained from the transport equations for $\rho$ and $s$
\begin{subequations}
                                                                                                                                    \label{Eq_PDTEFCTPCF_s_BEqsDNSC_ss_TTV_001}
\begin{alignat}{8}
\dfrac{Dp}{Dt}&\;=\;&\left(\dfrac{\partial p}{\partial\rho}\right)_s\;&\dfrac{D\rho}{Dt}&\;+\;&\left(\dfrac{\partial p}{\partial s}\right)_\rho\;&\dfrac{D s}{Dt}
                                                                                                                                    \label{Eq_PDTEFCTPCF_s_BEqsDNSC_ss_TTV_001a}\\
\dfrac{DT}{Dt}&\;=\;&\left(\dfrac{\partial T}{\partial\rho}\right)_s\;&\dfrac{D\rho}{Dt}&\;+\;&\left(\dfrac{\partial T}{\partial s}\right)_\rho\;&\dfrac{D s}{Dt}
              &\;=\;&\left(\dfrac{\partial T}{\partial   p}\right)_s\;&\dfrac{D   p}{Dt}&\;+\;&\left(\dfrac{\partial T}{\partial s}\right)_p   \;&\dfrac{D s}{Dt}
                                                                                                                                    \label{Eq_PDTEFCTPCF_s_BEqsDNSC_ss_TTV_001b}\\
\dfrac{Dh}{Dt}&\;=\;&\left(\dfrac{\partial h}{\partial\rho}\right)_s\;&\dfrac{D\rho}{Dt}&\;+\;&\left(\dfrac{\partial h}{\partial s}\right)_\rho\;&\dfrac{D s}{Dt}
              &\;=\;&\left(\dfrac{\partial h}{\partial   p}\right)_s\;&\dfrac{D   p}{Dt}&\;+\;&\left(\dfrac{\partial h}{\partial s}\right)_p   \;&\dfrac{D s}{Dt}
                                                                                                                                    \label{Eq_PDTEFCTPCF_s_BEqsDNSC_ss_TTV_001c}
\end{alignat}
\end{subequations}
where $D_t(\cdot):=\partial_t(.)+u_\ell\partial_{x_\ell}(.)$ is the substantial derivative \citep[p.~13]{Pope_2000a}.
For the perfect-gas thermodynamics \eqref{Eq_PDTEFCTPCF_s_BEqsDNSC_ss_FM_001d}
used in the present \tsn{DNS} calculations, the thermodynamic derivatives in \eqref{Eq_PDTEFCTPCF_s_BEqsDNSC_ss_TTV_001}
can be expressed in terms of $R_g$ and $\gamma$ as \citep{Liepmann_Roshko_1957a,
                                                          Bridgman_1961a,
                                                          Kestin_1979a}
\begin{alignat}{10}
\eqref{Eq_PDTEFCTPCF_s_BEqsDNSC_ss_FM_001d}\;\;\Longrightarrow\quad
\left\{\begin{array}{lclclcl} \left(\dfrac{\partial p}{\partial\rho}\right)_s & \;=\; & a^2               =   \gamma\dfrac{p}{\rho}       &\;\quad\qquad\;& \left(\dfrac{\partial p}{\partial s}\right)_\rho & \;=\; & (\gamma-1)\rho T \\     
                              \left(\dfrac{\partial T}{\partial\rho}\right)_s & \;=\; &(\gamma-1)     \dfrac{T}{\rho}                     &\;\quad\qquad\;& \left(\dfrac{\partial T}{\partial s}\right)_\rho & \;=\; & \dfrac{T}{c_v}   \\
                              \left(\dfrac{\partial T}{\partial   p}\right)_s & \;=\; &\dfrac{        1}{\rho c_p}                        &\;\quad\qquad\;& \left(\dfrac{\partial T}{\partial s}\right)_p    &\;=\; &\dfrac{T}{c_p} \\
                              \left(\dfrac{\partial h}{\partial\rho}\right)_s & \;=\; &\dfrac{a^2}{\rho} =   \gamma\dfrac{p}{\rho^2}      &\;\quad\qquad\;& \left(\dfrac{\partial h}{\partial s}\right)_\rho &\;=\; & \gamma T \\
                              \left(\dfrac{\partial h}{\partial   p}\right)_s & \;=\; &\dfrac{1}{\rho}                                    &\;\quad\qquad\;& \left(\dfrac{\partial h}{\partial s}\right)_p    &\;=\; & T \\ \end{array}\right.
                                                                                                                                    \label{Eq_PDTEFCTPCF_s_BEqsDNSC_ss_TTV_002}
\end{alignat}
where $a:=\sqrt{\partial_\rho p}\stackrel{\eqref{Eq_PDTEFCTPCF_s_BEqsDNSC_ss_TTV_002}}{=}\sqrt{\gamma p \rho^{-1}}\stackrel{\eqref{Eq_PDTEFCTPCF_s_BEqsDNSC_ss_FM_001d}}{=}\sqrt{\gamma R_g T}$
is the sound speed. Combining the equations for $D_t\rho$ \eqref{Eq_PDTEFCTPCF_s_BEqsDNSC_ss_FM_001a}
and $D_t s$, obtained by \eqrefsab{Eq_PDTEFCTPCF_s_BEqsDNSC_ss_FM_001b}
                                  {Eq_PDTEFCTPCF_s_BEqsDNSC_ss_FM_001c},
with \eqrefsab{Eq_PDTEFCTPCF_s_BEqsDNSC_ss_TTV_002}
              {Eq_PDTEFCTPCF_s_BEqsDNSC_ss_TTV_001} yields
\begin{subequations}
                                                                                                                                    \label{Eq_PDTEFCTPCF_s_BEqsDNSC_ss_TTV_003}
\begin{alignat}{4}
  \frac{D\rho}{Dt}      &\;=\;&-\rho    \dfrac{\partial u_\ell}{\partial x_\ell}
                                                                                                                                    \label{Eq_PDTEFCTPCF_s_BEqsDNSC_ss_TTV_003a}\\
\rho    \frac{Ds}{Dt}   &\;=\;&
                        &\; \;&\dfrac{1}{T}
                        &\left(\tau_{m\ell}S_{m\ell}-\frac{\partial q_\ell}{\partial x_\ell}\right)
                                                                                                                                    \label{Eq_PDTEFCTPCF_s_BEqsDNSC_ss_TTV_003b}\\
        \frac{Dp}{Dt}   &\;=\;&-\gamma p\dfrac{\partial u_\ell}{\partial x_\ell}
                        &\;+\;&(\gamma-1)
                        &\left(\tau_{m\ell}S_{m\ell}-\frac{\partial q_\ell}{\partial x_\ell}\right)
                                                                                                                                    \label{Eq_PDTEFCTPCF_s_BEqsDNSC_ss_TTV_003c}\\
\rho    \frac{DT}{Dt}   &\;=\;&-\dfrac{1}{c_v}         p\dfrac{\partial u_\ell}{\partial x_\ell}
                        &\;+\;&\dfrac{1}{c_v}
                        &\left(\tau_{m\ell}S_{m\ell}-\frac{\partial q_\ell}{\partial x_\ell}\right)
                                                                                                                                    \label{Eq_PDTEFCTPCF_s_BEqsDNSC_ss_TTV_003d}\\
                        &\;=\;&\dfrac{1}{c_p}\dfrac{Dp}{Dt}
                        &\;+\;&\dfrac{1}{c_p}
                        &\left(\tau_{m\ell}S_{m\ell}-\frac{\partial q_\ell}{\partial x_\ell}\right)
                                                                                                                                    \label{Eq_PDTEFCTPCF_s_BEqsDNSC_ss_TTV_003e}\\
\rho    \frac{Dh}{Dt}   &\;=\;&-\gamma p\dfrac{\partial u_\ell}{\partial x_\ell}
                        &\;+\;&\gamma
                        &\left(\tau_{m\ell}S_{m\ell}-\frac{\partial q_\ell}{\partial x_\ell}\right)
                                                                                                                                    \label{Eq_PDTEFCTPCF_s_BEqsDNSC_ss_TTV_003f}\\
                        &\;=\;&\dfrac{Dp}{Dt}
                        &\;+\;&
                        &\left(\tau_{m\ell}S_{m\ell}-\frac{\partial q_\ell}{\partial x_\ell}\right)
                                                                                                                                    \label{Eq_PDTEFCTPCF_s_BEqsDNSC_ss_TTV_003g}
\end{alignat}
\end{subequations}
These transport equations \eqref{Eq_PDTEFCTPCF_s_BEqsDNSC_ss_TTV_003} are the starting point for the analysis of the dynamics
of the turbulent fluctuations of thermodynamic quantities. The material derivatives of
pressure $p$ \eqref{Eq_PDTEFCTPCF_s_BEqsDNSC_ss_TTV_003c}, temperature $T$ \eqref{Eq_PDTEFCTPCF_s_BEqsDNSC_ss_TTV_003d} and enthalpy $h$ \eqref{Eq_PDTEFCTPCF_s_BEqsDNSC_ss_TTV_003f},
are expressed as the nonlinearly weighted sum of 2 terms, the first containing the dilatation-rate $\Theta:=\partial_{x_\ell}u_\ell$ \eqref{Eq_PDTEFCTPCF_s_BEqsDNSC_ss_FM_001e}
which represents the opposite of the material derivative of relative density-variations of a fluid-particle $-\rho^{-1}D_t\rho$ \eqref{Eq_PDTEFCTPCF_s_BEqsDNSC_ss_TTV_003a},
and the second containing the viscous dissipation/conduction term  $(\tau_{m\ell}S_{m\ell}-\partial_{x_\ell}q_\ell)$ which is responsible for the entropy variation of a fluid-particle
$\rho T D_t s$ \eqref{Eq_PDTEFCTPCF_s_BEqsDNSC_ss_TTV_003b}.
In this respect, density and entropy are used as the independent thermodynamic variables. Therefore, the transport equations \eqref{Eq_PDTEFCTPCF_s_BEqsDNSC_ss_TTV_003} clearly illustrate the coupling
of entropy production $\rho T D_t s$ \eqref{Eq_PDTEFCTPCF_s_BEqsDNSC_ss_TTV_003b} with the substantial derivatives of the other thermodynamic quantities $\{p,\rho,T\}$.           
This coupling may be nonnegligible in regions of the flow where viscous effects \eqref{Eq_PDTEFCTPCF_s_BEqsDNSC_ss_TTV_003b} are important, presumably
in the near-wall region, and is related to the second-order theory of mode interaction \citep{Chu_Kovasznay_1958a}.

Of course, under the assumption of perfect-gas thermodynamics \eqref{Eq_PDTEFCTPCF_s_BEqsDNSC_ss_FM_001d}
the transport equations for temperature \eqref{Eq_PDTEFCTPCF_s_BEqsDNSC_ss_TTV_003d} and enthalpy \eqref{Eq_PDTEFCTPCF_s_BEqsDNSC_ss_TTV_003f} are identical \citep{Liepmann_Roshko_1957a,
                                                                                                                                                                    Bridgman_1961a,
                                                                                                                                                                    Kestin_1979a},
under the constant proportionality coefficient $c_p$ \eqref{Eq_PDTEFCTPCF_s_BEqsDNSC_ss_FM_001d}, \ie $D_t h=c_p D_t T$ \eqref{Eq_PDTEFCTPCF_s_BEqsDNSC_ss_FM_001d}.
Some authors \citep{LeRibault_Friedrich_1997a,
                    Canuto_1997a},
when studying the temperature-variance and turbulent heat-fluxes, prefer using the alternative equations \eqref{Eq_PDTEFCTPCF_s_BEqsDNSC_ss_TTV_003e} for temperature
or \eqref{Eq_PDTEFCTPCF_s_BEqsDNSC_ss_TTV_003g} for enthalpy, which contain the material derivative of pressure $D_t p$ in lieu of the dilatation $\Theta$ \eqrefsab{Eq_PDTEFCTPCF_s_BEqsDNSC_ss_TTV_003d}
                                                                                                                                                                    {Eq_PDTEFCTPCF_s_BEqsDNSC_ss_TTV_003f}.

Recall also that for the thermodynamics \eqref{Eq_PDTEFCTPCF_s_BEqsDNSC_ss_FM_001d} assumed in the present work, integration of \eqref{Eq_PDTEFCTPCF_s_BEqsDNSC_ss_TTV_001b},
using the expressions \eqref{Eq_PDTEFCTPCF_s_BEqsDNSC_ss_TTV_002} for $(\partial_p T)_s$ and $(\partial_s T)_p$,
yields the expression of entropy as a function of $(p,T)$ or $(\rho,T)$
\begin{alignat}{4}
\eqref{Eq_PDTEFCTPCF_s_BEqsDNSC_ss_FM_001d}\Longrightarrow\dfrac{s-s_\tsn{ISA}}{R_g}=\dfrac{\gamma-1}{\gamma}\ln\dfrac{T}{T_\tsn{ISA}}-\ln\dfrac{p}{p_\tsn{ISA}}
                                                                                    =\dfrac{1}{\gamma-1}\ln\dfrac{T}{T_\tsn{ISA}}-\ln\dfrac{\rho}{\rho_\tsn{ISA}}       
                                                                                                                                    \label{Eq_PDTEFCTPCF_s_BEqsDNSC_ss_TTV_004}
\end{alignat}
with respect to an arbitrary reference condition $(\cdot)_\tsn{ISA}$ (which was chosen here as the sea-level conditions of the international standard atmosphere, $T_\tsn{ISA} = 288.15\;\mathrm{K}$ and $p_\tsn{ISA} = 101325\;\mathrm{Pa}$),
implying by \eqref{Eq_PDTEFCTPCF_s_BEqsDNSC_ss_FM_001d} $\rho_\tsn{ISA}\approxeq 1.225055$.

%
%
%
%
%
\subsection{Statistics and averaging}\label{PDTEFCTPCF_s_BEqsDNSC_ss_SA}
%
%
%
%
%

Using \citep[(2.1), p. 188]{Huang_Coleman_Bradshaw_1995a} Reynolds-decomposition for $p$ and $\rho$
and Favre-decomposition for $u_i$, $s$, $T$ and $h$, in \eqrefsab{Eq_PDTEFCTPCF_s_BEqsDNSC_ss_FM_001b}
                                                                 {Eq_PDTEFCTPCF_s_BEqsDNSC_ss_TTV_003},
we can work out transport equations for the variances ($\overline{\rho'^2}$, $\overline{p'^2}$, $\overline{\rho s''^2}$ and $\overline{\rho h''^2}\stackrel{\eqref{Eq_PDTEFCTPCF_s_BEqsDNSC_ss_FM_001d}}{=}c_p\;\overline{\rho T''^2}$)
and the fluxes ($\overline{\rho'u_i'}$, $\overline{p'u_i'}$, $\overline{\rho s''u_i''}$ and $\overline{\rho h''u_i''}\stackrel{\eqref{Eq_PDTEFCTPCF_s_BEqsDNSC_ss_FM_001d}}{=}c_p\;\overline{\rho T''u_i''}$)
of the turbulent fluctuations of the thermodynamic state-variables.
The algebra for the development of the equations albeit straightforward is quite lengthy,
but can be formulated in a systematic way \citep{Gerolymos_Vallet_JFluidMech_01_news} which reduces analytical effort.

%
\subsubsection{Reynolds and Favre decomposition}\label{PDTEFCTPCF_s_BEqsDNSC_ss_SA_sss_RFD}
%

The usual decompositions of the fluctuating quantities \citep[(2.1), p. 188]{Huang_Coleman_Bradshaw_1995a} are used
\begin{subequations}
                                                                                                                                    \label{Eq_PDTEFCTPCF_s_BEqsDNSC_ss_SA_sss_RFD_001}
\begin{alignat}{4}
(\cdot)=\overline{(\cdot)}+(\cdot)'=\widetilde{(\cdot)}+(\cdot)''
                                                                                                                                    \label{Eq_PDTEFCTPCF_s_BEqsDNSC_ss_SA_sss_RFD_001a}
\end{alignat}
consistently throughout the paper. In \eqref{Eq_PDTEFCTPCF_s_BEqsDNSC_ss_SA_sss_RFD_001a}, $(\cdot)$ denotes any flow quantity,
$\overline{(\cdot)}$ denotes Reynolds (ensemble) averaging with fluctuations $(\cdot)'$ satisfying
\begin{alignat}{4}
\overline{(\cdot)'}=0
                                                                                                                                    \label{Eq_PDTEFCTPCF_s_BEqsDNSC_ss_SA_sss_RFD_001b}
\end{alignat}
and $\widetilde{(\cdot)}$ denotes Favre averaging with fluctuations $(\cdot)''$ defined by
\begin{alignat}{4}
\left.
\begin{array}{lcl}
\widetilde{(\cdot)}&:=&\dfrac{1}{\bar\rho}\;\overline{\rho\;(\cdot)}\\
                                                                    \\
(\cdot)''          &:=&(\cdot)-\widetilde{(\cdot)}                  \\\end{array}\right\}
\stackrel{\eqrefsab{Eq_PDTEFCTPCF_s_BEqsDNSC_ss_SA_sss_RFD_001a}
                   {Eq_PDTEFCTPCF_s_BEqsDNSC_ss_SA_sss_RFD_001b}}{\Longrightarrow}\left\{\begin{array}{lcl}\overline{\rho\;(\cdot)''}&=&0                                                                                      \\
                                                                                                                                                                                                                             \\
                                                                                                           \overline{(\cdot)''}    &=&-\dfrac{1}{\bar\rho}\;\overline{\rho'\;(\cdot)'}=\overline{(\cdot)}-\widetilde{(\cdot)}\\
                                                                                                                                                                                                                             \\
                                                                                                           (\cdot)''               &=&(\cdot)'+\overline{(\cdot)''}                                                          \\\end{array}\right.
                                                                                                                                    \label{Eq_PDTEFCTPCF_s_BEqsDNSC_ss_SA_sss_RFD_001c}
\end{alignat}
Recall also that any 2 flow quantities $[\cdot]$ and $(\cdot)$ satisfy the important identity \citep{Sarkar_Erlebacher_Hussaini_Kreiss_1991a,
                                                                                                     Canuto_1997a}
\begin{alignat}{4}
\overline{[\cdot]'\;(\cdot)''}=\overline{[\cdot]''\;(\cdot)'}=\overline{[\cdot]'\;(\cdot)'}
                                                                                                                                    \label{Eq_PDTEFCTPCF_s_BEqsDNSC_ss_SA_sss_RFD_001d}
\end{alignat}
which is readily obtained
by replacing \smash{$(\cdot)''\stackrel{\eqref{Eq_PDTEFCTPCF_s_BEqsDNSC_ss_SA_sss_RFD_001c}}{=}(\cdot)'+\overline{(\cdot)''}$}
and \smash{$[\cdot]''\stackrel{\eqref{Eq_PDTEFCTPCF_s_BEqsDNSC_ss_SA_sss_RFD_001c}}{=}[\cdot]'+\overline{[\cdot]''}$} in \eqref{Eq_PDTEFCTPCF_s_BEqsDNSC_ss_SA_sss_RFD_001d}.
\end{subequations}

%
\subsubsection{Need for additional notation}\label{PDTEFCTPCF_s_BEqsDNSC_ss_SA_sss_NAN}
%

Favre averaging defined by \eqref{Eq_PDTEFCTPCF_s_BEqsDNSC_ss_SA_sss_RFD_001c} does not commute with differentiation, contrary to Reynolds averaging,
because
\begin{subequations}
                                                                                                                                    \label{Eq_PDTEFCTPCF_s_BEqsDNSC_ss_SA_sss_NAN_001}
\begin{alignat}{4}
\widetilde{\left(\dfrac{\partial(\cdot)}{\partial x_j}\right)}&\stackrel{\eqref{Eq_PDTEFCTPCF_s_BEqsDNSC_ss_SA_sss_RFD_001c}}{:=}&
\dfrac{1}{\bar\rho}\overline{\left(\rho\dfrac{\partial(\cdot)}{\partial x_j}\right)}\stackrel{\eqref{Eq_PDTEFCTPCF_s_BEqsDNSC_ss_SA_sss_RFD_001a}}{:=}
\dfrac{1}{\bar\rho}\overline{\left(\rho\dfrac{\partial\widetilde{(\cdot)}}{\partial x_j}
                                  +\rho\dfrac{\partial(\cdot)''}{\partial x_j}\right)}&=&
\dfrac{\partial\widetilde{(\cdot)}}{\partial x_j}+\dfrac{1}{\bar\rho}\overline{\left(\rho\dfrac{\partial(\cdot)''}{\partial x_j}\right)}
                                                                                                                                    \label{Eq_PDTEFCTPCF_s_BEqsDNSC_ss_SA_sss_NAN_001a}\\
&=&
\dfrac{\partial\widetilde{(\cdot)}}{\partial x_j}+\dfrac{1}{\bar\rho}\left(\dfrac{\partial\overline{\rho\;\;(\cdot)''}}{\partial x_j}
                                                                          -\overline{(\cdot)''\;\dfrac{\partial\rho}{\partial x_j}}\right)&\stackrel{\eqref{Eq_PDTEFCTPCF_s_BEqsDNSC_ss_SA_sss_RFD_001c}}{:=}&
\dfrac{\partial\widetilde{(\cdot)}}{\partial x_j}-\overline{\dfrac{(\cdot)''}{\bar\rho}\dfrac{\partial\rho}{\partial x_j}}
                                                                                                                                    \label{Eq_PDTEFCTPCF_s_BEqsDNSC_ss_SA_sss_NAN_001b}
\end{alignat}
In the same way, the Favre-fluctuation operator $(\cdot)''$ \eqref{Eq_PDTEFCTPCF_s_BEqsDNSC_ss_SA_sss_RFD_001c} does not commute with differentiation, and we have by straightforward computation
\begin{alignat}{4}
\dfrac{\partial(\cdot)''}{\partial x_j}\stackrel{\eqref{Eq_PDTEFCTPCF_s_BEqsDNSC_ss_SA_sss_RFD_001a}}{=}\dfrac{\partial(\cdot)}{\partial x_j}
                                                                                                       -\dfrac{\partial\widetilde{(\cdot)}}{\partial x_j}
\neq\left(\dfrac{\partial(\cdot)}{\partial x_j}\right)''\stackrel{\eqref{Eq_PDTEFCTPCF_s_BEqsDNSC_ss_SA_sss_RFD_001c}}{:=}&\dfrac{\partial(\cdot)}{\partial x_j}-\widetilde{\left(\dfrac{\partial(\cdot)}{\partial x_j}\right)}
                                                                                                                                  \notag\\
                                                        \stackrel{\eqrefsab{Eq_PDTEFCTPCF_s_BEqsDNSC_ss_SA_sss_RFD_001a}
                                                                           {Eq_PDTEFCTPCF_s_BEqsDNSC_ss_SA_sss_NAN_001a}}{=}&\dfrac{\partial(\cdot)''}{\partial x_j}-\dfrac{1}{\bar\rho}\overline{\left(\rho\dfrac{\partial(\cdot)''}{\partial x_j}\right)}
                                                                                                                                    \label{Eq_PDTEFCTPCF_s_BEqsDNSC_ss_SA_sss_NAN_001c}\\
                                                        \stackrel{\eqrefsab{Eq_PDTEFCTPCF_s_BEqsDNSC_ss_SA_sss_RFD_001a}
                                                                           {Eq_PDTEFCTPCF_s_BEqsDNSC_ss_SA_sss_NAN_001b}}{=}&\dfrac{\partial(\cdot)''}{\partial x_j} +\overline{\dfrac{(\cdot)''}{\bar\rho}\dfrac{\partial\rho}{\partial x_j}}
                                                                                                                                    \label{Eq_PDTEFCTPCF_s_BEqsDNSC_ss_SA_sss_NAN_001d}
\end{alignat}
\end{subequations}
Often in the development of transport equations appear terms of the form $\partial_{x_j}\widetilde{(\cdot)}$ or $\partial_{x_j}(\cdot)''$, especially for the strain-rate tensor $S_{ij}$ \eqref{Eq_PDTEFCTPCF_s_BEqsDNSC_ss_FM_001e}
or for the dilatation-rate $\Theta:=\partial_{x_\ell}u_\ell=S_{\ell\ell}$ \eqref{Eq_PDTEFCTPCF_s_BEqsDNSC_ss_FM_001e}. To avoid the inconsistent, yet widespread, notation $\tilde S_{ij}$ or $S_{ij}''$ ($\tilde\Theta$ or $\Theta''$) for such terms
it is useful to introduce a new decomposition
\begin{subequations}
                                                                                                                                    \label{Eq_PDTEFCTPCF_s_BEqsDNSC_ss_SA_sss_NAN_002}
\begin{alignat}{8}
S_{ij}&=&&\breve{S}_{ij}+S_{ij}^{\backprime\backprime}&\quad;\quad&
\breve{S}_{ij}&:=&\tfrac{1}{2}\left(\dfrac{\partial\tilde{u}_i}{\partial x_j}
                                   +\dfrac{\partial\tilde{u}_j}{\partial x_i}\right)&\quad;\quad&
S_{ij}^{\backprime\backprime}&:=&\tfrac{1}{2}\left(\dfrac{\partial u_i''}{\partial x_j}
                                                  +\dfrac{\partial u_j''}{\partial x_i}\right)
                                                                                                                                    \label{Eq_PDTEFCTPCF_s_BEqsDNSC_ss_SA_sss_NAN_002a}\\
\Theta&=&&\breve{\Theta}+\Theta^{\backprime\backprime}&\quad;\quad&
\breve{\Theta}&:=&\dfrac{\partial\tilde{u}_\ell}{\partial x_\ell}&\quad;\quad&
\Theta^{\backprime\backprime}&:=&\dfrac{\partial u''_\ell}{\partial x_\ell}
                                                                                                                                    \label{Eq_PDTEFCTPCF_s_BEqsDNSC_ss_SA_sss_NAN_002b}
\end{alignat}
Quantities $\breve{(\cdot)}$ are not averages, but functions of averaged quantities \citep{Gerolymos_Vallet_1996a}, and are different from Favre averages, because by \eqref{Eq_PDTEFCTPCF_s_BEqsDNSC_ss_SA_sss_NAN_001b}
\begin{alignat}{4}
\tilde{S}_{ij}&\stackrel{\eqrefsabc{Eq_PDTEFCTPCF_s_BEqsDNSC_ss_FM_001e}
                                  {Eq_PDTEFCTPCF_s_BEqsDNSC_ss_SA_sss_NAN_001b}
                                  {Eq_PDTEFCTPCF_s_BEqsDNSC_ss_SA_sss_NAN_002a}}{=}&&\breve{S}_{ij}-\tfrac{1}{2}\left(\overline{\dfrac{u_i''}{\bar\rho}\dfrac{\partial\rho}{\partial x_j}}
                                                                                                                     +\overline{\dfrac{u_j''}{\bar\rho}\dfrac{\partial\rho}{\partial x_i}}\right)
                                                                                                                                    \label{Eq_PDTEFCTPCF_s_BEqsDNSC_ss_SA_sss_NAN_002c}\\
\tilde{\Theta}&\stackrel{\eqrefsab{Eq_PDTEFCTPCF_s_BEqsDNSC_ss_SA_sss_NAN_001b}
                                  {Eq_PDTEFCTPCF_s_BEqsDNSC_ss_SA_sss_NAN_002a}}{=}&&\breve{\Theta}-\overline{\dfrac{u_\ell''}{\bar\rho}\dfrac{\partial\rho}{\partial x_\ell}}
                                                                                                                                    \label{Eq_PDTEFCTPCF_s_BEqsDNSC_ss_SA_sss_NAN_002d}
\end{alignat}
Finally, notice that identity \eqref{Eq_PDTEFCTPCF_s_BEqsDNSC_ss_SA_sss_RFD_001d} also holds for $S_{ij}^{\backprime\backprime}$ and $\Theta^{\backprime\backprime}$, \ie
\begin{alignat}{4}
\overline{(.)'[.]^{\backprime\backprime}}=\overline{(.)^{\backprime\backprime}[.]'}=\overline{(.)'[.]'}\qquad;\qquad
\overline{(\cdot)'S_{ij}^{\backprime\backprime}}=\overline{(\cdot)'S_{ij}'}\qquad;\qquad\overline{(\cdot)'\Theta^{\backprime\backprime}}=\overline{(\cdot)'\Theta'}
                                                                                                                                    \label{Eq_PDTEFCTPCF_s_BEqsDNSC_ss_SA_sss_NAN_002d}
\end{alignat}
as can be verified by replacing \smash{$S_{ij}^{\backprime\backprime}\stackrel{\eqrefsab{Eq_PDTEFCTPCF_s_BEqsDNSC_ss_SA_sss_RFD_001a}
                                                                                        {Eq_PDTEFCTPCF_s_BEqsDNSC_ss_SA_sss_NAN_002a}}{=}\bar{S}_{ij}-\breve{S}_{ij}+S_{ij}'$}
and \smash{$\Theta^{\backprime\backprime}\stackrel{\eqrefsab{Eq_PDTEFCTPCF_s_BEqsDNSC_ss_SA_sss_RFD_001a}
                                                            {Eq_PDTEFCTPCF_s_BEqsDNSC_ss_SA_sss_NAN_002b}}{=}\bar{\Theta}-\breve{\Theta}+\Theta'$} in \eqref{Eq_PDTEFCTPCF_s_BEqsDNSC_ss_SA_sss_NAN_002d}.
\end{subequations}

%
%
%
%
%
%
%
%
%
\section{Scalings}\label{PDTEFCTPCF_s_S}
%
%
%
%
%
%
%
%
%

In addition to $Re$-scaling which uniquely defines incompressible plane channel flow \citep{Moser_Kim_Mansour_1999a},
$M$-scaling with respect to a representative Mach number is necessary in the compressible flow case \citep{Coleman_Kim_Moser_1995a,
                                                                                                            Huang_Coleman_Bradshaw_1995a}.
%
%
%
%
%
\subsection{Compressible channel flow}\label{PDTEFCTPCF_s_S_ss_CCF}
%
%
%
%
%
In \citet{Coleman_Kim_Moser_1995a}, the \tsn{DNS} operating point was defined by a bulk Reynolds number $Re_{\tsn{B}_w}$ and a bulk Mach number $M_{\tsn{B}_w}$, defined as
\begin{subequations}
                                                                                                                                    \label{Eq_PDTEFCTPCF_s_S_ss_CCF_001}
\begin{align}
Re_{\tsn{B}_w}:=\dfrac{\bar\rho_\tsn{B}\bar{u}_\tsn{B}\delta}
                      {\bar\mu_w                            }
\qquad;\qquad
M_{\tsn{B}_w}:=\dfrac{\bar{u}_\tsn{B}}
                     {\bar a_w       }
                                                                                                                                    \label{Eq_PDTEFCTPCF_s_S_ss_CCF_001a}
\end{align}
where $\bar\mu_w$ is the dynamic viscosity at the wall, $\bar a_w$ is the sound-speed at the wall (in the isothermal wall case studied here, $T_w=\bar T_w=\breve T_w=\const$,
so that $\bar\mu_w\stackrel{\eqref{Eq_PDTEFCTPCF_s_BEqsDNSC_ss_FM_001g}}{=}\mu(T_w)$ and $\bar a_w\stackrel{\eqref{Eq_PDTEFCTPCF_s_BEqsDNSC_ss_FM_001d}}{=}a(T_w)$),
and the subscript $(\cdot)_\tsn{B}$ denotes volume-averaging over the entire computational domain \citep[(46), p. 791]{Gerolymos_Senechal_Vallet_2010a}.
\citet{Huang_Coleman_Bradshaw_1995a} found that the standard wall-coordinates and friction-Reynolds-number
\begin{align}
y^+:=\dfrac{\bar\rho_w\sqrt{\dfrac{\bar\tau_w}{\bar\rho_w}}(y-y_w)}{\bar\mu_w}
\qquad;\qquad
Re_{\tau_w}:=\delta^+
                                                                                                                                    \label{Eq_PDTEFCTPCF_s_S_ss_CCF_001b}
\end{align}
did not correctly represent the effect of rapid wall-normal variation $\bar\rho(y)$ and $\bar\mu(y)$ near the wall, and did a poor job in collapsing profiles of different variables.
They suggested instead an empirical mixed scaling
\begin{align}
y^\star:=\dfrac{\bar\rho(y)\sqrt{\dfrac{\bar\tau_w}{\bar\rho(y)}}(y-y_w)}{\bar\mu(y)}\stackrel{\eqref{Eq_PDTEFCTPCF_s_S_ss_CCF_001b}}{=}\dfrac{\sqrt{\dfrac{\bar\rho(y)}{\bar\rho_w}}}{\dfrac{\bar\mu(y)}{\bar\mu_w}}y^+
\qquad;\qquad
Re_{\tau^\star}:=\delta^\star=\dfrac{\sqrt{\dfrac{\bar\rho_\tsn{CL}}{\bar\rho_w}}}{\dfrac{\bar\mu_\tsn{CL}}{\bar\mu_w}}\delta^+
                                                                                                                                    \label{Eq_PDTEFCTPCF_s_S_ss_CCF_001c}
\end{align}
that we will call hereafter \tsn{HCB}-scaling \citep{Huang_Coleman_Bradshaw_1995a}, and which has proven useful in approximately collapsing the profiles of the Reynolds-stresses \citep{Morinishi_Tamano_Nakabayashi_2004a,
                                                                                                                                                                                         Foysi_Sarkar_Friedrich_2004a}
for different values of $(Re_{\tsn{B}_w},M_{\tsn{B}_w})$.
\begin{figure}
\begin{center}
\begin{picture}(450,280)
\put(-70,-225){\includegraphics[angle=0,width=525pt]{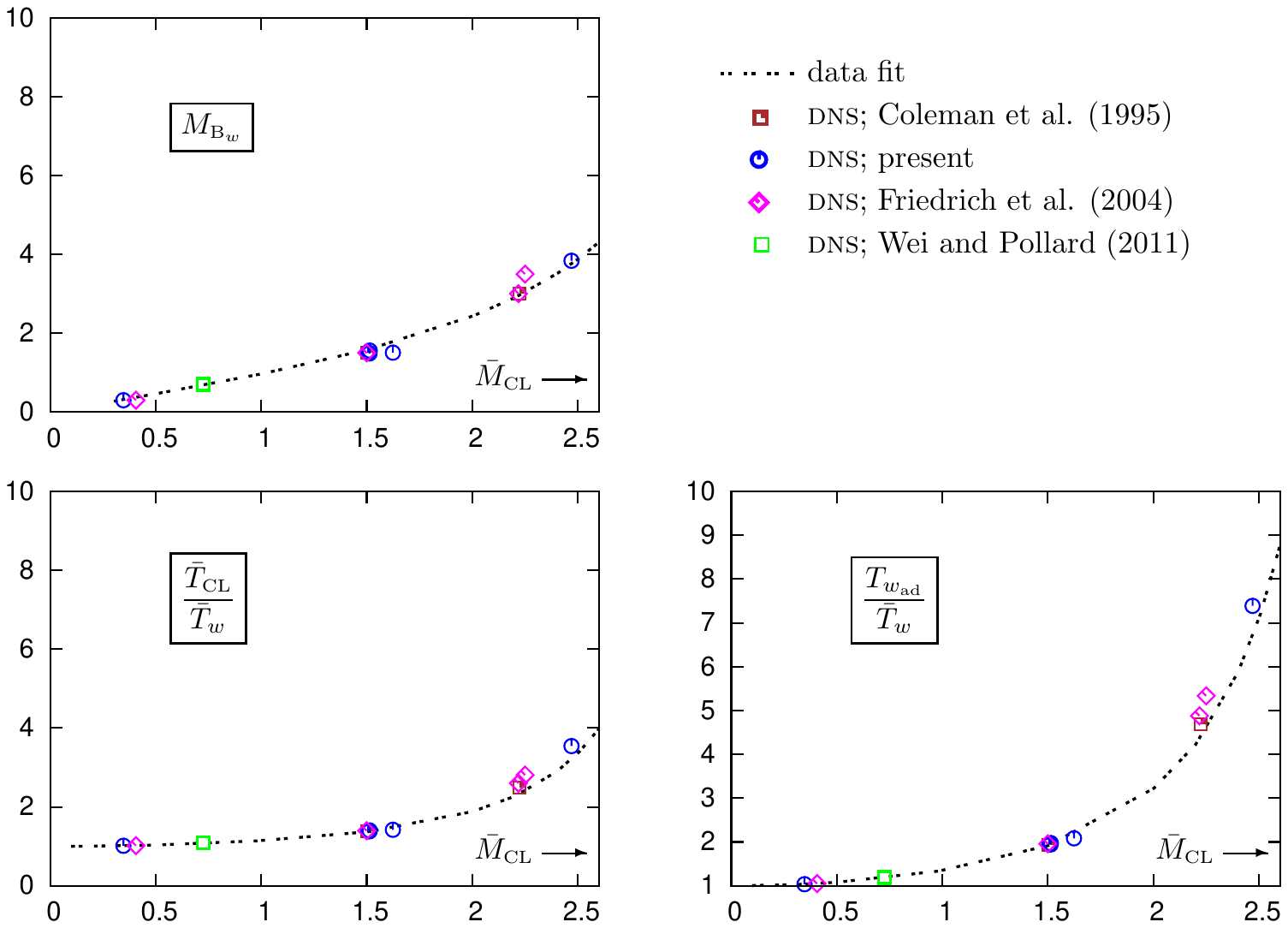}}
\end{picture}
\end{center}
\caption{Evolution of bulk Mach number $M_{\tsn{B}_w}$ \eqref{Eq_PDTEFCTPCF_s_S_ss_CCF_001a}, ratio of centerline-to-wall static temperatures,
and wall-to-theoretical-adiabatic-wall temperatures \eqref{Eq_PDTEFCTPCF_s_S_ss_CCF_001e}, as a function of centerline Mach number $\bar M_\tsn{CL}$ \eqref{Eq_PDTEFCTPCF_s_S_ss_CCF_001d},
for various Reynolds numbers, using the present large-box \tsn{DNS} results \tabref{Tab_PDTEFCTPCF_s_S_ss_CCF_001}
and other available \tsn{DNS} data \citep{Coleman_Kim_Moser_1995a,
                                          Friedrich_Foysi_Sesterhenn_2006a,
                                          Wei_Pollard_2011a,
                                          Wei_Pollard_2011b}.}
\label{Fig_PDTEFCTPCF_s_S_ss_CCF_001}
\end{figure}
%
\begin{table}
\begin{center}
\input{JFluidMech_01_Tab_DNS_global_parameters}
\caption{Global nondimensional parameters of the large-box ($L_x\times L_y\times L_z=8\pi\delta\times2\delta\times4\pi\delta$) computations \tabref{Tab_PDTEFCTPCF_s_BEqsDNSC_ss_DNSCs_001}
         [$Re_{\tau_\star}$ is the friction Reynolds number in \tsn{HCB}-scaling \eqref{Eq_PDTEFCTPCF_s_S_ss_CCF_001c};
          $\bar M_\tsn{CL}$ \eqref{Eq_PDTEFCTPCF_s_S_ss_CCF_001d} is the centerline Mach number;
          $M_{\tsn{B}_w}$ \eqref{Eq_PDTEFCTPCF_s_S_ss_CCF_001a} is the bulk Mach number;
          $M_{\tau_w}:=\bar a_w^{-1} u_\tau$ is the friction Mach number based on $u_\tau$ \eqref{Eq_PDTEFCTPCF_s_S_ss_WUs_001b} and sound-velocity at wall conditions;
          $Re_{\tau_w}:=\bar\rho_w u_\tau \delta\;\bar\mu_w^{-1}$ \eqref{Eq_PDTEFCTPCF_s_S_ss_CCF_001b} is the friction Reynolds number;
          $Re_{\tsn{B}_w}$ \eqref{Eq_PDTEFCTPCF_s_S_ss_CCF_001a} is the bulk Reynolds number;
          $Re_{\theta_w}:=\bar\rho_w\bar u_\tsn{CL}\theta\;\bar\mu_w^{-1}$ is the momentum-thickness ($\theta$) Reynolds number at wall conditions;
          $Re_{\theta_\tsn{CL}}:=\bar\rho_\tsn{CL}\bar u_\tsn{CL}\theta\;\bar\mu_\tsn{CL}^{-1}$ is the momentum-thickness Reynolds number at centerline conditions;
          $Re_{\theta_{\tsn{CL}w}}:=\bar\rho_\tsn{CL}\bar u_\tsn{CL}\theta\;\bar\mu_w^{-1}$ is the momentum-thickness Reynolds number with mixed scaling ($\bar\rho_\tsn{CL}$ and $\bar\mu_w$);
          $\bar T_\tsn{CL}$ is the centerline temperature;
          $\bar T_w$ is the wall temperature;
          $B_{q_w}:=(\bar\rho_w u_\tau h_w)^{-1}\bar q_w$ is the nondimensional wall heat-flux \citep{Coleman_Kim_Moser_1995a};
          $\max_y[\bar\rho^{-1}\;\rho'_\mathrm{rms}]$ is the maximum rms level of relative density fluctuations].}
\label{Tab_PDTEFCTPCF_s_S_ss_CCF_001}
\end{center}
\end{table}

Since the wall is colder than the centerline region ($\bar T_\tsn{CL}>\bar T_w$) in the compressible plane channel case \citep[Tab. 3, p. 169]{Coleman_Kim_Moser_1995a},
the centerline Mach number
\begin{align}
\bar M_\tsn{CL}:=\overline{\left(\dfrac{u_\tsn{CL}}{a_\tsn{CL}}\right)}
\qquad;\qquad
\breve M_\tsn{CL}:=\dfrac{\tilde u_\tsn{CL}}{\breve a_\tsn{CL}}
                                                                                                                                    \label{Eq_PDTEFCTPCF_s_S_ss_CCF_001d}
\end{align}
where \smash{$\breve a_\tsn{CL}\stackrel{\eqref{Eq_PDTEFCTPCF_s_BEqsDNSC_ss_FM_001d}}{:=}\sqrt{\gamma R_g\tilde T_\tsn{CL}}$},
is systematically smaller than the bulk Mach number $M_{\tsn{B}_w}$ \eqref{Eq_PDTEFCTPCF_s_S_ss_CCF_001a}, mainly \tabref{Tab_PDTEFCTPCF_s_S_ss_CCF_001} because of the difference in the sound-speeds appearing in \eqrefsab{Eq_PDTEFCTPCF_s_S_ss_CCF_001a}
                                                                                                                                                                                                                              {Eq_PDTEFCTPCF_s_S_ss_CCF_001d}.
Both defintions in \eqref{Eq_PDTEFCTPCF_s_S_ss_CCF_001d} can be used and are numerically very close.

The nondimensional form \citep[(3), p. 161]{Coleman_Kim_Moser_1995a} of the static temperature equation \eqref{Eq_PDTEFCTPCF_s_BEqsDNSC_ss_TTV_003e} suggests that viscous heating
$\overline{\tau_{m\ell}S_{m\ell}}$ increases with Mach number, thus increasing the centerline temperature $\bar T_\tsn{CL}$ compared
to wall-temperature $\bar T_w$ \figref{Fig_PDTEFCTPCF_s_S_ss_CCF_001}. This increase is nonlinear with $\bar M_\tsn{CL}$ and the ratio $[\bar T_w^{-1}\;\bar T_\tsn{CL}]$ rises sharply
when $\bar M_\tsn{CL}\gtrapprox2$ \figref{Fig_PDTEFCTPCF_s_S_ss_CCF_001}.
Since the sound speed varies as $\sqrt{T}$ \eqref{Eq_PDTEFCTPCF_s_BEqsDNSC_ss_FM_001d},
the ratio of $\breve a_\tsn{CL}$ to $\bar a_w$ increases, and so does the ratio of $M_{\tsn{B}_w}$ \eqref{Eq_PDTEFCTPCF_s_S_ss_CCF_001a}
to $\bar M_\tsn{CL}$ \eqref{Eq_PDTEFCTPCF_s_S_ss_CCF_001d}, nonlinearly with $\bar M_\tsn{CL}$ \figref{Fig_PDTEFCTPCF_s_S_ss_CCF_001}.
Thus, for the present computations, $\bar M_\tsn{CL}\approxeq2.47$ corresponds to $M_{\tsn{B}_w}\approxeq3.83$ \figref{Fig_PDTEFCTPCF_s_S_ss_CCF_001}.
More carefull examination of the data \tabref{Tab_PDTEFCTPCF_s_S_ss_CCF_001} indicates also a slight dependence on Reynolds number.
Therefore $M$-scalings with $M_{\tsn{B}_w}$ \eqref{Eq_PDTEFCTPCF_s_S_ss_CCF_001a} or $\bar M_\tsn{CL}$ \eqref{Eq_PDTEFCTPCF_s_S_ss_CCF_001d} are not equivalent,
and, for instance, the coefficients of variation of thermodynamic fluctuations ($[\bar\rho^{-1} \rho']_{\rm rms}$, $[\bar p^{-1} p']_{\rm rms}$ and $[\bar T^{-1} T']_{\rm rms}$)
may scale with some power of either $\bar M_{\tsn{B}_w}$ or $\bar M_\tsn{CL}$, but not both.
It will be shown in the following \parref{PDTEFCTPCF_s_S_ss_HCBMCL} that, contrary to usual practice \citep{Huang_Coleman_Bradshaw_1995a,
                                                                                                            Friedrich_2007a},
the centerline Mach number $\bar M_\tsn{CL}$ scales better the data.

Obviously, the wall is colder than the centerline region \figref{Fig_PDTEFCTPCF_s_S_ss_CCF_001}, as was the case of the boundary-layer computations of \citet[Tab. 2, p. 31]{Duan_Martin_2011a}.
However, in the the plane channel case, the viscous heating that is responsible for the high temperatures in the centerline region is part of the solution, and temperature ratio $\bar T_w^{-1}\;\bar T_\tsn{CL}$
increases \tabref{Tab_PDTEFCTPCF_s_S_ss_CCF_001}, the wall getting colder relative to the centerline region, with increasing $\bar M_\tsn{CL}$. This can be better quantified by comparing $\bar T_w$ with
the theoretical adiabatic wall temperature \citep[(23.35--23.36), pp. 713--714]{Schlichting_1979a}
\begin{align}
T_{w_\mathrm{ad}}:=\bar T_\tsn{CL}\left(1+r_f\;\dfrac{\gamma-1}{2}\bar M^2_\tsn{CL}\right)
\qquad;\qquad
r_f:=0.89
                                                                                                                                    \label{Eq_PDTEFCTPCF_s_S_ss_CCF_001e}
\end{align}
where the value of the recovery factor $r_f$ \eqref{Eq_PDTEFCTPCF_s_S_ss_CCF_001e} is appropriate for airflow. The ratio $\bar T_w^{-1} T_{w_\mathrm{ad}}$ increases rapidly with $\bar M_\tsn{CL}$,
from $\bar T_w^{-1} T_{w_\mathrm{ad}}\approxeq2$ at $\bar M_\tsn{CL}=1.5$ to $\bar T_w^{-1} T_{w_\mathrm{ad}} \approxeq 7.5$ at $\bar M_\tsn{CL}=2.5$, which corresponds to the boundary-layer temperature ratios studied in \citet[Tab. 2, p. 31]{Duan_Martin_2011a}.
\end{subequations}
Notice that all of the \tsn{DNS} computations, reported in the present work, apply the same constant wall-temperature $T_w=\bar{T}_w=\tilde{T}_w=298 $K \tabref{Tab_PDTEFCTPCF_s_S_ss_CCF_001}, so that the highest local temperature
$\bar{T}_{\tsn{CL}}(Re_{\tau^\star}=111,\bar M_\tsn{CL}=2.47)\approxeq 1055$ K is far from dissociation limits \citep[Fig.~1, p.~57]{Hansen_1958a} which would invalidate the perfect-gas equation-of-state \eqref{Eq_PDTEFCTPCF_s_BEqsDNSC_ss_FM_001d} approximation.
On the other hand, the variation of $c_p(T)$ in this temperature range $\bar{T}\in [298,1055]$ is $\sim 10\%$ \citep[pp.~64--66, 780]{Eckert_Drake_1972a} and should be taken into account in future studies.

%
%
%
%
%
\subsection{Wall-units}\label{PDTEFCTPCF_s_S_ss_WUs}
%
%
%
%
%

In line with standard incompressible \citep{Kim_Moin_Moser_1987a} and compressible \citep{Coleman_Kim_Moser_1995a} wall-bounded flow analysis,
the independent variables, in wall-units, are defined as
\begin{subequations}
                                                                                                                                    \label{Eq_PDTEFCTPCF_s_S_ss_WUs_001}
\begin{alignat}{6}
t^+:=\dfrac{t}{\left(\dfrac{\breve\nu_w}{u_\tau^2}\right)}
\qquad;\qquad
x_i^+:=\dfrac{x_i-\delta_{iy}y_w}{\left(\dfrac{\breve\nu_w}{u_\tau}\right)}
                                                                                                                                    \label{Eq_PDTEFCTPCF_s_S_ss_WUs_001a}
\end{alignat}
where the term $\delta_{iy}y_w$ serves to position the wall at $y^+=0$, and
\begin{alignat}{6}
u_\tau:=\sqrt{\dfrac{\bar\tau_w}{\bar\rho_w}}
\qquad;\qquad
\breve\nu_w:=\dfrac{\bar\mu_w}{\bar\rho_w}
                                                                                                                                    \label{Eq_PDTEFCTPCF_s_S_ss_WUs_001b}
\end{alignat}
A consistent scaling is chosen for the nondimensionalisation of the thermodynamic variables,\ie one for which
the flow equations \eqrefsab{Eq_PDTEFCTPCF_s_BEqsDNSC_ss_FM_001}
                            {Eq_PDTEFCTPCF_s_BEqsDNSC_ss_TTV_003}
retain exactly the same form when written in wall-units, without the appearance of nondimensional numbers (Mach, Reynolds, $\cdots$). Therefore,
by \eqrefsatob{Eq_PDTEFCTPCF_s_BEqsDNSC_ss_FM_001a}
              {Eq_PDTEFCTPCF_s_BEqsDNSC_ss_FM_001c},
\begin{alignat}{6}
u_i^+:=\dfrac{u_i}{u_\tau}
\;\;;\;\;
\rho^+:=\dfrac{\rho}{\bar\rho_w}
\;\;;\;\;
p^+:=\dfrac{p}{\bar\tau_w}
\;\;;\;\;
\tau_{ij}^+:=\dfrac{\tau_{ij}}{\bar\tau_w}
\;\;;\;\;
h^+:=\dfrac{h}{u_\tau^2}
\;\;;\;\;
q_i^+:=\dfrac{q_i}{\bar\rho_w u_\tau^3}
                                                                                                                                    \label{Eq_PDTEFCTPCF_s_S_ss_WUs_001c}
\end{alignat}
\begin{figure}
\begin{center}
\begin{picture}(450,440)
\put(-65,-125){\includegraphics[angle=0,width=475pt]{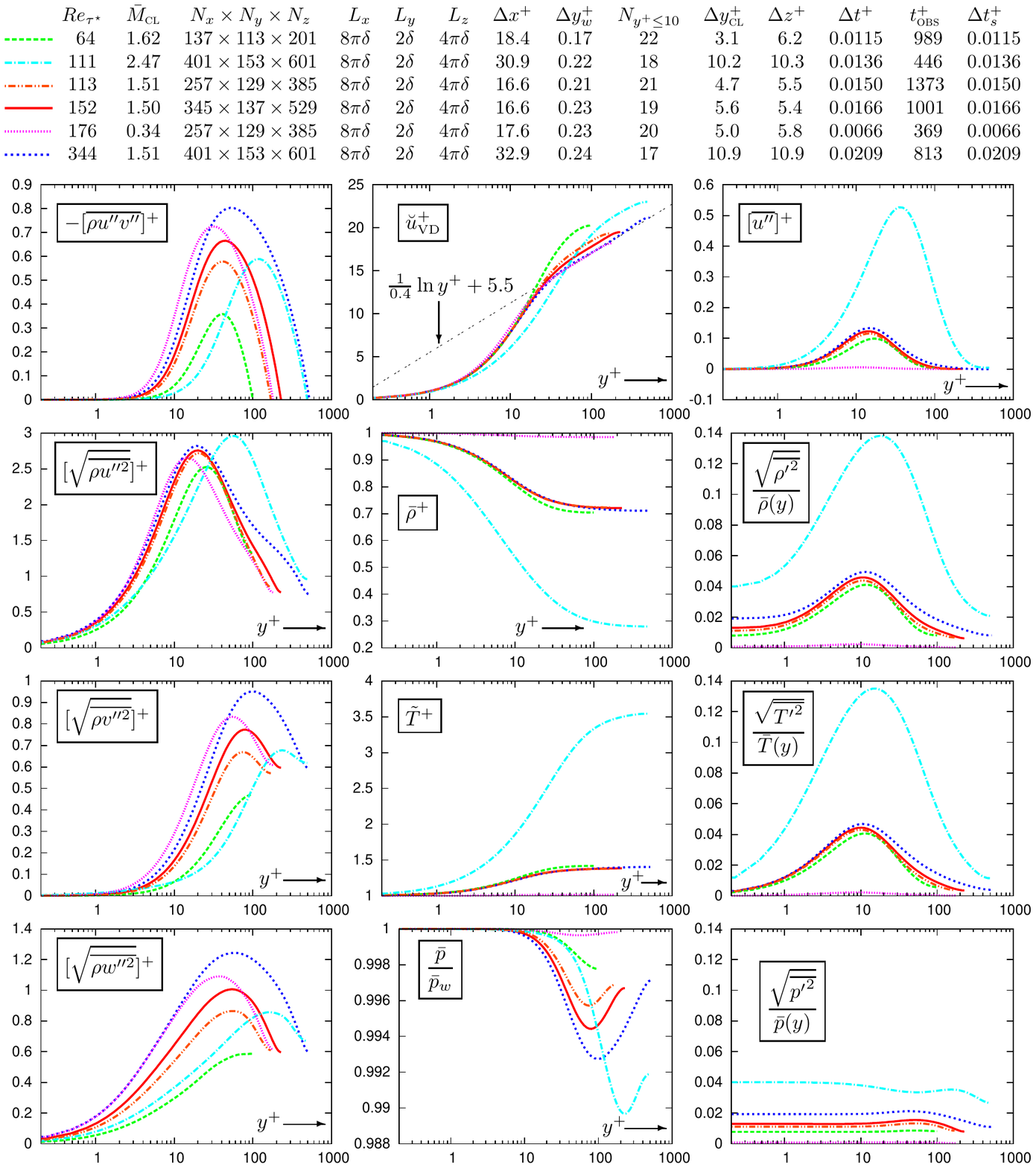}}
\end{picture}
\end{center}
\caption{\tsn{DNS} \tabref{Tab_PDTEFCTPCF_s_BEqsDNSC_ss_DNSCs_001} results for the profiles of Reynolds-stresses in wall units $[\overline{\rho u_i''u_j''}]^+$, of the van Driest-transformed
 \eqref{Eq_PDTEFCTPCF_s_S_ss_HCBMCL_001} mean velocity $\breve{u}_{\tsn{VD}}^+$, of the  evolution of mean thermodynamic variables relative to their wall values
($\bar\rho_w^{-1}\bar\rho\stackrel{\eqref{Eq_PDTEFCTPCF_s_S_ss_WUs_001c}}{=}\bar\rho^+$,
$\tilde{T}_w^{-1}\tilde{T}\stackrel{\eqref{Eq_PDTEFCTPCF_s_S_ss_WUs_001d}}{=}\tilde{T}^+$ and $\bar{p}_w^{-1}\bar{p}$), of the Reynolds average of the Favre fluctuations of the streamwise velocity in wall
units $[\overline{u''}]^+$, and of the relative rms-levels of thermodynamic variables  ($\bar\rho^{-1}\;\rho'_{\rm rms}$, $\bar T^{-1}\;T'_{\rm rms}$ and $\bar p^{-1}\;p'_{\rm rms}$), plotted against $y^+$ \eqref{Eq_PDTEFCTPCF_s_S_ss_CCF_001b}.}
\label{Fig_PDTEFCTPCF_s_S_ss_CCF_002}
\end{figure}

Obviously, the choices \eqref{Eq_PDTEFCTPCF_s_S_ss_WUs_001c} for the thermal quantities, $h$ and $q_i$, are dictated by the scaling of the dynamic field.
Choosing $\tilde T_w$ as the wall-unit for $T$ \citep[p. 193]{Gatski_Bonnet_2009a}, and requiring that \eqrefsab{Eq_PDTEFCTPCF_s_BEqsDNSC_ss_FM_001}
                                                                                                                {Eq_PDTEFCTPCF_s_BEqsDNSC_ss_TTV_003}
hold in wall-units without the appearance of nondimensional numbers, yields
\begin{alignat}{6}
T^+:=\dfrac{T}{\tilde T_w}
\;\;;\;\;
R_g^+:=\dfrac{R_g}{\left(\dfrac{\bar\tau_w}{\bar\rho_w\tilde T_w}\right)}\stackrel{\eqref{Eq_PDTEFCTPCF_s_BEqsDNSC_ss_FM_001d}}{=}\bar p_w^+
\;\;;\;\;
c_p^+:=\dfrac{\gamma}{\gamma-1}R_g^+
\;\;;\;\;
\gamma^+:=\gamma
                                                                                                                                    \label{Eq_PDTEFCTPCF_s_S_ss_WUs_001d}
\end{alignat}
\begin{alignat}{6}
s^+:=\dfrac{s}{\left(\dfrac{\bar\tau_w}{\bar\rho_w\tilde T_w}\right)}\stackrel{\eqref{Eq_PDTEFCTPCF_s_S_ss_WUs_001d}}{\Longrightarrow}\dfrac{s^+}{R_g^+}=\dfrac{s}{R_g}
\;\;;\;\;
\mu^+:=\dfrac{\mu}{\bar\mu_w}
\;\;;\;\;
\lambda^+:=\dfrac{\lambda}{\left(\dfrac{\bar\mu_w u_\tau^2}{\tilde T_w}\right)}
                                                                                                                                    \label{Eq_PDTEFCTPCF_s_S_ss_WUs_001e}
\end{alignat}
where the relation for $R_g^+$ is obtained from the state-equation \eqref{Eq_PDTEFCTPCF_s_BEqsDNSC_ss_FM_001d},
the relation for $c_p^+$ to satisfy the definition $c_p:=(\partial_T h)_p$ \citep[(2.43b), p. 552]{Kestin_1979a},
the relation for $s^+$ to satisfy the definition $Tds=dh-\rho^{-1}dp$ of entropy \citep[Tab.~12.1, p. 536]{Kestin_1979a},
the definition of $\mu^+$ to satisfy the stress-strain relation \eqref{Eq_PDTEFCTPCF_s_BEqsDNSC_ss_FM_001e},
and the definition of $\lambda^+$ to satisfy Fourier's law \eqref{Eq_PDTEFCTPCF_s_BEqsDNSC_ss_FM_001f}.
\end{subequations}

%
%
%
%
%
\subsection{HCB-scaling and $\bar M_\tsn{CL}$-scaling}\label{PDTEFCTPCF_s_S_ss_HCBMCL}
%
%
%
%
%

Examination of the profiles, plotted against $y^+$ \eqref{Eq_PDTEFCTPCF_s_S_ss_CCF_001b}, of various flow statistics \figref{Fig_PDTEFCTPCF_s_S_ss_CCF_002}, obtained from the 
\tsn{DNS} computations \tabref{Tab_PDTEFCTPCF_s_BEqsDNSC_ss_DNSCs_001}, illustrates the influence of (${Re}_{\tau^\star}$,$\bar M_\tsn{CL}$), or equivalently \tabref{Tab_PDTEFCTPCF_s_S_ss_CCF_001} of 
(${Re}_{\tau_w}$,$M_{\tsn{B}_w}$). The deficit from unity of the maximum shear Reynolds-stress in wall units, $\max_y[-\overline{\rho u''v''}]^+$, quantifies the importance of viscous effects for each configuration, since the averaged streamwise-momentum
equation \eqref{Eq_PDTEFCTPCF_s_BEqsDNSC_ss_FM_001b} reads $\bar\tau_{xy}^+-[\overline{\rho u''v''}]^+=1-{Re}_{\tau_w}^{-1}y^+=1-{Re}_{\tau^\star}^{-1} y^\star$.
Obviously \figref{Fig_PDTEFCTPCF_s_S_ss_CCF_002}, ${Re}_{\tau^\star}$ correlates well with the relative importance of viscous effects, contrary to ${Re}_{\tau_w}$ \tabref{Tab_PDTEFCTPCF_s_S_ss_CCF_001}, in agreement with the arguments in justification
of the \tsn{HCB}-scaling \citep{Huang_Coleman_Bradshaw_1995a}. The (${Re}_{\tau^\star}$,$\bar M_\tsn{CL}$)=(64,1.62) channel is a very-low-$Re$ configuration, where the viscous shear-stress dominates the flow, since $\max_y[-\overline{\rho u''v''}]^+\approxeq0.35$
\figref{Fig_PDTEFCTPCF_s_S_ss_CCF_002}, comparable to incompressible very-low-$Re$ computations \citep{Hu_Morfey_Sandham_2006a}. Futhermore, the (${Re}_{\tau^\star}$,$\bar M_\tsn{CL}$)$\in\{(113,1.51),(111,2.47)\}$ channels
which have different ${Re}_{\tau_w}$ (169 and 492 respectively; \tabrefnp{Tab_PDTEFCTPCF_s_S_ss_CCF_001}), but approximately equal ${Re}_{\tau^\star}$ reach similar levels of $\max_y[-\overline{\rho u''v''}]^+$ \figref{Fig_PDTEFCTPCF_s_S_ss_CCF_002}.

Consideration \figref{Fig_PDTEFCTPCF_s_S_ss_CCF_002} of the van Driest transformed velocity \citep[(6), p.174]{Coleman_Kim_Moser_1995a}
\begin{eqnarray}
\breve{u}_{\tsn{VD}}^+=\int_0^{y^+}\sqrt{\dfrac{\bar\rho}{\bar\rho_w}}\dfrac{du^+}{dY^+}dY^+   \label{Eq_PDTEFCTPCF_s_S_ss_HCBMCL_001}
\end{eqnarray}
confirms that ${Re}_{\tau^\star}$ is a representative Reynolds number in comparing flows with different $\bar M_\tsn{CL}$. As ${Re}_{\tau^\star}$ increases \figref{Fig_PDTEFCTPCF_s_S_ss_CCF_002}, $\breve{u}_{\tsn{VD}}^+$ approaches a logarithmic
zone, the (${Re}_{\tau^\star}$,$\bar M_\tsn{CL}$)=(344,1.51) data following reasonably well $\tfrac{1}{0.4}\ln y^+ +5.5$ for $y^+\gtrapprox 20$ and so do the (${Re}_{\tau^\star}$,$\bar M_\tsn{CL}$)=(176,0.34) data 
 \figref{Fig_PDTEFCTPCF_s_S_ss_CCF_002}. Notice a similar ${Re}_{\tau_w}$-behaviour of the incompressible \tsn{DNS} data of \citet[Fig.~(1.a), p.~1543]{Hu_Morfey_Sandham_2006a} .
The low-$Re$ effects in the (${Re}_{\tau^\star}$,$\bar M_\tsn{CL}$)$\in\{(64,1.62),(111,2.47)\}$ channels
are clearly seen, in the $\breve{u}_{\tsn{VD}}^+$ plots \figref{Fig_PDTEFCTPCF_s_S_ss_CCF_002}. Mean density $\bar\rho^+$ decreases and mean temperature $\tilde{T}^+$ increases with $y^+$ \figref{Fig_PDTEFCTPCF_s_S_ss_CCF_002}, with 
a clear $\bar M_\tsn{CL}$-dependence in these distributions, while the distribution of mean pressure $\bar{p}_w^{-1}\bar{p}$ is influenced by both $\bar M_\tsn{CL}$ and ${Re}_{\tau^\star}$ \figref{Fig_PDTEFCTPCF_s_S_ss_CCF_002}. Concerning the streamwise mass-flux
$[\overline{u''}]^+\stackrel{\eqref{Eq_PDTEFCTPCF_s_BEqsDNSC_ss_SA_sss_RFD_001c}}{=}-[\bar\rho^{-1}\overline{\rho'u'}]^+$, there is, expectedly, a clear $\bar M_\tsn{CL}$-dependence which is also observed \figref{Fig_PDTEFCTPCF_s_S_ss_CCF_002}
in the coefficients of variation of density and temperature ($[\bar\rho^{-1}\rho']_{\rm rms}$ and $[\bar T^{-1} T']_{\rm rms}$), while the coefficient of variation of pressure ($[\bar{p}^{-1}p']_{\rm rms}$) shows strong dependence on both parameters,
${Re}_{\tau^\star}$ and $\bar M_\tsn{CL}$ \figref{Fig_PDTEFCTPCF_s_S_ss_CCF_002}.
\begin{figure}
\begin{center}
\begin{picture}(450,320)
\put(-23,-235){\includegraphics[angle=0,width=470pt]{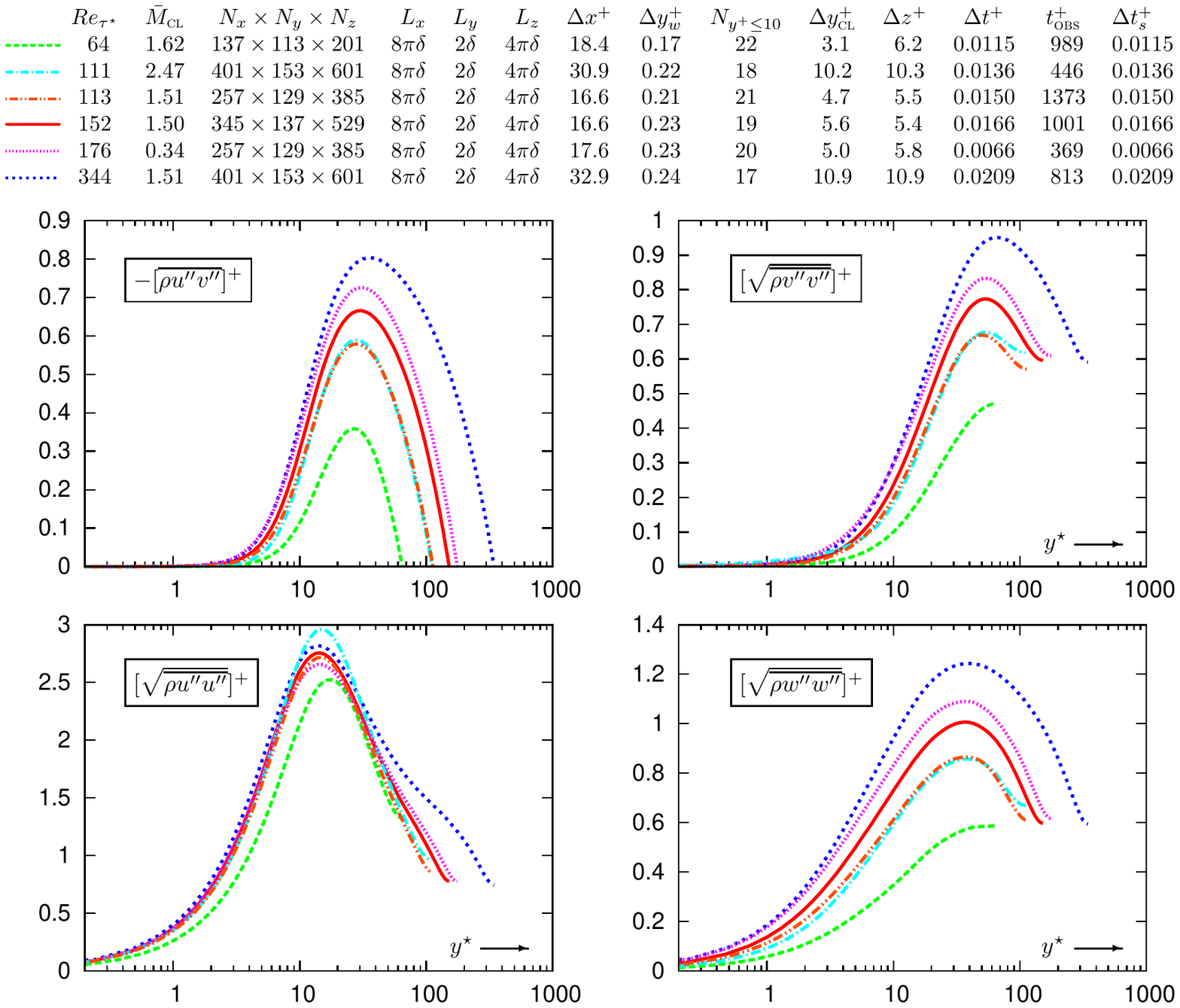}}
\end{picture}
\end{center}
\caption{\tsn{DNS} \tabref{Tab_PDTEFCTPCF_s_BEqsDNSC_ss_DNSCs_001} results for the profiles of Reynolds-stresses in wall units $[\overline{\rho u_i''u_j''}]^+$ (which corresponds to \tsn{HCB}-scaling),
plotted against $y^\star$ \eqref{Eq_PDTEFCTPCF_s_S_ss_CCF_001c}.}
\label{Fig_PDTEFCTPCF_s_S_ss_HCBMCL_001}
\end{figure}

The behaviour of the Reynolds-stress \figref{Fig_PDTEFCTPCF_s_S_ss_CCF_002} is better understood  \figref{Fig_PDTEFCTPCF_s_S_ss_HCBMCL_001} by considering semi-local \tsn{HCB}-scaling \citep{Huang_Coleman_Bradshaw_1995a}.
The Reynolds stresses in wall-units are defined as $[-\overline{\rho u_i''u_j''}]^+\stackrel{\eqref{Eq_PDTEFCTPCF_s_S_ss_WUs_001c}}{:=}-\bar\tau_w^{-1}\overline{\rho u_i''u_j''}$, and are therefore already \tsn{HCB}-scaled.
The shear Reynolds-stress $[-\overline{\rho u''v''}]^+$, plotted against $y^\star$ \eqref{Eq_PDTEFCTPCF_s_S_ss_CCF_001c}, scales with ${Re}_{\tau^\star}$ and is independent of $\bar M_\tsn{CL}$, as evidenced by the data for 
(${Re}_{\tau^\star}$,$\bar M_\tsn{CL}$)$\in\{(113,1.51),(111,2.47)\}$ which are practically indistinguishable \figref{Fig_PDTEFCTPCF_s_S_ss_HCBMCL_001}, thus confirming the validity of \tsn{HCB}-scaling for $[-\overline{\rho u''v''}]^+$.
The same validity of \tsn{HCB}-scaling \citep{Huang_Coleman_Bradshaw_1995a} applies to the wall-normal $[\overline{\rho v''v''}]^+$ and to the spanwise $[\overline{\rho w''w''}]^+$ components \figref{Fig_PDTEFCTPCF_s_S_ss_HCBMCL_001}, despite a 
small difference between the (${Re}_{\tau^\star}$,$\bar M_\tsn{CL}$)$\in\{(113,1.51),(111,2.47)\}$  distributions in the centerline region. Regarding the streamwise component $[\overline{\rho u''u''}]^+$, all of the data, with the exception
of the very-low-$Re$ (${Re}_{\tau^\star}$,$\bar M_\tsn{CL}$)=(64,1.62) case, collapse on a single curve \figref{Fig_PDTEFCTPCF_s_S_ss_HCBMCL_001} between the wall and the maximum peak ($0\lessapprox y^\star \lessapprox 15$), and then decrease to 
approximately the same centerline value.  There is, nonetheless, a small difference in the peak value of $[\overline{\rho u''u''}]^+$ between the  (${Re}_{\tau^\star}$,$\bar M_\tsn{CL}$)$\in\{(113,1.51),(111,2.47)\}$  distributions.
As a conclusion, \tsn{HCB}-scaling filters out reasonably well any influence of $\bar M_\tsn{CL}$ on the Reynolds-stresses $[-\overline{\rho u_i''u_j''}]^+$ \figref{Fig_PDTEFCTPCF_s_S_ss_HCBMCL_001}, except perhaps for the 
near-wall peak of the streamwise component which seems to be weakly $\bar M_\tsn{CL}$-dependent, although more \tsn{DNS} data at $\bar M_\tsn{CL}\geq 2$ and at higher ${Re}_{\tau^\star}$ are required to fully substantiate and quantify this observation.

\begin{figure}
\begin{center}
\begin{picture}(450,440)
\put(-65,-125){\includegraphics[angle=0,width=475pt]{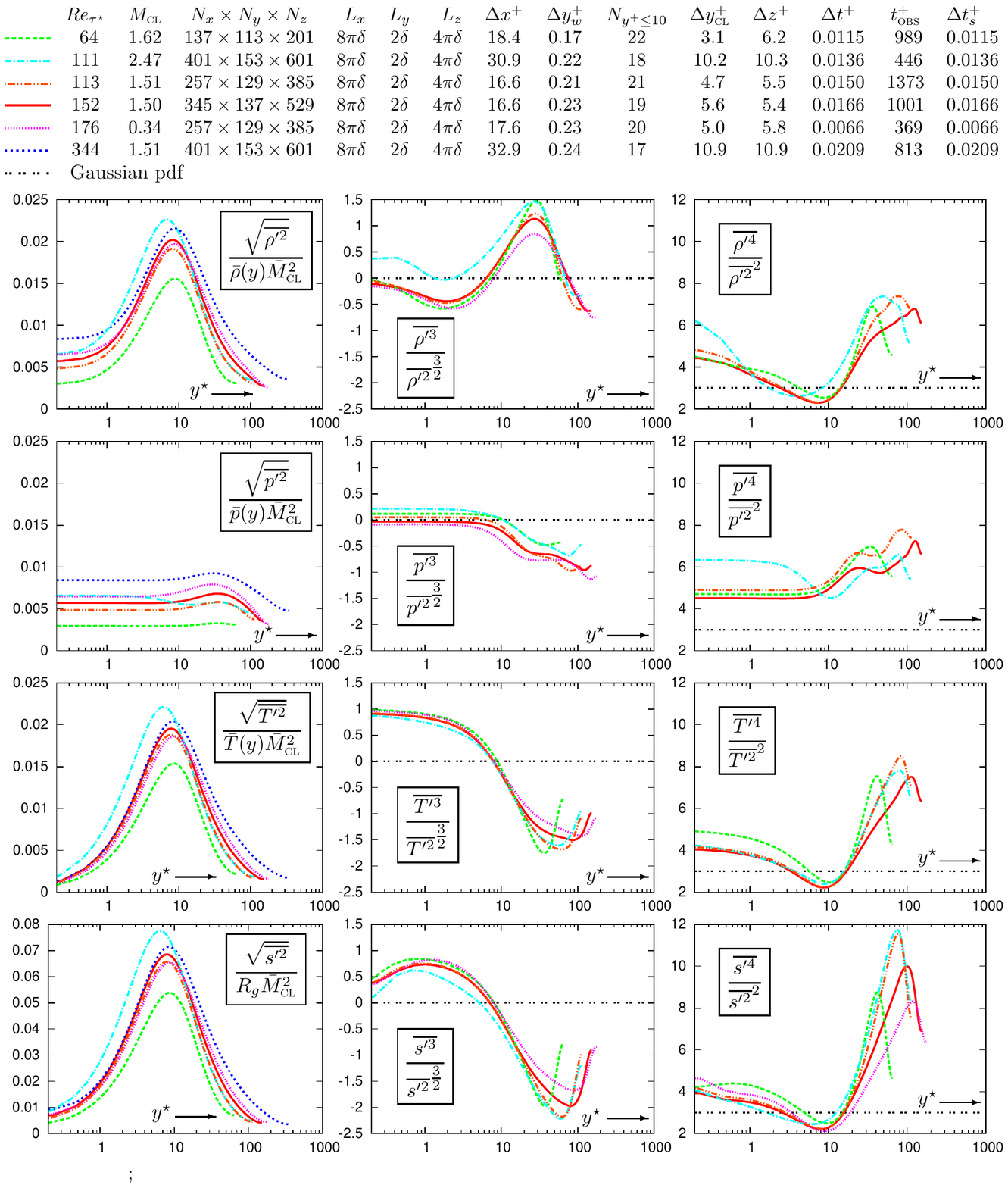}}
\end{picture}
\end{center}
\caption{\tsn{DNS} \tabref{Tab_PDTEFCTPCF_s_BEqsDNSC_ss_DNSCs_001} results for the profiles of the relative fluctuation levels of thermodynamic quantities 
($[\bar\rho^{-1}\rho']_{\rm rms}$, $[\bar p^{-1}p']_{\rm rms}$, $[\bar T^{-1}T']_{\rm rms}$ and $[R_g^{-1}s']_{\rm rms}$), scaled by $\bar M_\tsn{CL}^2$, and of the skewness and flatness of the
thermodynamic fluctuations, plotted (inner scaling) against $y^\star$ \eqref{Eq_PDTEFCTPCF_s_S_ss_CCF_001c}.} 
\label{Fig_PDTEFCTPCF_s_S_ss_HCBMCL_002}
\end{figure}

Expectedly, the coefficients of variation of the thermodynamic variables ($[\bar\rho^{-1} \rho']_{\rm rms}$, $[\bar T^{-1} T']_{\rm rms}$ and $[\bar p^{-1} p']_{\rm rms}$) exhibit strong $\bar M_\tsn{CL}$-dependence \figref{Fig_PDTEFCTPCF_s_S_ss_CCF_002},
and closer examination of the data suggests that they scale reasonably well as  $\bar M_\tsn{CL}^2$ \figref{Fig_PDTEFCTPCF_s_S_ss_HCBMCL_002}. As discussed in \parrefnp{PDTEFCTPCF_s_S_ss_CCF}, this implies that they do not scale with some power of 
$M_{\tsn{B}_w}$  \tabref{Tab_PDTEFCTPCF_s_S_ss_CCF_001}, suggesting that $\bar M_\tsn{CL}$ and not $M_{\tsn{B}_w}$ is the correct choice of representative Mach number. Actually, $M_{\tsn{B}_w}$ is widely used \citep{Huang_Coleman_Bradshaw_1995a,Friedrich_2007a}
simply because the constraints and boundary-conditions applied in the \tsn{DNS} calculations \parref{PDTEFCTPCF_s_BEqsDNSC_ss_DNSCs} fix the value of $M_{\tsn{B}_w}$ \eqref{Eq_PDTEFCTPCF_s_S_ss_CCF_001a}. Before discussing in detail the
$y^\star$-distributions of the $\bar M_\tsn{CL}^2$-scaled coefficients of variation of the thermodynamic fluctuations, it is useful to summarize the implications of the strictly isothermal wall boundary-condition
($T_w=\bar{T}_w=\tilde{T}_w=\const \; \Longrightarrow \; T'_w=0$) used in the present \tsn{DNS} computations, which implies, because of the equation-of-state \eqref{Eq_PDTEFCTPCF_s_BEqsDNSC_ss_FM_001d}
\begin{eqnarray}
T'_w=0\; \stackrel{\eqref{Eq_PDTEFCTPCF_s_BEqsDNSC_ss_FM_001d}}{\Longrightarrow} 
\left\{\begin{array}{l}  a_w=\bar{a}_w=\breve{a}_w=\sqrt{\gamma R_g\bar{T}_w} \\
                         \gamma\bar p_w=\bar{a}_w^2\bar\rho_w                 \\
                         \gamma p'_w=\bar{a}_w^2 \rho'_w                      \\\end{array}\right.
                                                                                                                                    \label{Eq_PDTEFCTPCF_s_S_ss_HCBMCL_001}
\end{eqnarray}
%
\begin{figure}
\begin{center}
\begin{picture}(450,440)
\put(-65,-125){\includegraphics[angle=0,width=475pt]{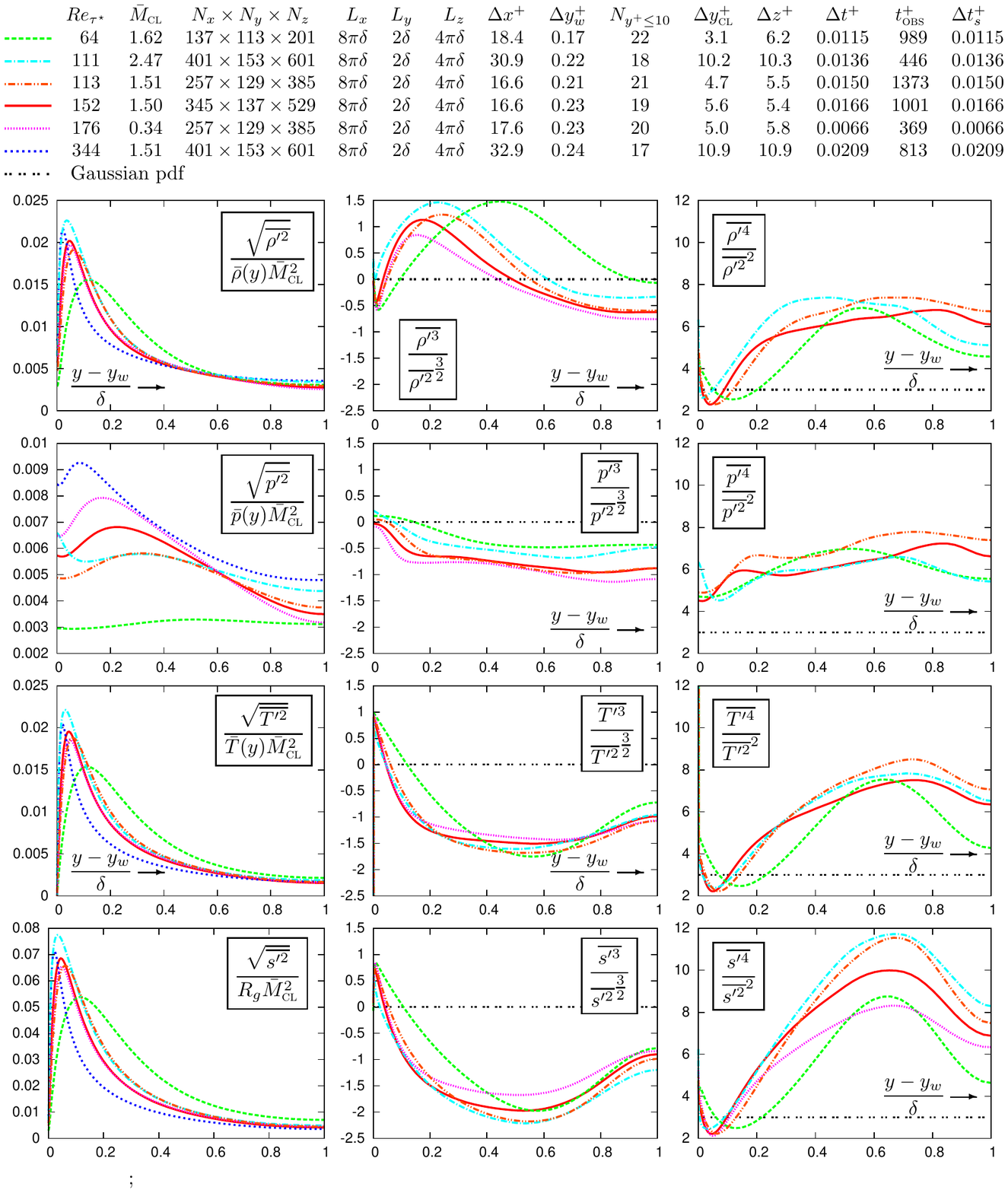}}
\end{picture}
\end{center}
\caption{\tsn{DNS} \tabref{Tab_PDTEFCTPCF_s_BEqsDNSC_ss_DNSCs_001} results for the profiles of the relative fluctuation levels of thermodynamic quantities 
($[\bar\rho^{-1}\rho']_{\rm rms}$, $[\bar p^{-1}p']_{\rm rms}$, $[\bar T^{-1}T']_{\rm rms}$ and $[R_g^{-1}s']_{\rm rms}$), scaled by $\bar M_\tsn{CL}^2$, and of the skewness and flatness of the
thermodynamic fluctuations, plotted (outer scaling) against $\delta^{-1}(y-y_w)$.} 
\label{Fig_PDTEFCTPCF_s_S_ss_HCBMCL_003}
\end{figure}

Plotted against the inner-scaled wall-distance $y^\star$ \eqref{Eq_PDTEFCTPCF_s_S_ss_CCF_001c},
$[\bar M_\tsn{CL}^{-2}\bar p^{-1} p']_{\rm rms}$ \figref{Fig_PDTEFCTPCF_s_S_ss_HCBMCL_002} shows clearly a ${Re}_{\tau^\star}$-influence, in line with incompressible flow data for $[p']^+_{\rm rms}$
\citep[Fig.~6, p.~1546]{Hu_Morfey_Sandham_2006a}. Notice, nonetheless, the specific behaviour of the (${Re}_{\tau^\star}$,$\bar M_\tsn{CL}$)=(111,2.47) data, which are quasi-identical with the (${Re}_{\tau^\star}$,$\bar M_\tsn{CL}$)=(113,1.51) data for
$y^\star\gtrapprox 10$, but differ very near the wall ($y^\star\lessapprox 10$), where the $\bar M_\tsn{CL}$=2.47  data indicate that $[\bar p^{-1}p']_{{\rm rms},w}$ is higher than the peak of $[\bar p^{-1} p']_{\rm rms}$ at $y^\star\approxeq 35$,
contrary to the other cases \figref{Fig_PDTEFCTPCF_s_S_ss_HCBMCL_002}.
When plotted against the outer-scaled wall-distance $\delta^{-1}(y-y_w)$, $[\bar M_\tsn{CL}^{-2}\bar p^{-1} p']_{\rm rms}$ \figref{Fig_PDTEFCTPCF_s_S_ss_HCBMCL_003} depends on both parameters (${Re}_{\tau^\star}$,$\bar M_\tsn{CL}$),
the centerline value increasing with increasing ${Re}_{\tau^\star}$ (at constant $\bar M_\tsn{CL}$, \eg $({Re}_{\tau^\star},\bar M_\tsn{CL})\in\{(64,1.62),(113,1.51),(152,1.50),(344,1.51)\}$)
or with increasing $\bar M_\tsn{CL}$ (at constant ${Re}_{\tau^\star}$, \eg (${Re}_{\tau^\star},\bar M_\tsn{CL})\in\{(113,1.51),(111,2.47)\}$).
On the other hand, comparison of the $[\bar M_\tsn{CL}^{-2}\bar p^{-1} p']_{\rm rms}$ plots, against inner-scaled \figref{Fig_PDTEFCTPCF_s_S_ss_HCBMCL_002} and outer-scaled \figref{Fig_PDTEFCTPCF_s_S_ss_HCBMCL_003}
wall-distance, highlights the behaviour of the near-wall peak, whose level is clearly a function of ${Re}_{\tau^\star}$ only \figrefsab{Fig_PDTEFCTPCF_s_S_ss_HCBMCL_002}
                                                                                                                                       {Fig_PDTEFCTPCF_s_S_ss_HCBMCL_003}.
In particular \figref{Fig_PDTEFCTPCF_s_S_ss_HCBMCL_003}, the difference of the near-wall peak-value of $[\bar M_\tsn{CL}^{-2}\bar p^{-1} p']_{\rm rms}$ from the centerline value
increases with increasing ${Re}_{\tau^\star}$, the very-low-$Re$ (${Re}_{\tau^\star}$,$\bar M_\tsn{CL}$)=(64,1.62) case having a nearly constant $[\bar M_\tsn{CL}^{-2}\bar p^{-1} p']_{\rm rms}\;\forall y$\figrefsab{Fig_PDTEFCTPCF_s_S_ss_HCBMCL_002}
                                                                                                                                                                                                                      {Fig_PDTEFCTPCF_s_S_ss_HCBMCL_003}. 

The plots against the inner-scaled wall-distance $y^\star$ \eqref{Eq_PDTEFCTPCF_s_S_ss_CCF_001c} of the $\bar M_\tsn{CL}^2$-scaled relative rms-levels
of the other thermodynamic fluctuations ($[\bar M_\tsn{CL}^{-2}\bar\rho^{-1}\rho']_{\rm rms}$, $[\bar M_\tsn{CL}^{-2}\bar T^{-1}T']_{\rm rms}$ and $[\bar M_\tsn{CL}^{-2} R_g^{-1}s']_{\rm rms}$),
indicate reasonable superposition for $y^\star\gtrapprox 10$ \figref{Fig_PDTEFCTPCF_s_S_ss_HCBMCL_002}, with a ${Re}_{\tau^\star}$-effect which is particularly visible for the very-low-$Re$ (${Re}_{\tau^\star}$,$\bar M_\tsn{CL}$)=(64,1.62) case
\figref{Fig_PDTEFCTPCF_s_S_ss_HCBMCL_002}. Again, near the wall ($y^\star\lessapprox 10$), the $\bar M_\tsn{CL}$=2.47  data indicate higher fluctuation levels. Notice that the perfect correlation \eqref{Eq_PDTEFCTPCF_s_S_ss_HCBMCL_001} between 
$\rho'_w$ and $p'_w$ imposed by the strictly isothermal wall condition introduces a ${Re}_{\tau^\star}$-dependence of $[\bar M_\tsn{CL}^{-2} \bar\rho^{-1} \rho']_{\rm rms}$ in the viscous sublayer which does not extend beyond $y^\star\approxeq 3$
\figref{Fig_PDTEFCTPCF_s_S_ss_HCBMCL_002}.
Finally, by \eqref{Eq_PDTEFCTPCF_s_BEqsDNSC_ss_TTV_004}, entropy fluctuations nondimensionalized by the gas-constant $R_g$ \eqref{Eq_PDTEFCTPCF_s_BEqsDNSC_ss_FM_001d} are related to the relative fluctuations of $T$ and $\rho$ by
\begin{alignat}{4}
\!\!\dfrac{s'}{R_g}\stackrel{\eqref{Eq_PDTEFCTPCF_s_BEqsDNSC_ss_TTV_004}}{=}\dfrac{1}{\gamma-1}\ln\left (1\!\!+\!\!\dfrac{T'}{\bar{T}}\right)-\ln\left(1\!\!+\!\!\dfrac{\rho'}{\bar{\rho}}\right)
                                                                             -\left[\dfrac{\bar{s}-s_\tsn{ISA}}{R_g}-\left (\dfrac{1}{\gamma-1}\ln\dfrac{\bar{T}}{T_\tsn{ISA}}-\ln\dfrac{\bar\rho}{\rho_\tsn{ISA}}\right )\right ]
                                                                                                                                  \label{Eq_PDTEFCTPCF_s_S_ss_HCBMCL_002}
\end{alignat}
explaining the similarity between the $y^\star$-distributions of $[\bar M_\tsn{CL}^{-2} R_g^{-1} s']_{\rm rms}$ and $[\bar M_\tsn{CL}^{-2} \bar{T}^{-1} T']_{\rm rms}$, as well as the slight ${Re}_{\tau^\star}$-dependence for $y^\star\lessapprox 3$ related
to $[\bar M_\tsn{CL}^{-2} \bar\rho^{-1} \rho']_{\rm rms}$ \figref{Fig_PDTEFCTPCF_s_S_ss_HCBMCL_002}.
This relation \eqref{Eq_PDTEFCTPCF_s_S_ss_HCBMCL_002} shows that $R_g^{-1}s'$ is a function of the relative fluctuations
$\bar\rho^{-1}\rho'$ and $\bar T^{-1}T'$, thereby  explaining the choice of plotting $[\bar M_\tsn{CL}^{-2} R_g^{-1}s']_{\rm rms}$, instead of the coefficient of variation of entropy,
because by \eqref{Eq_PDTEFCTPCF_s_BEqsDNSC_ss_TTV_004} only entropy variations have physical meaning.
Notice the nonfluctuating term $[\cdot]$ in \eqref{Eq_PDTEFCTPCF_s_S_ss_HCBMCL_002} which represents the difference between the Reynolds-averaged entropy and the expression \eqref{Eq_PDTEFCTPCF_s_BEqsDNSC_ss_TTV_004} evaluated at mean temperature and density.
When plotted against the outer-scaled wall-distance $\delta^{-1}(y-y_w)$, $\{[\bar M_\tsn{CL}^{-2}\bar\rho^{-1}\rho']_{\rm rms}, [\bar M_\tsn{CL}^{-2}\bar T^{-1}T']_{\rm rms}, [\bar M_\tsn{CL}^{-2} R_g^{-1}s']_{\rm rms}\}$
collapse \figref{Fig_PDTEFCTPCF_s_S_ss_HCBMCL_003} reasonably well on a single curve, except perhaps for $[\bar M_\tsn{CL}^{-2} R_g^{-1}s']_{\rm rms}$ in the very-low-$Re$ (${Re}_{\tau^\star}$,$\bar M_\tsn{CL}$)=(64,1.62) case.

The skewness $[\overline{(.)'^{2}}^{\;-\tfrac{3}{2}}\overline{(.)'^{3}}]$ and flatness $[\overline{(.)'^{2}}^{\;-2}\overline{(.)'^{4}}]$ of the thermodynamic fluctuations quantify the departure of the corresponding pdfs
from the Gaussian-distribution \citep{Donzis_Jagannathan_2013a}, which has 0 skewness and a flatness of 3 \citep{Jimenez_1998a}.
Near the wall ($y^\star\lessapprox 10$), the skewness of $p'$ is close to 0 but increases
from slightly negative values for the quasi incompressible $\bar M_\tsn{CL}=0.34$ case to slightly positive values with increasing $\bar M_\tsn{CL}$ \figref{Fig_PDTEFCTPCF_s_S_ss_HCBMCL_002},
invariably decreasing to negative values towards the centerline region \figref{Fig_PDTEFCTPCF_s_S_ss_HCBMCL_003}.
The flatness of $p'$ is generally larger than Gaussian \figref{Fig_PDTEFCTPCF_s_S_ss_HCBMCL_002}, in agreement with other \tsn{DNS} results \citep[Fig.~12, p.~95]{Wei_Pollard_2011a} with the $\bar M_\tsn{CL}$=2.47  data indicating higher 
flatness in the near-wall region ($y^\star\lessapprox 10$; \figrefnp{Fig_PDTEFCTPCF_s_S_ss_HCBMCL_002}), and increases from the wall towards the centerline \figref{Fig_PDTEFCTPCF_s_S_ss_HCBMCL_003}.
The skewness and flatness of $T'$ show little influence on ${Re}_{\tau^\star}$ or $\bar M_\tsn{CL}$ \figrefsab{Fig_PDTEFCTPCF_s_S_ss_HCBMCL_002}
                                                                                                              {Fig_PDTEFCTPCF_s_S_ss_HCBMCL_003}.
The skewness of $T'$ \figref{Fig_PDTEFCTPCF_s_S_ss_HCBMCL_002} is positive near the wall ($y^\star\lessapprox 10$), reaching a value of 1 as $y^\star\to 0$, and negative skewness further away ($y^\star\gtrapprox 10$),
remaining approximately constant in a large part of the channel ($y-y_w\gtrapprox\tfrac{3}{10}\delta$; \figrefnp{Fig_PDTEFCTPCF_s_S_ss_HCBMCL_003}).
The flatness of $T'$ is higher than Gaussian very near the wall ($y^\star\lessapprox 1$; \figrefnp{Fig_PDTEFCTPCF_s_S_ss_HCBMCL_002}),
decreases to a value of $\sim2$ at $y^\star\approxeq9$ \figref{Fig_PDTEFCTPCF_s_S_ss_HCBMCL_002}, and then increases to values higher than Gaussian towards the centerline region \figrefsab{Fig_PDTEFCTPCF_s_S_ss_HCBMCL_002}
                                                                                                                                                                                            {Fig_PDTEFCTPCF_s_S_ss_HCBMCL_003}.
The wall-normal evolution of the skewness and flatness of $s'$ is quite similar to that of $T'$ \figrefsab{Fig_PDTEFCTPCF_s_S_ss_HCBMCL_002}
                                                                                                          {Fig_PDTEFCTPCF_s_S_ss_HCBMCL_003},
although higher values of flatness are observed in the centerline region \figref{Fig_PDTEFCTPCF_s_S_ss_HCBMCL_003}.
The wall-normal evolution of the flatness of $\rho'$ is quite similar to that of $T'$ \figrefsab{Fig_PDTEFCTPCF_s_S_ss_HCBMCL_002}
                                                                                                {Fig_PDTEFCTPCF_s_S_ss_HCBMCL_003}
in agreement to other \tsn{DNS} data \citep[Fig.~13, p.~96]{Wei_Pollard_2011a}. When plotted against the inner-scaled wall-distance $y^\star$ \eqref{Eq_PDTEFCTPCF_s_S_ss_CCF_001c},
the wall-normal evolution \figref{Fig_PDTEFCTPCF_s_S_ss_HCBMCL_002} of the skewness of $\rho'$ also shows little influence
on ${Re}_{\tau^\star}$ or $\bar M_\tsn{CL}$, with the exception of the $\bar M_\tsn{CL}$=2.47 data near the wall ($y^\star\lessapprox 25$; \figrefnp{Fig_PDTEFCTPCF_s_S_ss_HCBMCL_002}),
but this is no longer the cases when it is plotted against the outer-scaled wall-distance $\delta^{-1}(y-y_w)$ \figref{Fig_PDTEFCTPCF_s_S_ss_HCBMCL_003}.
Probably the most important observation \figref{Fig_PDTEFCTPCF_s_S_ss_HCBMCL_002} is that, near the wall ($y^\star\lessapprox30$),
with a few exceptions for the $\bar M_\tsn{CL}$=2.47 case, the skewness and flatness $y^\star$-distributions of the thermodynamic fluctuations are not very sensitive to ${Re}_{\tau^\star}$ or $\bar M_\tsn{CL}$.

%
%
%
%
%
\subsection{Correlation coefficients}\label{PDTEFCTPCF_s_S_ss_CCs}
%
%
%
%
%

The correlation coefficient between any 2 flow quantities $[\cdot]$ and $(\cdot)$ is defined by
\begin{subequations}
                                                                                                                                    \label{Eq_PDTEFCTPCF_s_S_ss_CCs_001}
\begin{alignat}{6}
c_{(\cdot)'[\cdot]'}:=\dfrac{\overline{(\cdot)'[\cdot]'}}
                            {\sqrt{\overline{(\cdot)'^2}}\sqrt{\overline{[\cdot]'^2}}}
                                                                                                                                    \label{Eq_PDTEFCTPCF_s_S_ss_CCs_001a}
\end{alignat}
This definition can be extended to higher-order correlations as
\begin{alignat}{6}
c_{(\cdot)'\cdots[\cdot]'}:=\dfrac{\overline{(\cdot)'\cdots[\cdot]'}}
                            {\sqrt{\overline{(\cdot)'^2}}\cdots\sqrt{\overline{[\cdot]'^2}}}
                                                                                                                                    \label{Eq_PDTEFCTPCF_s_S_ss_CCs_001b}
\end{alignat}
\end{subequations}
Correlations between fluctuations are of particular importance both to understand the complex interactions between different quantities and to model unclosed terms~\citep{Taulbee_vanOsdol_1991a}. For instance, the exact relations
\begin{subequations}
                                                                                                                                    \label{Eq_PDTEFCTPCF_s_S_ss_CCs_002}
\begin{alignat}{6}
\dfrac{\overline{T''}}{\bar T}&\stackrel{\eqref{Eq_PDTEFCTPCF_s_BEqsDNSC_ss_SA_sss_RFD_001c}}{=}&-\dfrac{\overline{\rho'T'}}{\bar\rho\;\bar T}
                              &\stackrel{\eqref{Eq_PDTEFCTPCF_s_S_ss_CCs_001}}{=}               &&-c_{\rho'T'}\dfrac{\sqrt{\overline{\rho'^2}}}{\bar\rho}\;
                                                                                                             \dfrac{\sqrt{\overline{   T'^2}}}{\bar   T}
                                                                                                                                    \label{Eq_PDTEFCTPCF_s_S_ss_CCs_002a}\\
\dfrac{\overline{u''}}{\bar u}&\stackrel{\eqref{Eq_PDTEFCTPCF_s_BEqsDNSC_ss_SA_sss_RFD_001c}}{=}&-\dfrac{\overline{\rho'u'}}{\bar\rho\;\bar u}
                              &\stackrel{\eqref{Eq_PDTEFCTPCF_s_S_ss_CCs_001}}{=}               &&-c_{\rho'u'}\dfrac{\sqrt{\overline{\rho'^2}}}{\bar\rho}\;
                                                                                                             \dfrac{\sqrt{\overline{   u'^2}}}{\bar   u}
                                                                                                                                    \label{Eq_PDTEFCTPCF_s_S_ss_CCs_002b}\\
\dfrac{\overline{v''}}{\bar u}&\stackrel{\eqref{Eq_PDTEFCTPCF_s_BEqsDNSC_ss_SA_sss_RFD_001c}}{=}&-\dfrac{\overline{\rho'v'}}{\bar\rho\;\bar u}
                              &\stackrel{\eqref{Eq_PDTEFCTPCF_s_S_ss_CCs_001}}{=}               &&-c_{\rho'v'}\dfrac{\sqrt{\overline{   v'^2}}}{\sqrt{\overline{   u'^2}}}\;
                                                                                                             \dfrac{\sqrt{\overline{\rho'^2}}}{\bar\rho}\;
                                                                                                             \dfrac{\sqrt{\overline{   u'^2}}}{\bar   u}
                                                                                                                                    \label{Eq_PDTEFCTPCF_s_S_ss_CCs_002c}
\end{alignat}
\end{subequations}
expressing the Reynolds averages of Favre fluctuations contain correlation coefficients ($c_{\rho'T'}$, $c_{\rho'u'}$, $c_{\rho'v'}$), coefficients of variation 
($[\bar\rho^{-1}\rho']_{\rm rms}$, $[\bar{T}^{-1}T']_{\rm rms}$, $[\bar{u}^{-1}u']_{\rm rms}$) and structure parameters ($[u']_{\rm rms}^{-1} [v']_{\rm rms}$).
\begin{figure}
\begin{center}
\begin{picture}(450,250)
\put(-65,-310){\includegraphics[angle=0,width=475pt]{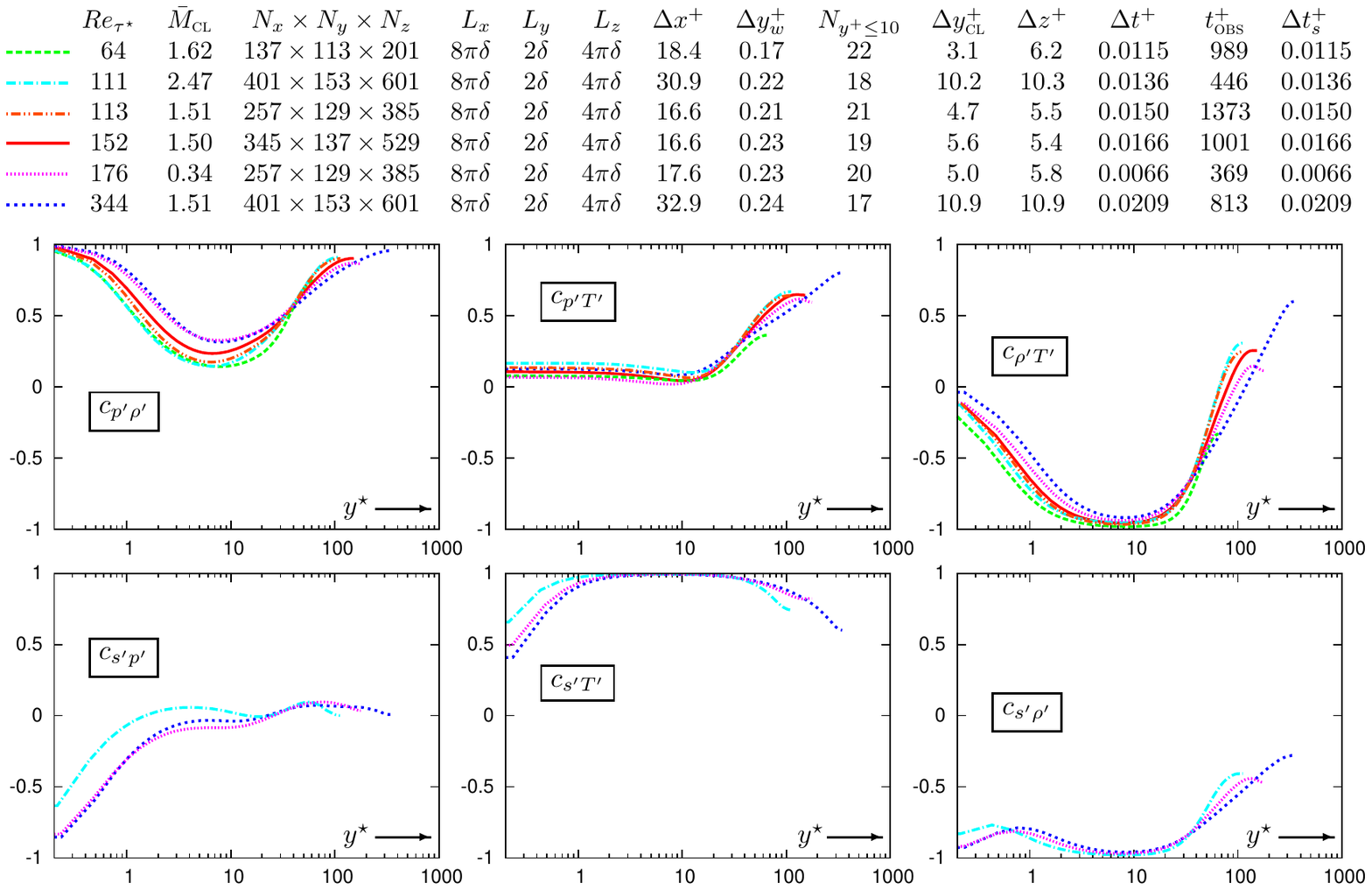}}
\end{picture}
\end{center}
\caption{\tsn{DNS} \tabref{Tab_PDTEFCTPCF_s_BEqsDNSC_ss_DNSCs_001} results for the profiles of the correlation coefficients \eqref{Eq_PDTEFCTPCF_s_S_ss_CCs_001a} between thermodynamic fluctuations
($c_{p'\rho'}$, $c_{p'T'}$, $c_{\rho'T'}$, $c_{s'p'}$, $c_{s'T'}$, $c_{s'\rho'}$; coefficients containing $s'$ were not acquired for all of the \tsn{DNS} computations), plotted (inner scaling) against $y^\star$ \eqref{Eq_PDTEFCTPCF_s_S_ss_CCF_001c}.}
\label{Fig_PDTEFCTPCF_s_S_ss_CCs_001}
%
\begin{center}
\begin{picture}(450,250)
\put(-65,-310){\includegraphics[angle=0,width=475pt]{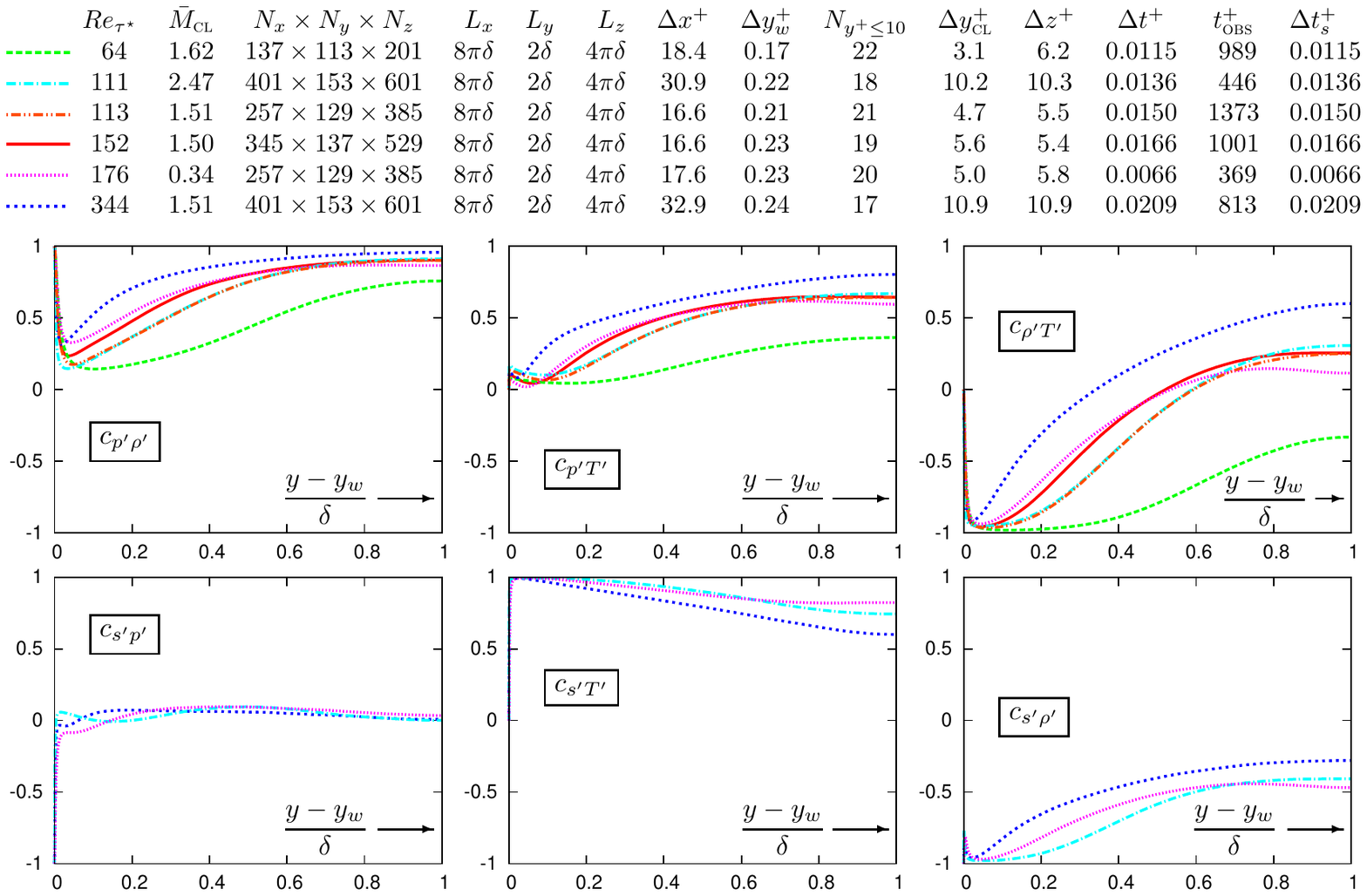}}
\end{picture}
\end{center}
\caption{\tsn{DNS} \tabref{Tab_PDTEFCTPCF_s_BEqsDNSC_ss_DNSCs_001} results for the profiles of the correlation coefficients \eqref{Eq_PDTEFCTPCF_s_S_ss_CCs_001a} between thermodynamic fluctuations
($c_{p'\rho'}$, $c_{p'T'}$, $c_{\rho'T'}$, $c_{s'p'}$, $c_{s'T'}$, $c_{s'\rho'}$; coefficients containing $s'$ were not acquired for all of the \tsn{DNS} computations), plotted (outer scaling) against $\delta^{-1}(y-y_w)$.}
\label{Fig_PDTEFCTPCF_s_S_ss_CCs_002}
\end{figure}

Consideration of the correlation coefficients between thermodynamic fluctuations ($c_{p'\rho'}$, $c_{p'T'}$, $c_{\rho'T'}$, $c_{s'p'}$, $c_{s'T'}$, $c_{s'\rho'}$),
plotted \figref{Fig_PDTEFCTPCF_s_S_ss_CCs_001} against the inner-scaled wall-distance $y^\star$ \eqref{Eq_PDTEFCTPCF_s_S_ss_CCF_001c},
indicates weak influence  of $Re_{\tau^\star}$ or $\bar M_\tsn{CL}$, except very close to the wall ($y^\star\lessapprox3$), where the wall boundary-condition \eqref{Eq_PDTEFCTPCF_s_S_ss_HCBMCL_001}
dominates the behaviour of the correlation coefficients \figref{Fig_PDTEFCTPCF_s_S_ss_CCs_001}.
Pressure and density, are perfectly correlated at the wall \eqref{Eq_PDTEFCTPCF_s_S_ss_HCBMCL_001}, implying $[c_{p'\rho'}]_w=1$ \figref{Fig_PDTEFCTPCF_s_S_ss_CCs_001}.
The correlation coefficient $c_{p'\rho'}$ decreases from this maximum value with distance from the wall \figref{Fig_PDTEFCTPCF_s_S_ss_CCs_001}, to a value around $\sim0.25$, in the
neighbourhood of $y^\star\approxeq10$, which is the location of the $[\bar\rho^{-1}\rho']_{\rm rms}$ peak \figref{Fig_PDTEFCTPCF_s_S_ss_HCBMCL_002},
and then increases again approaching $1$ in the centerline region \figref{Fig_PDTEFCTPCF_s_S_ss_CCs_001}.
When plotted against the outer-scaled wall-distance $\delta^{-1}(y-y_w)$ \figref{Fig_PDTEFCTPCF_s_S_ss_CCs_002},
$c_{p'\rho'}$ exhibits a $Re_{\tau^\star}$-dependence, with the centerline value being closer to 1 as $Re_{\tau^\star}$ increases \figref{Fig_PDTEFCTPCF_s_S_ss_CCs_002}.
Since, at the wall, $\rho'$ is perfectly correlated with $p'$ \eqref{Eq_PDTEFCTPCF_s_S_ss_HCBMCL_001}, it is weakly correlated with $T'$, so that $c_{\rho'T'}$ has values close to $0$ as $y^\star\to0$ \figref{Fig_PDTEFCTPCF_s_S_ss_CCs_001}.
Further away from the wall, $c_{\rho'T'}$ rapidly reaches values close to $-1$ \figref{Fig_PDTEFCTPCF_s_S_ss_CCs_001}, in the region $y^\star\in[3,20]$,
\ie in the neighbourhood of $y^\star\approxeq10$ where are located the peaks of $[\bar\rho^{-1}\rho']_{\rm rms}$ and $[\bar{T}^{-1}T']_{\rm rms}$ \figref{Fig_PDTEFCTPCF_s_S_ss_HCBMCL_002},
and then increases again, reaching slightly positive values at the centerline \figref{Fig_PDTEFCTPCF_s_S_ss_CCs_001}, provided that $Re_{\tau^\star}$ is sufficiently high.
When plotted against the outer-scaled wall-distance $\delta^{-1}(y-y_w)$, $c_{\rho'T'}$ exhibits $Re_{\tau^\star}$-dependence \figref{Fig_PDTEFCTPCF_s_S_ss_CCs_002}.
Finally the coefficient $c_{p'T'}$ is low near the wall ($y^\star\lessapprox 10$) and then increases towards the centerline region. Notice that both $c_{p'\rho'}>0\;\forall y^\star$ and $c_{p'T'}>0\;\forall y^\star$, while 
$c_{\rho'T'}<0\; \forall y^\star\lessapprox 60$, then rising towards $[c_{\rho'T'}]_{\tsn{CL}}>0$,
provided that $Re_{\tau^\star}$ is sufficiently high \figref{Fig_PDTEFCTPCF_s_S_ss_CCs_002}. Hence, independently of $(Re_{\tau^\star},\bar M_\tsn{CL})$, $p'$ is always positively correlated with $\rho'$ and $T'$, although this correlation may be 
weak (close to 0; \figrefnp{Fig_PDTEFCTPCF_s_S_ss_CCs_001}). On the other hand $\rho'$ is negatively correlated with $T'$ in the wall region and positively correlated near the centerline \figref{Fig_PDTEFCTPCF_s_S_ss_CCs_001},
provided that $Re_{\tau^\star}$ is sufficiently high. Notice that
in sustained \tsn{HIT} $c_{\rho'T'}>0$ also \citep[Tab.~1, p.~225]{Donzis_Jagannathan_2013a}, highlighting the strong anisotropy that prevails when approaching the wall,
and also that in the $Re_{\tau^\star}\approxeq64$ case wall effects dominate the entire flow up to the centerline \figrefsab{Fig_PDTEFCTPCF_s_S_ss_CCs_001}
                                                                                                                            {Fig_PDTEFCTPCF_s_S_ss_CCs_002}.

The fluctuating entropy $s'$ is positively correlated with $T'$ ($c_{s'T'}>0$) and negatively correlated with $\rho'$ ($c_{\rho'T'}>0$), and in both cases the correlation or anticorrelation is generally strong \figrefsab{Fig_PDTEFCTPCF_s_S_ss_CCs_001}
                                                                                                                                                                                                                            {Fig_PDTEFCTPCF_s_S_ss_CCs_002}.
The correlation $c_{s'T'}$ \figref{Fig_PDTEFCTPCF_s_S_ss_CCs_001} decreases very near the wall ($y^\star\lessapprox 1$) but is nearly 1 for $y^\star\;\in\;[1,50]$, and then decreases towards centerline \figref{Fig_PDTEFCTPCF_s_S_ss_CCs_002}.
The generally strong correlation of $s'$ with $T'$ was mentioned in the early works of \citet{Kovasznay_1953a} and \citet{Morkovin_1962a}.
On the other hand, the correlation $c_{s'\rho'}<0$ \figrefsab{Fig_PDTEFCTPCF_s_S_ss_CCs_001}
                                                             {Fig_PDTEFCTPCF_s_S_ss_CCs_002},
approaching -1 in the region $y^\star\;\in\;[3,20]$ where $c_{\rho'T'}\approxeq -1$ and $c_{s'T'}\approxeq 1$
\figref{Fig_PDTEFCTPCF_s_S_ss_CCs_001}. Again, in the major part of the flow, the influence of $(Re_{\tau^\star},\bar M_\tsn{CL})$ on the correlation coefficients ($c_{s'\rho'}$, $c_{s'p'}$, $c_{s'T'}$) is weak,
especially with inner scaling of the wall-distance ($y^\star$; \figrefnp{Fig_PDTEFCTPCF_s_S_ss_CCs_001}).
Comparison of the plots of the correlation coefficients ($c_{p'\rho'}$, $c_{p'T'}$, $c_{\rho'T'}$, $c_{s'p'}$, $c_{s'T'}$, $c_{s'\rho'}$),
with inner ($y^\star$; \figrefnp{Fig_PDTEFCTPCF_s_S_ss_CCs_001}) and outer ($\delta^{-1}(y-y_w)$; \figrefnp{Fig_PDTEFCTPCF_s_S_ss_CCs_002}) scaling of the wall-distance,
suggests that, for the range of $(Re_{\tau^\star},\bar M_\tsn{CL})$ that was investigated \tabref{Tab_PDTEFCTPCF_s_BEqsDNSC_ss_DNSCs_001},
inner scaling ($y^\star$) provides reasonable collapse of the data \figref{Fig_PDTEFCTPCF_s_S_ss_CCs_001}, except very near the wall ($y^\star\lessapprox3$),
and that outer scaling ($\delta^{-1}(y-y_w)$) does not. Notice also that $c_{s'p'}\approxeq0$ in the major part of the flow \figrefsab{Fig_PDTEFCTPCF_s_S_ss_CCs_001}
                                                                                                                                      {Fig_PDTEFCTPCF_s_S_ss_CCs_002},
except very near the wall ($y^\star\approxeq3$; \figrefnp{Fig_PDTEFCTPCF_s_S_ss_CCs_001}). Furthermore, the fluctuating entropy $s'$ is not only strongly correlated with the fluctuating temperature $T'$,
but is equally strongly anticorrelated with the fluctuating density $\rho'$ \figrefsab{Fig_PDTEFCTPCF_s_S_ss_CCs_001}
                                                                                      {Fig_PDTEFCTPCF_s_S_ss_CCs_002}.

Probably the most interesting observation from the above results \figrefsabc{Fig_PDTEFCTPCF_s_S_ss_CCF_002}{Fig_PDTEFCTPCF_s_S_ss_HCBMCL_002}{Fig_PDTEFCTPCF_s_S_ss_CCs_001} is that, invariably,
\begin{alignat}{6}
O\left(\dfrac{\sqrt{\overline{   T'^2}}}{\bar   T}\right)=
O\left(\dfrac{\sqrt{\overline{\rho'^2}}}{\bar\rho}\right)=
O\left(\dfrac{\sqrt{\overline{   p'^2}}}{\bar   p}\right)\qquad\forall y^\star\qquad \forall (Re_{\tau^\star},\bar M_\tsn{CL})
                                                                                                                                    \label{Eq_PDTEFCTPCF_s_S_ss_CCs_002}
\end{alignat}
are of the same order-of-magnitude everywhere in the channel, and for all $(Re_{\tau^\star},\bar M_\tsn{CL})$ that were investigated \figrefsab{Fig_PDTEFCTPCF_s_S_ss_CCF_002}
                                                                                                                                               {Fig_PDTEFCTPCF_s_S_ss_HCBMCL_002},
even at the low Mach-number limit, and that, the correlation coefficients \figref{Fig_PDTEFCTPCF_s_S_ss_CCs_001}
show little dependence on $\bar M_\tsn{CL}$, except perhaps very near the wall. The order-of-magnitude relation \eqref{Eq_PDTEFCTPCF_s_S_ss_CCs_002}
is also valid in sustained compressible \tsn{HIT} \citep[Tab.~1, p.~225]{Donzis_Jagannathan_2013a}. The reason for this is that
the basic thermodynamic variables ($p$, $\rho$, $T$) are related by the equation-of-state \eqref{Eq_PDTEFCTPCF_s_BEqsDNSC_ss_FM_001d}, implying
\begin{subequations}
                                                                                                                                    \label{Eq_PDTEFCTPCF_s_S_ss_CCs_004}
\begin{alignat}{6}
\bar p\stackrel{\eqrefsab{Eq_PDTEFCTPCF_s_BEqsDNSC_ss_FM_001d}
                         {Eq_PDTEFCTPCF_s_BEqsDNSC_ss_SA_sss_RFD_001c}}{=}\bar\rho R_g \tilde T
       \stackrel{\eqref{Eq_PDTEFCTPCF_s_BEqsDNSC_ss_SA_sss_RFD_001c}}{=}\bar\rho R_g(\bar T-\overline{T''})
                                                                                                                                    \label{Eq_PDTEFCTPCF_s_S_ss_CCs_004a}
\end{alignat}
\begin{alignat}{6}
\left(1-\dfrac{\overline{T''}}{\bar T}\right)\dfrac{p'}{\bar p}\stackrel{\eqrefsab{Eq_PDTEFCTPCF_s_BEqsDNSC_ss_FM_001d}
                                                                                  {Eq_PDTEFCTPCF_s_S_ss_CCs_004a}}{=}\left(\dfrac{\rho'}{\bar\rho}
                                                                                                                          +\dfrac{   T'}{\bar   T}
                                                                                                                          +\dfrac{\rho'T'}{\bar\rho\bar T}
                                                                                                                          +\dfrac{\overline{T''}}{\bar T}\right)
                                                                                                                                    \label{Eq_PDTEFCTPCF_s_S_ss_CCs_004b}
\end{alignat}
\end{subequations}
Multiplying \eqref{Eq_PDTEFCTPCF_s_S_ss_CCs_004b} by $p'$, $\rho'$ or $T'$, we obtain upon averaging the exact relations
\begin{subequations}
                                                                                                                                    \label{Eq_PDTEFCTPCF_s_S_ss_CCs_005}
\begin{alignat}{6}
\!\!\!\!\!\!\!\!\left(1+c_{\rho'T'}\dfrac{\sqrt{\overline{\rho'^2}}}{\bar\rho}\;
                   \dfrac{\sqrt{\overline{   T'^2}}}{\bar   T}\right)\dfrac{\sqrt{\overline{p'^2}}}{\bar p}\stackrel{\eqref{Eq_PDTEFCTPCF_s_S_ss_CCs_004b}}{=}
                                                                                              c_{p'\rho'  }\dfrac{\sqrt{\overline{\rho'^2}}}{\bar\rho}
                                                                                             +c_{p'     T'}\dfrac{\sqrt{\overline{   T'^2}}}{\bar   T}
                                                                                             +c_{p'\rho'T'}\dfrac{\sqrt{\overline{\rho'^2}}}{\bar\rho}
                                                                                                           \dfrac{\sqrt{\overline{   T'^2}}}{\bar   T}
                                                                                                                                    \label{Eq_PDTEFCTPCF_s_S_ss_CCs_005a}\\
\!\!\!\!\!\!\!\!\left(1+c_{\rho'T'}\dfrac{\sqrt{\overline{\rho'^2}}}{\bar\rho}\;
                   \dfrac{\sqrt{\overline{   T'^2}}}{\bar   T}\right)c_{p'\rho'  }\dfrac{\sqrt{\overline{p'^2}}}{\bar p}\stackrel{\eqref{Eq_PDTEFCTPCF_s_S_ss_CCs_004b}}{=}
                                                                                                              \dfrac{\sqrt{\overline{\rho'^2}}}{\bar\rho}
                                                                                             +c_{     \rho'T'}\dfrac{\sqrt{\overline{   T'^2}}}{\bar   T}
                                                                                             +c_{\rho'\rho'T'}\dfrac{\sqrt{\overline{\rho'^2}}}{\bar\rho}
                                                                                                           \dfrac{\sqrt{\overline{   T'^2}}}{\bar   T}
                                                                                                                                    \label{Eq_PDTEFCTPCF_s_S_ss_CCs_005b}\\
\!\!\!\!\!\!\!\!\left(1+c_{\rho'T'}\dfrac{\sqrt{\overline{\rho'^2}}}{\bar\rho}\;
                   \dfrac{\sqrt{\overline{   T'^2}}}{\bar   T}\right)c_{p'T'}\dfrac{\sqrt{\overline{p'^2}}}{\bar p}\stackrel{\eqref{Eq_PDTEFCTPCF_s_S_ss_CCs_004b}}{=}
                                                                                              c_{\rho'T'  }\dfrac{\sqrt{\overline{\rho'^2}}}{\bar\rho}
                                                                                             +             \dfrac{\sqrt{\overline{   T'^2}}}{\bar   T}
                                                                                             +c_{\rho'T'T'}\dfrac{\sqrt{\overline{\rho'^2}}}{\bar\rho}
                                                                                                           \dfrac{\sqrt{\overline{   T'^2}}}{\bar   T}
                                                                                                                                    \label{Eq_PDTEFCTPCF_s_S_ss_CCs_005c}
\end{alignat}
\end{subequations}
where we used the definitions \eqref{Eq_PDTEFCTPCF_s_S_ss_CCs_001} of the correlation coefficients and the expression \eqref{Eq_PDTEFCTPCF_s_S_ss_CCs_002a} for $\overline{T''}$.
Furthermore, the triple correlation $c_{p'\rho'T'}$, appearing in \eqref{Eq_PDTEFCTPCF_s_S_ss_CCs_005a}, can be expressed, by multiplying \eqref{Eq_PDTEFCTPCF_s_S_ss_CCs_004b} by $\rho'T'$ and averaging,
and using again \eqrefsab{Eq_PDTEFCTPCF_s_S_ss_CCs_001}
                         {Eq_PDTEFCTPCF_s_S_ss_CCs_002a}, as
\begin{alignat}{6}
\left(1+c_{\rho'T'}\dfrac{\sqrt{\overline{\rho'^2}}}{\bar\rho}\;
                   \dfrac{\sqrt{\overline{   T'^2}}}{\bar   T}\right)c_{p'\rho'T'}\dfrac{\sqrt{\overline{p'^2}}}{\bar p}\stackrel{\eqref{Eq_PDTEFCTPCF_s_S_ss_CCs_004b}}{=}&
                                                                                              c_{\rho'\rho'T'}\dfrac{\sqrt{\overline{\rho'^2}}}{\bar\rho}
                                                                                             +c_{\rho'   T'T'}\dfrac{\sqrt{\overline{   T'^2}}}{\bar   T}
                                                                                                                                  \notag\\
                                                                                            +&(c_{\rho'\rho'T'T'}-c_{\rho'T'}^2)\dfrac{\sqrt{\overline{\rho'^2}}}{\bar\rho}
                                                                                                                                \dfrac{\sqrt{\overline{   T'^2}}}{\bar   T}
                                                                                                                                    \label{Eq_PDTEFCTPCF_s_S_ss_CCs_006}
\end{alignat}
which implies that $c_{p'\rho'T'}$ is, because of the equation-of-state relation \eqref{Eq_PDTEFCTPCF_s_S_ss_CCs_004}, a function of the coefficients of variation of the
thermodynamic state-variables ($[\bar{\rho}^{-1}\rho']_{\rm rms}$, $[\bar{T}^{-1}T']_{\rm rms}$ and $[\bar{p}^{-1}p']_{\rm rms}$) and of correlation coefficicients involving only $\rho'$ and $T'$
($c_{\rho'\rho'T'}$, $c_{\rho'   T'T'}$, $c_{\rho'\rho'T'T'}$, $c_{\rho'T'}$).

It is usual in the analysis of compressible turbulence \citep{Kovasznay_1953a,Taulbee_vanOsdol_1991a} to consider the limiting case of small relative fluctuation
amplitudes, neglecting terms that are quadratic in relative fluctuation amplitudes. Applying this linearization to \eqref{Eq_PDTEFCTPCF_s_S_ss_CCs_005} yields
\begin{alignat}{6}
\left.\begin{array}{c}\dfrac{\sqrt{\overline{\rho'^2}}}{\bar\rho}\ll1\\
                      \dfrac{\sqrt{\overline{   T'^2}}}{\bar   T}\ll1\\\end{array}\right\}\stackrel{\eqrefsab{Eq_PDTEFCTPCF_s_S_ss_CCs_005}
                                                                                                             {Eq_PDTEFCTPCF_s_S_ss_CCs_002}}{\Longrightarrow}
\left\{\begin{array}{c}             \dfrac{\sqrt{\overline{p'^2}}}{\bar p}\approxeq c_{p'\rho'}\dfrac{\sqrt{\overline{\rho'^2}}}{\bar\rho}
                                                                                   +c_{p'   T'}\dfrac{\sqrt{\overline{   T'^2}}}{\bar   T}+O\left(\dfrac{\overline{\rho'^2}}{\bar\rho^2}\right)\\
                       c_{p'\rho'  }\dfrac{\sqrt{\overline{p'^2}}}{\bar p}\approxeq            \dfrac{\sqrt{\overline{\rho'^2}}}{\bar\rho}
                                                                                   +c_{\rho'T'}\dfrac{\sqrt{\overline{   T'^2}}}{\bar   T}+O\left(\dfrac{\overline{\rho'^2}}{\bar\rho^2}\right)\\
                       c_{p'T'     }\dfrac{\sqrt{\overline{p'^2}}}{\bar p}\approxeq c_{\rho'T'}\dfrac{\sqrt{\overline{\rho'^2}}}{\bar\rho}
                                                                                   +           \dfrac{\sqrt{\overline{   T'^2}}}{\bar   T}+O\left(\dfrac{\overline{\rho'^2}}{\bar\rho^2}\right)\\\end{array}\right.
                                                                                                                                    \label{Eq_PDTEFCTPCF_s_S_ss_CCs_007}
\end{alignat}
taking into account that, by definition \eqref{Eq_PDTEFCTPCF_s_S_ss_CCs_001b}, $|c_{p'\rho'T'}|\leq1$, $|c_{\rho'\rho'T'}|\leq1$ and $|c_{\rho'T'T'}|\leq1$.
The leading term of the approximation error in \eqref{Eq_PDTEFCTPCF_s_S_ss_CCs_007} comes from the terms containing the triple correlations ($c_{p'\rho'T'}$, $c_{\rho'\rho'T'}$, $c_{\rho'T'T'}$) in \eqref{Eq_PDTEFCTPCF_s_S_ss_CCs_005},
and is $O([\bar\rho^{-1}\rho']_{\rm rms}[\bar{T}^{-1}T']_{\rm rms})=O([\bar\rho^{-1}\rho']^2_{\rm rms})$, by \eqref{Eq_PDTEFCTPCF_s_S_ss_CCs_002}.
Notice that \eqref{Eq_PDTEFCTPCF_s_S_ss_CCs_007} are obtained by making the assumption that
coefficients of variation, $[\bar\rho^{-1}\rho']_{\rm rms}$ and $[\bar{T}^{-1}T']_{\rm rms}$, are small, which is less stringent than assuming that the instantaneous levels, $\bar\rho^{-1}\rho'$ and $\bar{T}^{-1}T'$ are small, as
in the standard linearized approximation \citep[(3.114), p.~72]{Gatski_Bonnet_2009a}
The system \eqref{Eq_PDTEFCTPCF_s_S_ss_CCs_007} can be solved for the correlation coefficients ($c_{p'\rho'}$, $c_{p'T'}$ and $c_{\rho'T'}$)
\begin{figure}
\begin{center}
\begin{picture}(450,250)
\put(-65,-310){\includegraphics[angle=0,width=475pt]{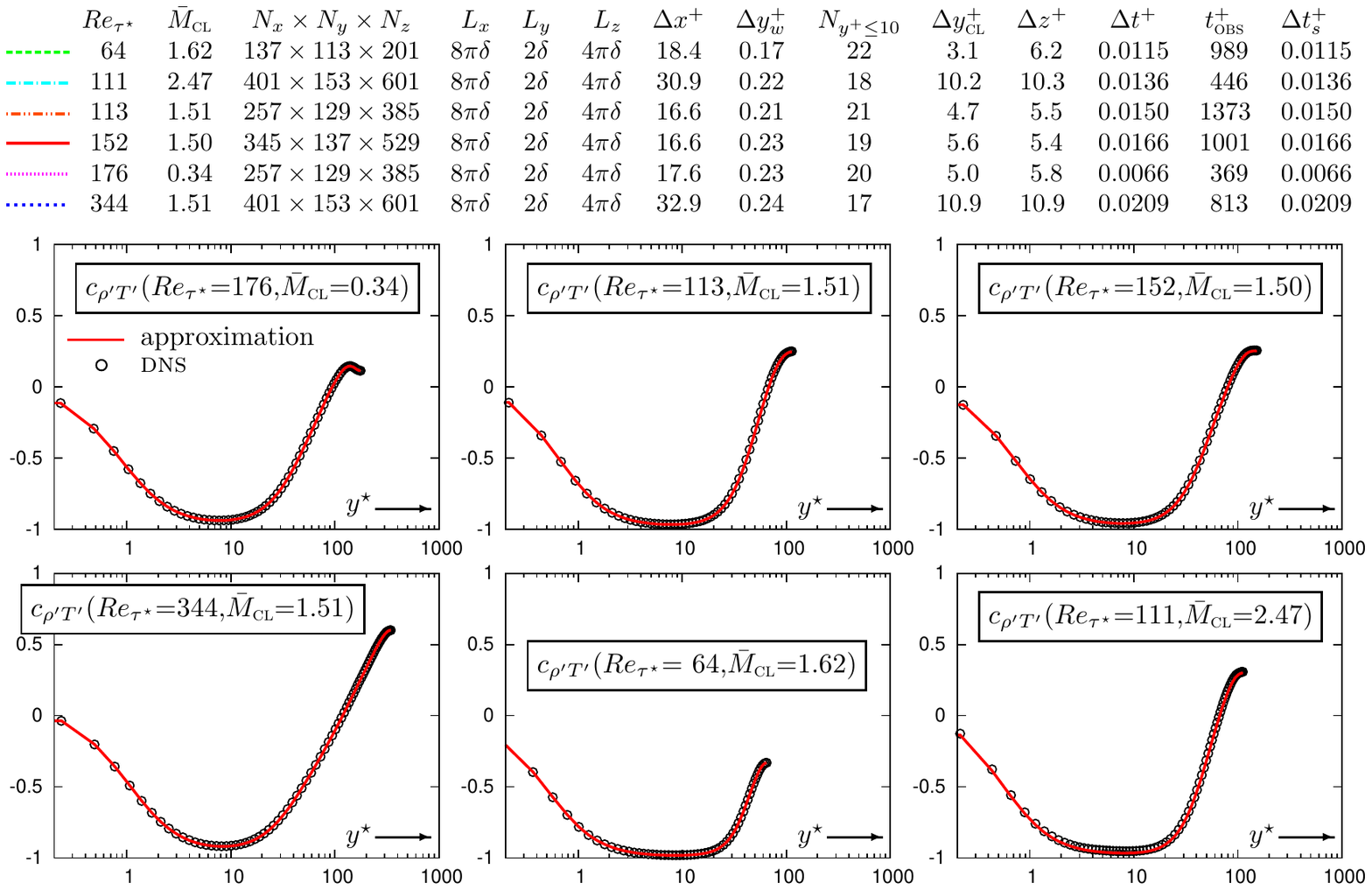}}
\end{picture}
\end{center}
\caption{Evaluation of the linearized approximation \eqref{Eq_PDTEFCTPCF_s_S_ss_CCs_008} expressing the correlation coefficient $c_{\rho'T'}$ as a unique function of the relative rms-levels of the thermodynamic fluctuations
($[\bar\rho^{-1}\rho']_{\rm rms}$, $[\bar T^{-1}T']_{\rm rms}$ and $[\bar p^{-1}p']_{\rm rms}$) by comparison with the present \tsn{DNS} data \tabref{Tab_PDTEFCTPCF_s_BEqsDNSC_ss_DNSCs_001}, plotted against $y^\star$ \eqref{Eq_PDTEFCTPCF_s_S_ss_CCF_001c},
for different values of ${Re}_{\tau^\star}$ and $\bar M_\tsn{CL}$ ({\color{red}{\large\bf{---}}} approximation, $\circ$ \tsn{DNS}).} 
\label{Fig_PDTEFCTPCF_s_S_ss_CCs_003}
%
\begin{center}
\begin{picture}(450,250)
\put(-65,-310){\includegraphics[angle=0,width=475pt]{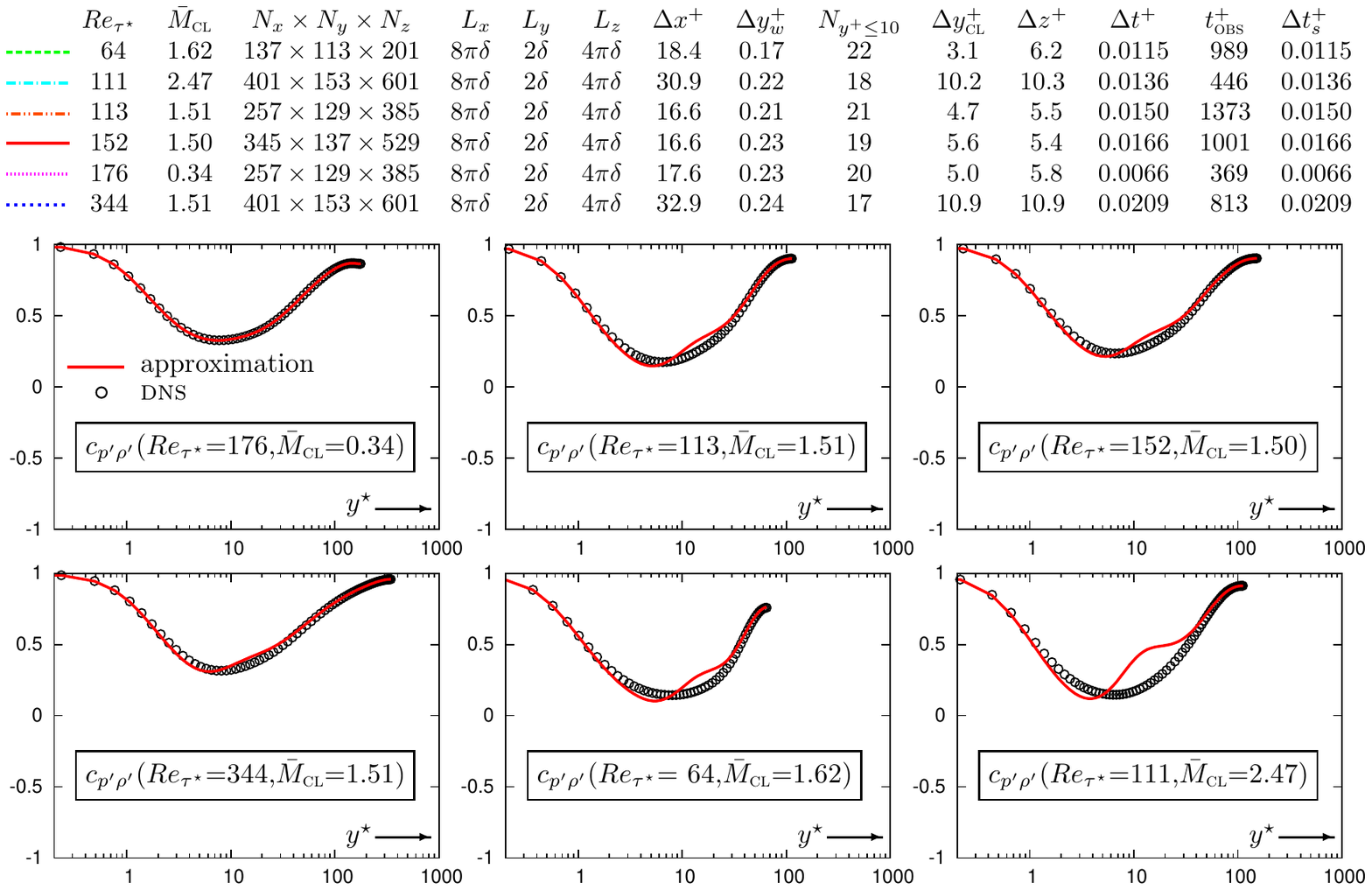}}
\end{picture}
\end{center}
\caption{Evaluation of the linearized approximation \eqref{Eq_PDTEFCTPCF_s_S_ss_CCs_008} expressing the correlation coefficient $c_{p'\rho'}$ as a unique function of the relative rms-levels of the thermodynamic fluctuations
($[\bar\rho^{-1}\rho']_{\rm rms}$, $[\bar T^{-1}T']_{\rm rms}$ and $[\bar p^{-1}p']_{\rm rms}$) by comparison with the present \tsn{DNS} data \tabref{Tab_PDTEFCTPCF_s_BEqsDNSC_ss_DNSCs_001}, plotted against $y^\star$ \eqref{Eq_PDTEFCTPCF_s_S_ss_CCF_001c},
for different values of ${Re}_{\tau^\star}$ and $\bar M_\tsn{CL}$ ({\color{red}{\large\bf{---}}} approximation, $\circ$ \tsn{DNS}).} 
\label{Fig_PDTEFCTPCF_s_S_ss_CCs_004}
\end{figure}
%
\begin{figure}
\begin{center}
\begin{picture}(450,250)
\put(-65,-310){\includegraphics[angle=0,width=475pt]{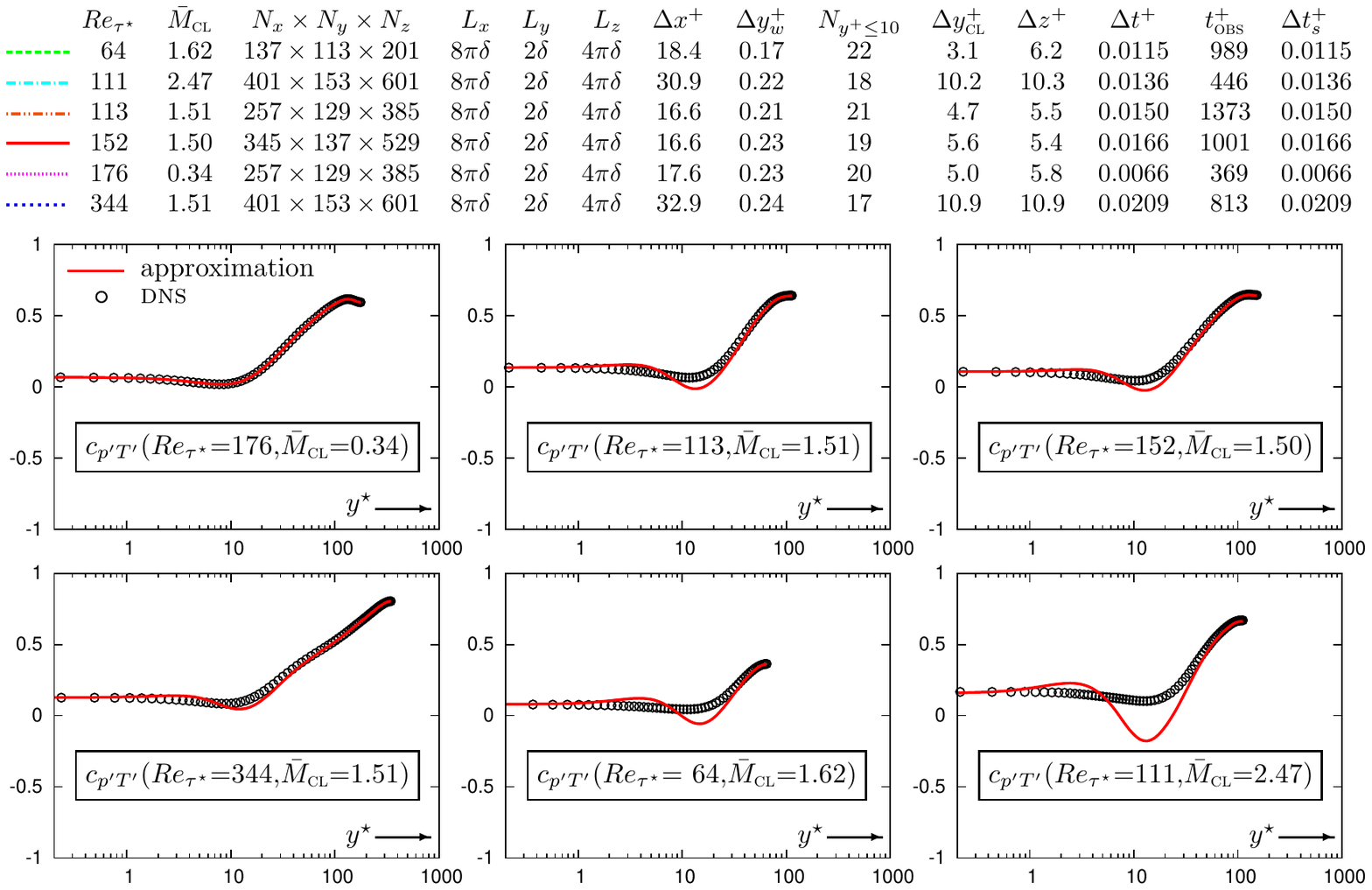}}
\end{picture}
\end{center}
\caption{Evaluation of the linearized approximation \eqref{Eq_PDTEFCTPCF_s_S_ss_CCs_008} expressing the correlation coefficient $c_{p'T'}$ as a unique function of the relative rms-levels of the thermodynamic fluctuations
($[\bar\rho^{-1}\rho']_{\rm rms}$, $[\bar T^{-1}T']_{\rm rms}$ and $[\bar p^{-1}p']_{\rm rms}$) by comparison with the present \tsn{DNS} data \tabref{Tab_PDTEFCTPCF_s_BEqsDNSC_ss_DNSCs_001}, plotted against $y^\star$ \eqref{Eq_PDTEFCTPCF_s_S_ss_CCF_001c},
for different values of ${Re}_{\tau^\star}$ and $\bar M_\tsn{CL}$ ({\color{red}{\large\bf{---}}} approximation, $\circ$ \tsn{DNS}).} 
\label{Fig_PDTEFCTPCF_s_S_ss_CCs_005}
\end{figure}
\begin{alignat}{6}
\!\!\!\!\!\!\!
\left.\begin{array}{c}\dfrac{\sqrt{\overline{\rho'^2}}}{\bar\rho}\ll1\\
                      \dfrac{\sqrt{\overline{   T'^2}}}{\bar   T}\ll1\\\end{array}\!\!\right\}\!\stackrel{\eqrefsab{Eq_PDTEFCTPCF_s_S_ss_CCs_007}
                                                                                                                   {Eq_PDTEFCTPCF_s_S_ss_CCs_002}}{\Longrightarrow}\!\!
\left\{\begin{array}{c}
\!
c_{p'\rho'}\approxeq-\dfrac{\left (\!\!\dfrac{\sqrt{\overline{   T'^2}}}{\bar   T}\!\right )^2\!\!-\!\left (\!\!\dfrac{\sqrt{\overline{\rho'^2}}}{\bar\rho}\!\right )^2\!\!-\!\left (\!\!\dfrac{\sqrt{\overline{p'^2}}}{\bar p}\!\right )^2}
                   {2\dfrac{\sqrt{\overline{p'^2}}}{\bar p}\dfrac{\sqrt{\overline{\rho'^2}}}{\bar\rho}}+O\left(\!\!\dfrac{\sqrt{\overline{\rho'^2}}}{\bar\rho}\!\right)\\
\\
\!
c_{p'T'}   \approxeq \dfrac{\left (\!\!\dfrac{\sqrt{\overline{   T'^2}}}{\bar   T}\!\right )^2\!\!-\!\left (\!\!\dfrac{\sqrt{\overline{\rho'^2}}}{\bar\rho}\!\right )^2\!\!+\!\left (\!\!\dfrac{\sqrt{\overline{p'^2}}}{\bar p}\!\right )^2}
                   {2\dfrac{\sqrt{\overline{p'^2}}}{\bar p}\dfrac{\sqrt{\overline{T'^2}}}{\bar{T}}}+O\left(\!\!\dfrac{\sqrt{\overline{\rho'^2}}}{\bar\rho}\!\right)\\
\\
\!
c_{\rho'T'}\approxeq-\dfrac{\left (\!\!\dfrac{\sqrt{\overline{   T'^2}}}{\bar   T}\!\right )^2\!\!+\!\left (\!\!\dfrac{\sqrt{\overline{\rho'^2}}}{\bar\rho}\!\right )^2\!\!-\!\left (\!\!\dfrac{\sqrt{\overline{p'^2}}}{\bar p}\!\right )^2}
                   {2\dfrac{\sqrt{\overline{\rho'^2}}}{\bar\rho}\dfrac{\sqrt{\overline{T'^2}}}{\bar{T}}}+O\left(\!\!\dfrac{\sqrt{\overline{\rho'^2}}}{\bar\rho}\!\right)\\\end{array}\right.                       
                                                                                                                                    \label{Eq_PDTEFCTPCF_s_S_ss_CCs_008}
\end{alignat}
implying that, if the relative rms-levels of fluctuation $[\bar{\rho}^{-1}\rho']_{\rm rms}$ and $[\bar{T}^{-1}T']_{\rm rms}$ are small, the correlation coefficients between thermodynamic fluctuations are uniquely defined by
the coefficients of variation of the thermodynamic variables ($[\bar{\rho}^{-1}\rho']_{\rm rms}$, $[\bar{T}^{-1}T']_{\rm rms}$ and $[\bar{p}^{-1}p']_{\rm rms}$).
As already stated regarding \eqref{Eq_PDTEFCTPCF_s_S_ss_CCs_007}, the leading term of the approximation error in \eqref{Eq_PDTEFCTPCF_s_S_ss_CCs_008}
comes from the terms containing the triple correlations ($c_{p'\rho'T'}$, $c_{\rho'\rho'T'}$, $c_{\rho'T'T'}$) in \eqref{Eq_PDTEFCTPCF_s_S_ss_CCs_005}, and
is $O([\bar\rho^{-1}\rho']_{\rm rms},[\bar{T}^{-1}T']_{\rm rms},[\bar{p}^{-1}p']_{\rm rms})$, \ie $O([\bar\rho^{-1}\rho']_{\rm rms})$ by \eqref{Eq_PDTEFCTPCF_s_S_ss_CCs_002}.
Notice that the error coming from the term $\bar{T}^{-1}\overline{T''}\stackrel{\eqref{Eq_PDTEFCTPCF_s_S_ss_CCs_002}}{=}-c_{\rho'T'}[\bar\rho^{-1}\rho']_{\rm rms}[\bar{T}^{-1}T']_{\rm rms}$ that appears in \eqref{Eq_PDTEFCTPCF_s_S_ss_CCs_005} is 
of higher order $O([\bar\rho^{-1}\rho']_{\rm rms}^2)$. The expression for $c_{\rho'T'}$ in \eqref{Eq_PDTEFCTPCF_s_S_ss_CCs_008} is similar with \citep[(3.4), p.~226]{Donzis_Jagannathan_2013a}, but the approximation of all
correlation coefficients \eqref{Eq_PDTEFCTPCF_s_S_ss_CCs_008} by the solution of the linear system \eqref{Eq_PDTEFCTPCF_s_S_ss_CCs_007} can be further exploited both to assess the approximation error and to obtain higher-order expansions in
powers of $[\bar\rho^{-1}\rho']_{\rm rms}$

In the present \tsn{DNS} computations \tabref{Tab_PDTEFCTPCF_s_BEqsDNSC_ss_DNSCs_001}, $[\bar{\rho}^{-1}\rho']_{\rm rms}\lessapprox 0.14$ and $[\bar{T}^{-1}T']_{\rm rms}\lessapprox 0.14$ \figref{Fig_PDTEFCTPCF_s_S_ss_CCF_002}.
These values prevail for $\bar M_\tsn{CL}=2.47$ \figref{Fig_PDTEFCTPCF_s_S_ss_CCF_002}, and are attained at the near-wall peaks located at $y^\star \approxeq 10$.
For $\bar M_\tsn{CL}\approxeq1.50$ these maximum values drop to $\sim0.04$ \figref{Fig_PDTEFCTPCF_s_S_ss_CCF_002}, since they scale with $\bar M_\tsn{CL}^2$ \figref{Fig_PDTEFCTPCF_s_S_ss_HCBMCL_002}.
The validity of the linearised approximation \eqref{Eq_PDTEFCTPCF_s_S_ss_CCs_008} was assessed by comparison with \tsn{DNS} results \tabref{Tab_PDTEFCTPCF_s_BEqsDNSC_ss_DNSCs_001},
for $c_{\rho'T'}$ \figref{Fig_PDTEFCTPCF_s_S_ss_CCs_003}, $c_{p'\rho'}$ \figref{Fig_PDTEFCTPCF_s_S_ss_CCs_004} and $c_{p'T'}$ \figref{Fig_PDTEFCTPCF_s_S_ss_CCs_005}.
The approximation \eqref{Eq_PDTEFCTPCF_s_S_ss_CCs_005} is very satisfactory for the $\bar M_\tsn{CL}=0.34$ case \figrefsatob{Fig_PDTEFCTPCF_s_S_ss_CCs_003}{Fig_PDTEFCTPCF_s_S_ss_CCs_005}, in agreement \figref{Fig_PDTEFCTPCF_s_S_ss_CCF_002} with the
$O([\bar\rho^{-1}\rho']_{\rm rms})$ estimate of the approximation error of \eqref{Eq_PDTEFCTPCF_s_S_ss_CCs_008}. The approximation of $c_{\rho'T'}$ by \eqref{Eq_PDTEFCTPCF_s_S_ss_CCs_005} is very satisfactory, $\forall y^\star$, even
for the higher Mach numbers \figref{Fig_PDTEFCTPCF_s_S_ss_CCs_003}. Regarding $c_{p'\rho'}$, the approximation \eqref{Eq_PDTEFCTPCF_s_S_ss_CCs_005} is globally satisfactory despite discrepancies in the region
$y^\star\in[8,30]$, which increase with $\bar M_\tsn{CL}$ \figref{Fig_PDTEFCTPCF_s_S_ss_CCs_004}. Finally, the approximation of $c_{p'T'}$ by \eqref{Eq_PDTEFCTPCF_s_S_ss_CCs_005} presents discrepancies in the same region $y^\star\in[8,30]$,
which increase with $\bar M_\tsn{CL}$ \figref{Fig_PDTEFCTPCF_s_S_ss_CCs_005}. The approximation \eqref{Eq_PDTEFCTPCF_s_S_ss_CCs_005} of $c_{p'T'}$ is not satisfactory for (${Re}_{\tau^\star}$,$\bar M_\tsn{CL}$)=(111,2.47) in the
region $y^\star\in[8,30]$, because, although it predicts correctly a weak correlation, it returns the wrong sign. Interestingly, possible discrepancies with \tsn{DNS} data decrease with increasing ${Re}_{\tau^\star}$, as shown 
\figrefsab{Fig_PDTEFCTPCF_s_S_ss_CCs_004}{Fig_PDTEFCTPCF_s_S_ss_CCs_005} by the $\bar M_\tsn{CL}\approxeq 1.5$ cases for ${Re}_{\tau^\star}\in\{113,152,344\}$.
\begin{figure}
\begin{center}
\begin{picture}(450,440)
\put(-65,-125){\includegraphics[angle=0,width=475pt]{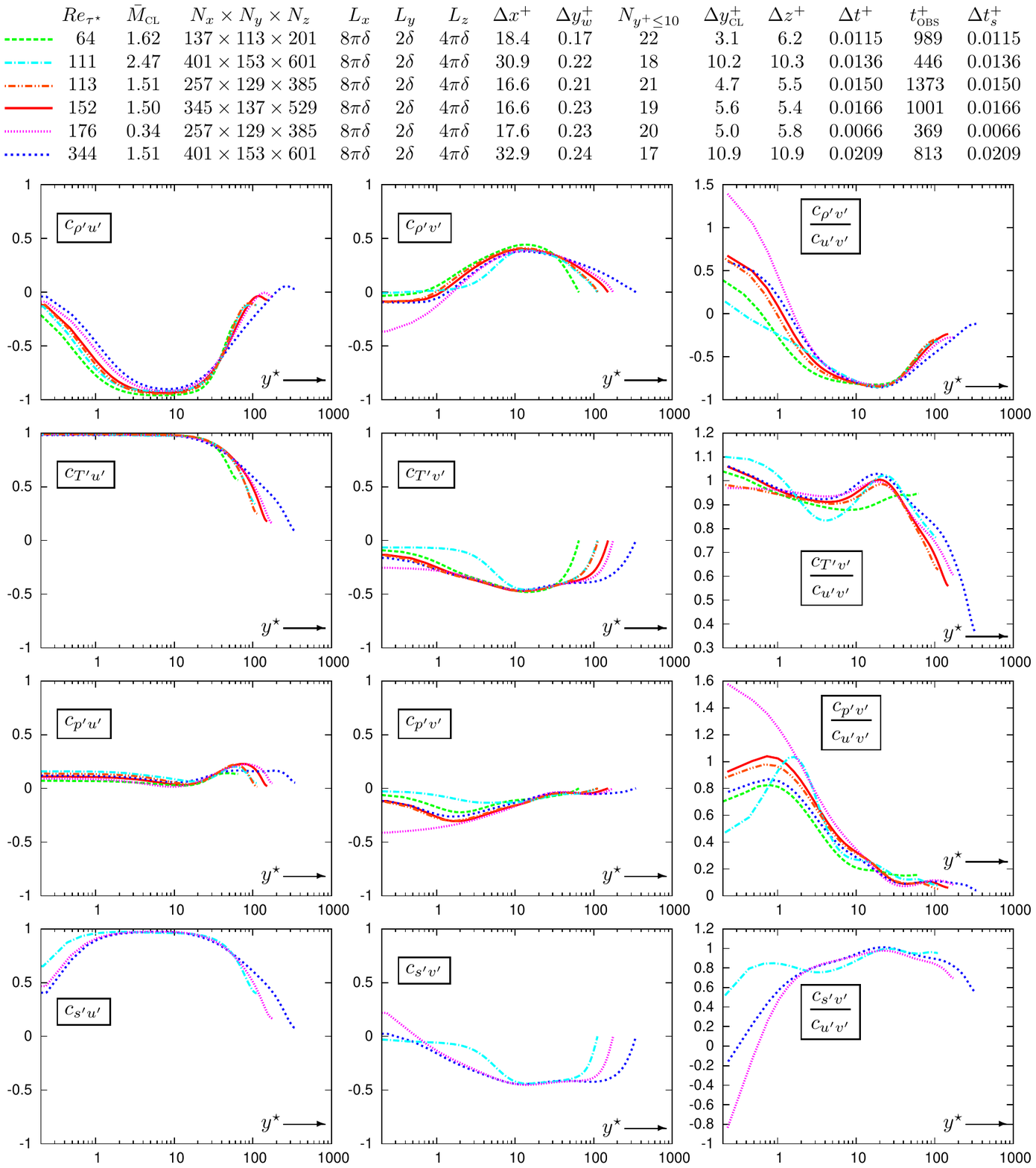}}
\end{picture}
\end{center}
\caption{\tsn{DNS} \tabref{Tab_PDTEFCTPCF_s_BEqsDNSC_ss_DNSCs_001} results for the profiles of the correlation coefficients \eqref{Eq_PDTEFCTPCF_s_S_ss_CCs_001a} between thermodynamic fluctuations and the streamwise
($c_{\rho'u'}$, $c_{   T'u'}$, $c_{   p'u'}$, $c_{   s'u'}$) and wall-normal ($c_{\rho'v'}$, $c_{   T'v'}$, $c_{   p'v'}$, $c_{   s'v'}$) velocity fluctuations, and of the ratio of 
the later (wall-normal transport of $\rho'$, $T'$, $p'$ and $s'$) to the shear momentum transport correlation coefficient $c_{u'v'}$, plotted (inner scaling)  against $y^\star$ \eqref{Eq_PDTEFCTPCF_s_S_ss_CCF_001c}.}
\label{Fig_PDTEFCTPCF_s_S_ss_CCs_006}
\end{figure}
%
\begin{figure}
\begin{center}
\begin{picture}(450,440)
\put(-65,-125){\includegraphics[angle=0,width=475pt]{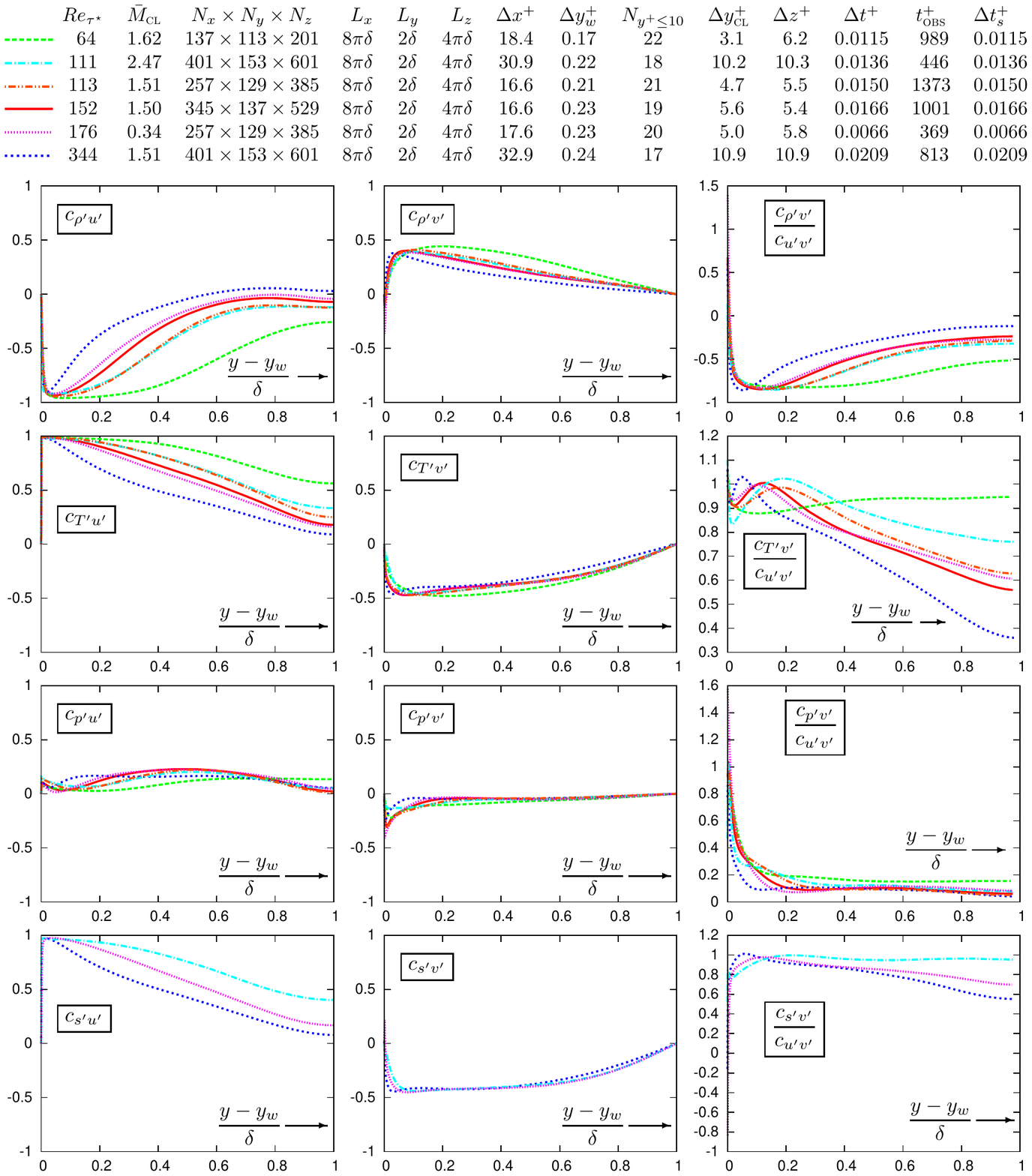}}
\end{picture}
\end{center}
\caption{\tsn{DNS} \tabref{Tab_PDTEFCTPCF_s_BEqsDNSC_ss_DNSCs_001} results for the profiles of the correlation coefficients \eqref{Eq_PDTEFCTPCF_s_S_ss_CCs_001a} between thermodynamic fluctuations and the streamwise
($c_{\rho'u'}$, $c_{   T'u'}$, $c_{   p'u'}$, $c_{   s'u'}$) and wall-normal ($c_{\rho'v'}$, $c_{   T'v'}$, $c_{   p'v'}$, $c_{   s'v'}$) velocity fluctuations, and of the ratio of 
the later (wall-normal transport of $\rho'$, $T'$, $p'$ and $s'$) to the shear momentum transport correlation coefficient $c_{u'v'}$, plotted (outer scaling) against $\delta^{-1}(y-y_w)$.}
\label{Fig_PDTEFCTPCF_s_S_ss_CCs_007}
\end{figure}
%
\begin{figure}
\begin{center}
\begin{picture}(450,340)
\put(-45,-165){\includegraphics[angle=0,width=465pt]{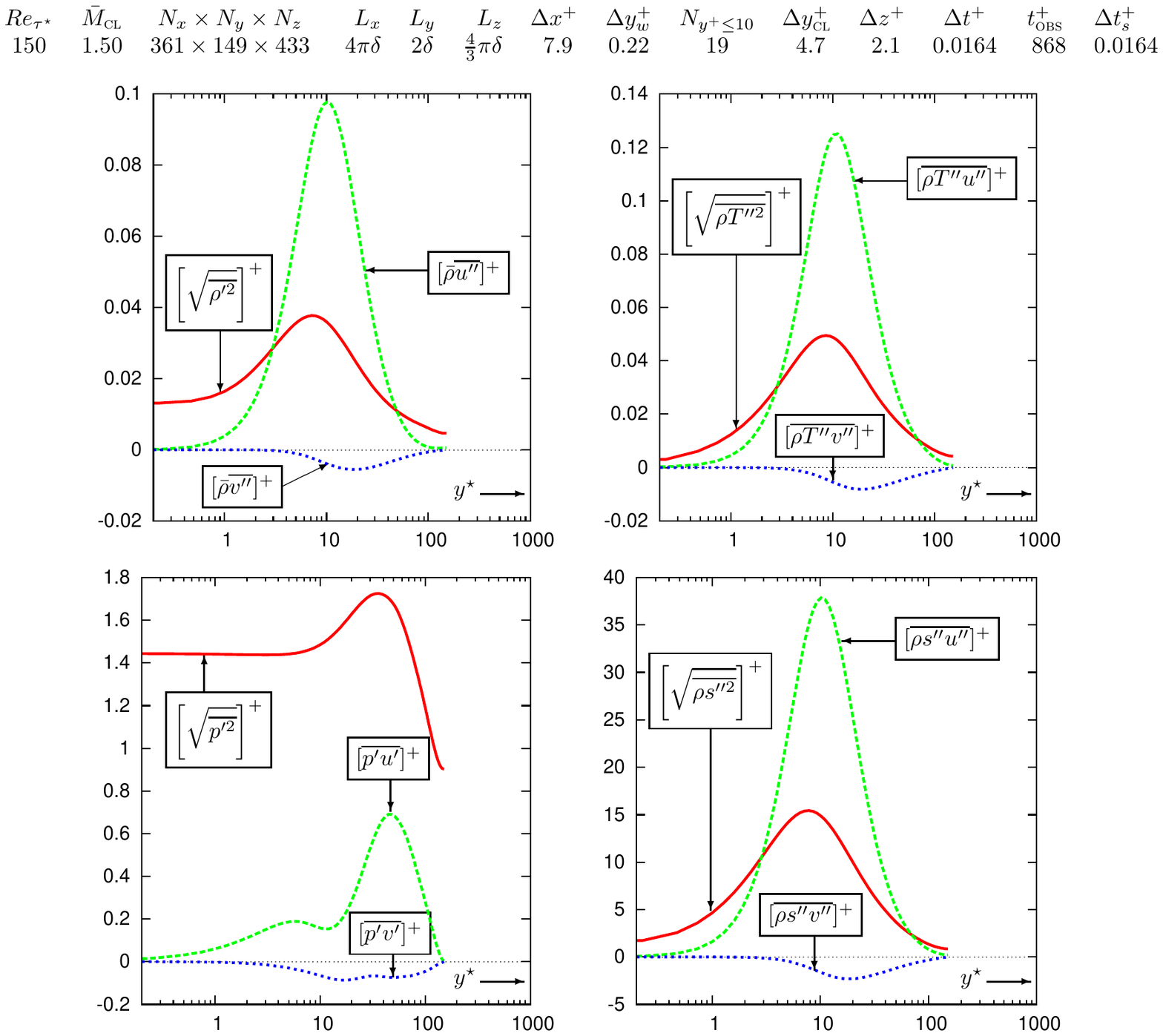}}
\end{picture}
\end{center}
\caption{\tsn{DNS} results (small box; \tabrefnp{Tab_PDTEFCTPCF_s_BEqsDNSC_ss_DNSCs_001}) at $(Re_{\tau^\star},\bar M_\tsn{CL})=(150,1.50)$ for the
variances ($\overline{\rho'^2}$, $\overline{p'^2}$, $\overline{\rho T''^2}$ and $\overline{\rho s''^2}$)
and the fluxes ($\bar\rho\overline{u''}\stackrel{\eqref{Eq_PDTEFCTPCF_s_BEqsDNSC_ss_SA_sss_RFD_001b}}{=}-\overline{\rho'u_i'}$, $\overline{p'u_i'}$, $\overline{\rho T''u_i''}$ and $\overline{\rho s''u_i''}$)
of the thermodynamic fluctuations (in wall-units; \parrefnp{PDTEFCTPCF_s_S_ss_WUs}), plotted against $y^\star$ \eqref{Eq_PDTEFCTPCF_s_S_ss_CCF_001c}.}
\label{Fig_PDTEFCTPCF_s_B_001}
\end{figure}
%
\begin{figure}
\begin{center}
\begin{picture}(450,360)
\put(-40,-190){\includegraphics[angle=0,width=465pt]{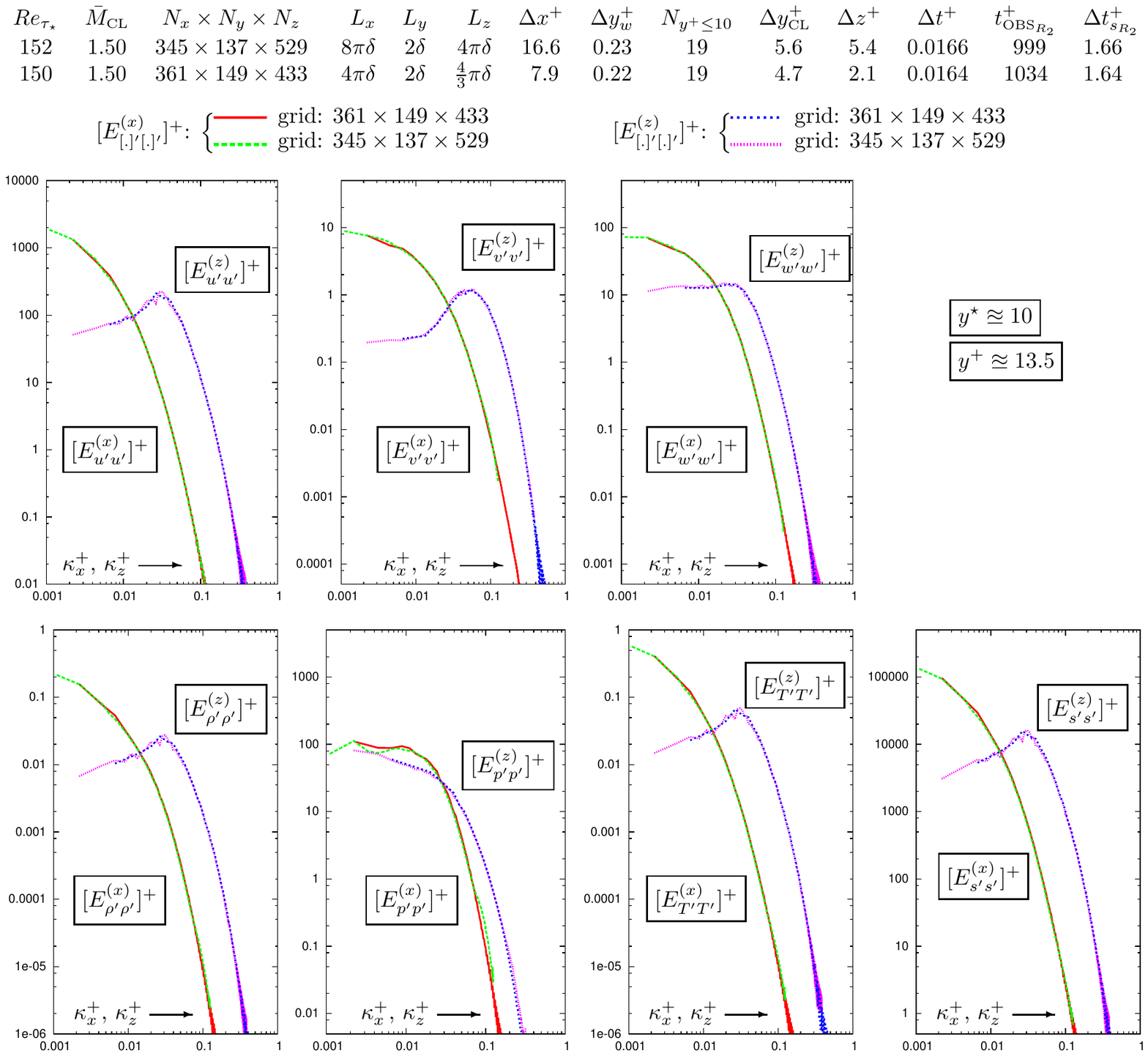}}
\end{picture}
\end{center}
\caption{Comparison in the neighbourhood \figref{Fig_PDTEFCTPCF_s_S_ss_HCBMCL_002} of the location of the maxima of thermodynamic fluctuations ($y^+\approxeq 10\;\; \Longleftrightarrow \; y^\star \approxeq 13.5$), between the large-box and the small-box
\tsn{DNS} results \tabref{Tab_PDTEFCTPCF_s_BEqsDNSC_ss_DNSCs_001}, of 1-D spectra (in wall units; \parrefnp{PDTEFCTPCF_s_S_ss_WUs}) in the homogeneous streamwise ($[E^{(x)}_{[.]'[.]'}]^+$) and spanwise ($[E^{(z)}_{[.]'[.]'}]^+$) directions, of the
fluctuating fields of velocity components ($u'$, $v'$ and $w'$) and of the thermodynamic state-variables ($p'$, $\rho'$, $T'$ and $s'$), plotted against the corresponding nondimensional wavenumbers ($\kappa_x^+$ or $\kappa_z^+$).}
\label{Fig_PDTEFCTPCF_s_B_002}
\end{figure}

The transport of thermodynamic fluctuations by the velocity field is represented by the streamwise ($c_{\rho'u'}$,$c_{T'u'}$,$c_{p'u'}$,$c_{s'u'}$) and wall-normal ($c_{\rho'v'}$,$c_{T'v'}$,$c_{p'v'}$,$c_{s'v'}$) correlation coefficients
\figrefsab{Fig_PDTEFCTPCF_s_S_ss_CCs_006}
          {Fig_PDTEFCTPCF_s_S_ss_CCs_007}.
The present flow being streamwise invariant in the mean ($\partial_x\overline{(.)}=0$), only wall-normal transport appears in various transport equations \parref{PDTEFCTPCF_s_B}, and in this respect it is interesting to consider the ratio of the
wall-normal correlation coefficients ($c_{\rho'v'}$,$c_{T'v'}$,$c_{p'v'}$,$c_{s'v'}$) to $c_{u'v'}$ which represents wall normal transport of the streamwise velocity fluctuation \figrefsab{Fig_PDTEFCTPCF_s_S_ss_CCs_006}
                                                                                                                                                                                            {Fig_PDTEFCTPCF_s_S_ss_CCs_007}.
The correlation coefficient of pressure transport, both streamwise $c_{p'u'}$ and wall-normal $c_{p'v'}$, which in the present flow is responsible for pressure diffusion \citep{Vallet_2007a,
                                                                                                                                                                                 Sauret_Vallet_2007a},
are very weak in the major part of the channel \figref{Fig_PDTEFCTPCF_s_S_ss_CCs_007}, except for $c_{p'v'}$ near the wall ($y^\star\lessapprox15$; \figrefnp{Fig_PDTEFCTPCF_s_S_ss_CCs_006}).
This is also observed regarding the ratio $c^{-1}_{u'v'}c_{p'v'}$, which is quite small ($\lessapprox0.1$) in the major part of the channel \figref{Fig_PDTEFCTPCF_s_S_ss_CCs_007},
except very near the wall ($y^\star\lessapprox30$; \figrefnp{Fig_PDTEFCTPCF_s_S_ss_CCs_006}). In the very-near-wall region ($y^\star\lessapprox3$; \figrefnp{Fig_PDTEFCTPCF_s_S_ss_CCs_006})
a very marked dependence on $\bar M_\tsn{CL}$ is obeserved, both for $c_{p'v'}$ and $c^{-1}_{u'v'}c_{p'v'}$. For the other thermodynamic variables, wall-normal transport correlation coefficients
($c_{\rho'v'}$, $c_{   T'v'}$, $c_{   s'v'}$) when plotted against the outer-scaled wall-distance $\delta^{-1}(y-y_w)$ \figref{Fig_PDTEFCTPCF_s_S_ss_CCs_007} appear to be quite independent of $(Re_{\tau^\star},\bar M_\tsn{CL}))$,
and this also true for inner-scaled wall-distance ($y^\star$), in the region $y^\star\in[10,100]$ \figref{Fig_PDTEFCTPCF_s_S_ss_CCs_006}.
The ratio of the wall-normal transport correlation coefficients ($c_{\rho'v'}$, $c_{   T'v'}$, $c_{   s'v'}$) on $c_{u'v'}$, it is reasonably independent of $(Re_{\tau^\star},\bar M_\tsn{CL})$ when plotted against
the inner-scaled wall-distance $y^\star$, for $y^\star\gtrapprox10$ \figref{Fig_PDTEFCTPCF_s_S_ss_CCs_006},
but shows a marked $Re_{\tau^\star}$-influence when plotted against the outer-scaled wall-distance $\delta^{-1}(y-y_w)$
(\figrefnp{Fig_PDTEFCTPCF_s_S_ss_CCs_007}; notice that the ratio is indeterminate at the centerline because, by symmetry, $[c_{u'v'}]_\tsn{CL}=0$). In particular,
the ratio $c^{-1}_{u'v'}c_{T'v'}$ remains \figref{Fig_PDTEFCTPCF_s_S_ss_CCs_006} nearly constant and close to $1\pm0.1$ from the wall up to $y^\star\approxeq 100\; \forall\; ({Re}_{\tau^\star},\bar M_\tsn{CL})$.
It is positive in the present flow because the wall is very cold, and we have $c_{u'T'}>0$ and $c_{v'T'}<0$, contrary to the case of {\em not-too-cold walls} \citep[p.~208]{Huang_Coleman_Bradshaw_1995a}, like in the boundary-layer \tsn{DNS} study
of \citet{Guarini_Moser_Shariff_Wray_2000a} where $c_{u'T'}<0$ and $c_{v'T'}>0$. It is easy to show, starting from the basic \tsn{HCB}-\tsn{SRA} relation \citep[(4.10), p.~208]{Huang_Coleman_Bradshaw_1995a}, that \tsn{HCB}-\tsn{SRA}, and
indeed all \tsn{SRA} extensions reviewed in \citet[pp.~207-208]{Huang_Coleman_Bradshaw_1995a} imply $|c_{u'T'}|\approxeq 1$ and $c_{v'T'}\approxeq c_{u'T'}c_{u'v'}$ which generalizes the relation given in \citet[(4.20), p.~20]{Guarini_Moser_Shariff_Wray_2000a}.
Finally, the streamwise correlation coefficients $c_{[.]'u'}$ ($c_{\rho'u'}$, $c_{   T'u'}$, $c_{   s'u'}$), when plotted against the inner-scaled wall-distance $y^\star$ \figref{Fig_PDTEFCTPCF_s_S_ss_CCs_006}, 
show little dependence on $(Re_{\tau^\star},\bar M_\tsn{CL}))$, except for $c_{\rho'u'}$ and $c_{s'u'}$ very near the wall ($y^\star\lessapprox 4$; \figrefnp{Fig_PDTEFCTPCF_s_S_ss_CCs_006}),
and a marked $Re_{\tau^\star}$-dependence when plotted against the outer-scaled wall-distance $\delta^{-1}(y-y_w)$ \figref{Fig_PDTEFCTPCF_s_S_ss_CCs_007}.

%
%
%
%
%
%
%
%
%
\section{Budgets}\label{PDTEFCTPCF_s_B}
%
%
%
%
%
%
%
%
%

Further insight into the dynamics of thermodynamic fluctuations and their interaction with the velocity-field, can be gained by studying the budgets of the transport equations
for the variances ($\overline{\rho'^2}$, $\overline{p'^2}$, $\overline{\rho s''^2}$ and $\overline{\rho h''^2}\stackrel{\eqref{Eq_PDTEFCTPCF_s_BEqsDNSC_ss_FM_001d}}{=}c_p\;\overline{\rho T''^2}$)
and the fluxes ($\overline{\rho'u_i'}$, $\overline{p'u_i'}$, $\overline{\rho s''u_i''}$ and $\overline{\rho h''u_i''}\stackrel{\eqref{Eq_PDTEFCTPCF_s_BEqsDNSC_ss_FM_001d}}{=}c_p\;\overline{\rho T''u_i''}$)
of the turbulent fluctuations of the thermodynamic state-variables \figref{Fig_PDTEFCTPCF_s_B_001}.
Notice \figref{Fig_PDTEFCTPCF_s_B_001} that, for $\{T'',s'',p'\}$, the streamwise fluxes are $\geq0$, while the wall-normal fluxes are $\leq0$, contrary to $\rho'$,
for which $\overline{\rho'u'}\stackrel{\eqref{Eq_PDTEFCTPCF_s_BEqsDNSC_ss_SA_sss_RFD_001c}}{=}-\overline{u''}\leq0$ and
$\overline{\rho'v'}\stackrel{\eqref{Eq_PDTEFCTPCF_s_BEqsDNSC_ss_SA_sss_RFD_001c}}{=}-\overline{v''}\geq0$ \figref{Fig_PDTEFCTPCF_s_B_001}.
In \parrefnp{PDTEFCTPCF_s_B_ss_DVMF} we study the transport equations for $\overline{u_i''}$ ($\overline{u''}\geq0$ and $\overline{v''}\leq0$; \figrefnp{Fig_PDTEFCTPCF_s_B_001}).
At the centerline all wall-normal fluxes are $=0$ by symmetry \figref{Fig_PDTEFCTPCF_s_B_001},
and the \tsn{DNS} results suggest that the streamwise fluxes are also very small at the centerline \figref{Fig_PDTEFCTPCF_s_B_001}.
The influence of $(Re_{\tau^\star},\bar M_\tsn{CL})$ on each term in the budgets of these transport equations is beyond the scope of the
present study, and will be the subject of future work. We concentrate instead on the detailed analysis of a highly resolved computation at $(Re_{\tau^\star},\bar M_\tsn{CL})\approxeq(150,1.5)$.

The large-box computational grid for the (${Re}_{\tau^\star}$,$\bar M_\tsn{CL}$)=(152,1.5) case has quite satisfactory resolution in wall-units \tabref{Tab_PDTEFCTPCF_s_BEqsDNSC_ss_DNSCs_001},
with repect to usual compressible channel \tsn{DNS} standards \citep{Coleman_Kim_Moser_1995a,
                                                                     Friedrich_Foysi_Sesterhenn_2006a,
                                                                     Tamano_Morinishi_2006a,
                                                                     Gerolymos_Senechal_Vallet_2010a,
                                                                     Wei_Pollard_2011b}.
Nonetheless, the transport equations for the variances and fluxes of thermodynamic variables contain complicated terms \parrefsatob{PDTEFCTPCF_s_B_ss_DVMF}
                                                                                                                                   {PDTEFCTPCF_s_B_ss_PVT},
including  triple
correlations of gradients of the fluctuating field and correlations containing the fluctuating dilatation $\Theta^{\backprime\backprime}$, which require higher resolution to achieve grid convergence,
compared to the basic 2-order moments \citep{Gerolymos_Senechal_Vallet_2010a}. For this
reason, the budgets of the transport equations were obtained on a finer grid with better resolution of the small scales ($\Delta x^+\approxeq 7.9$ and $\Delta z^+\approxeq 2.1$; \tabrefnp{Tab_PDTEFCTPCF_s_BEqsDNSC_ss_DNSCs_001}) but
in the smaller computational box \tabref{Tab_PDTEFCTPCF_s_BEqsDNSC_ss_DNSCs_001} used in the original computations of \citet{Coleman_Kim_Moser_1995a}. Comparison of 1-D spectra, in the homogeneous streamwise ($x$) and spanwise ($z$) directions,
for the fluctuations of the velocity components ($u'$, $v'$ and $w'$) and of the thermodynamic state-variables ($p'$, $\rho'$, $T'$ and $s'$), between the small-box and the large-box computations \tabref{Tab_PDTEFCTPCF_s_BEqsDNSC_ss_DNSCs_001},
shows very good agreement in the region of small wavenumbers \figref{Fig_PDTEFCTPCF_s_B_002} indicating  that the small-box is large enough.
The finer grid small-box computations resolve scales with a separation of at least 5 orders-of-magnitude for the streamwise spectra ($[E^{(x)}_{[.]'[.]'}]^+$; \figrefnp{Fig_PDTEFCTPCF_s_B_002}) 
and of at least 4 orders-of-magnitude for the spanwise spectra ($[E^{(z)}_{[.]'[.]'}]^+$; \figrefnp{Fig_PDTEFCTPCF_s_B_002}). Noticeable improvements in resolution  compared to the large-box computations are observed for $[E^{(x)}_{v'v'}]^+$
(almost 2 orders-of-magnitude; \figrefnp{Fig_PDTEFCTPCF_s_B_002}), for $[E^{(z)}_{v'v'}]^+$ (almost 1 order-of-magnitude; \figrefnp{Fig_PDTEFCTPCF_s_B_002}) and for $[E^{(x)}_{p'p'}]^+$ (almost 2 orders-of-magnitude; \figrefnp{Fig_PDTEFCTPCF_s_B_002}).
Notice that the 1-D spectra were computed \citep[p.~806]{Gerolymos_Senechal_Vallet_2010a} by taking the scaled \tsn{DFT} (discrete Fourier transform) of 2-point correlations obtained by sampling the computations every 100 iterations
($\Delta t^+_{s_{R_2}}=100 \Delta t^+$) over approximately the same observation time as for the single-point statistics ($t^+_{\tsn{OBS}_{R_2}}\approxeq t^+_{\tsn{OBS}}$) which were sampled at every iteration \tabref{Tab_PDTEFCTPCF_s_BEqsDNSC_ss_DNSCs_001}.

These single-point statisics were used to analyse \parrefsatob{PDTEFCTPCF_s_B_ss_DVMF}
                                                              {PDTEFCTPCF_s_B_ss_PVT}
the budgets of the transport equations for the variances and fluxes of thermodynamic fluctuations.
In the transport equations we use systematically the symbols $C_{(\cdot)}$ for convection, $d_{(\cdot)}$ for diffusion, $P_{(\cdot)}$ for production by mean-flow gradients,
$\Pi_{(\cdot)}$ for terms containing the fluctuating pressure-gradient $\partial_{x_i}p'$, $\varepsilon_{(\cdot)}$ for destruction by molecular mechanisms (viscosity or heat-conductivity),
$K_{(\cdot)}$ for direct compressibility effects proportional to $\overline{(\cdot)''}$, $B_{(\cdot)}$ for terms containing
the fluctuating dilatation $\Theta^{\backprime\backprime}$ \eqref{Eq_PDTEFCTPCF_s_BEqsDNSC_ss_SA_sss_NAN_002d} and $\Xi_{(\cdot)}$ for triple correlations.
In the present work, contrary to a previous investigation \citep{Senechal_2009a} analysing \tsn{DNS} data \citep{Gerolymos_Senechal_Vallet_2010a} obtained on coarser grids,
the fluctuating viscous stresses $\tau_{ij}'$ and heat-fluxes $q_i'$ were not decomposed in the transport equations by expanding the fluctuating part of \eqrefsab{Eq_PDTEFCTPCF_s_BEqsDNSC_ss_FM_001e}
                                                                                                                                                                 {Eq_PDTEFCTPCF_s_BEqsDNSC_ss_FM_001f},
and statistics containing $\tau_{ij}'$ or $q_i'$ were acquired directly using an onboard moving-averages technique \citep[\S4.4, p. 791]{Gerolymos_Senechal_Vallet_2010a}.

When considering streamwise invariant in the mean compressible turbulent plane channel flow,
the general forms of these transport equations can be  simplified, because of the specific symmetries
\begin{alignat}{6}
\left.
\begin{array}{lclclcl}\bar w                                        &=&\tilde w                                   &=&0\\
                                                                    & &                                           & & \\
                      \dfrac{\partial}{\partial t}\overline{(\cdot)}&=&\dfrac{\partial}{\partial t}\tilde{(\cdot)}&=&0\\
                                                                    & &                                           & & \\
                      \dfrac{\partial}{\partial x}\overline{(\cdot)}&=&\dfrac{\partial}{\partial x}\tilde{(\cdot)}&=&0\\
                                                                    & &                                           & & \\
                      \dfrac{\partial}{\partial z}\overline{(\cdot)}&=&\dfrac{\partial}{\partial z}\tilde{(\cdot)}&=&0\\\end{array}\right\}\stackrel{\eqrefsab{Eq_PDTEFCTPCF_s_BEqsDNSC_ss_FM_001a}
                                                                                                                                  {Eq_PDTEFCTPCF_s_BEqsDNSC_ss_SA_sss_RFD_001}}{\Longrightarrow}\left\{\begin{array}{l}\tilde v =0\\
                                                                                                                                   ~\\
                                                                                                                                   \breve\Theta=\bar\Theta-\overline{\Theta^{\backprime\backprime}}=0\\
                                                                                                                                                                                         \end{array}\right.
                                                                                                                                    \label{Eq_PDTEFCTPCF_s_B_001}
\end{alignat}
which imply that convection by the Favre-averaged mean velocities $C_{(\cdot)}\stackrel{\eqref{Eq_PDTEFCTPCF_s_B_001}}{=}0$.

%
%
%
%
%
\subsection{Density variance and mass-fluxes}\label{PDTEFCTPCF_s_B_ss_DVMF}
%
%
%
%
%

The transport equations for the density variance $\overline{\rho'^2}$ and for the mass-fluxes $\bar\rho\overline{u_i''}\stackrel{\eqref{Eq_PDTEFCTPCF_s_BEqsDNSC_ss_SA_sss_RFD_001c}}{=}-\overline{\rho'u_i'}$
were studied, in the context of compressible wall-turbulence, by \citet{Taulbee_vanOsdol_1991a}. The transport equation for the density variance is also central in the work of \citet{Yoshizawa_1992a} and
other related studies \citep{Hamba_Blaisdell_1997a,
                             Hamba_1999a,
                             Yoshizawa_Matsuo_Mizobuchi_2013a}. The retained form of the equations which are obtained from the fluctuating parts of the continuity
\eqref{Eq_PDTEFCTPCF_s_BEqsDNSC_ss_FM_001a} and momentum \eqref{Eq_PDTEFCTPCF_s_BEqsDNSC_ss_FM_001b} equations read
\begin{subequations}
                                                                                                                                    \label{Eq_PDTEFCTPCF_s_B_ss_DVMF_001}
\begin{alignat}{6}
                          \underbrace{
                          \dfrac{\partial\overline{\rho'^2}}{\partial t}+\tilde u_\ell\dfrac{\partial\overline{\rho'^2}}{\partial x_\ell}
                          }_{\displaystyle C_{(\rho'^2)}}
                       &\;=\;&\underbrace{-\dfrac{\partial\overline{\rho'^2u_\ell''}}{\partial x_\ell}}_{\displaystyle d_{(\rho'^2)}}\;
                              \underbrace{-2\overline{\rho'u_\ell'}\dfrac{\partial\bar\rho}{\partial x_\ell}
                                          -2\overline{\rho'^2}\breve\Theta}_{\displaystyle P_{(\rho'^2)}}\;
                              \underbrace{-2\bar\rho\overline{\rho'\Theta^{\backprime\backprime}}
                                          -\overline{\rho'^2\Theta^{\backprime\backprime}}}_{\displaystyle B_{(\rho'^2)}}
                                                                                                                                    \label{Eq_PDTEFCTPCF_s_B_ss_DVMF_001a}
\end{alignat}
\begin{alignat}{6}
& \underbrace{\dfrac{\partial\bar\rho\overline{u_i''}}
                    {\partial                       t}
             +\dfrac{\partial\bar\rho\overline{u_i''}\tilde u_\ell}
                    {\partial                               x_\ell}}_{\displaystyle C_{(u_i'')}}
= \underbrace{\dfrac{\partial       }
                    {\partial x_\ell}\left(\overline{\rho'u_i'' u_\ell''}
                                          +\overline{\tau'_{i\ell}\left(\dfrac{\bar\rho}
                                                                              {    \rho}\right)'}\right)}_{\displaystyle d_{(u_i'')}}
   \underbrace{
              -\bar\rho\overline{u_\ell''}\dfrac{\partial\tilde u_i}
                                                {\partial    x_\ell}
              +\overline{u_i''u_\ell''}\dfrac{\partial\bar\rho}
                                             {\partial  x_\ell}
              }_{\displaystyle P_{(u_i'')}}
  +\underbrace{\bar\rho\overline{u_i''\Theta^{\backprime\backprime}}}_{\displaystyle B_{(u_i'')}}
                                                                                                                                  \notag\\
  &\underbrace{-\overline{\left(\dfrac{\bar\rho}
                                      {\rho    }\right)'\dfrac{\partial p'}{\partial x_i}}}_{\displaystyle \Pi_{(u_i'')}}
  -\underbrace{\overline{\tau'_{i\ell}\dfrac{\partial}{\partial x_\ell}\left(\dfrac{\bar\rho}{\rho}\right)'}}_{\displaystyle \bar\rho\varepsilon_{(u_i'')}}      
  + \underbrace{
               \overline{\left(\dfrac{\bar\rho}
                                     {\rho    }-1\right)}\left(-\dfrac{\partial\bar p}
                                                                        {\partial x_i}
                                                               +\dfrac{\partial\bar\tau_{i\ell}}
                                                                      {\partial          x_\ell}\right)
              }_{\displaystyle K_{(u_i'')}}
   +\bar\rho\overline{f_{{\text{\sc v}}_i}''}
                                                                                                                                    \label{Eq_PDTEFCTPCF_s_B_ss_DVMF_001b}
\end{alignat}
\end{subequations}
Notice that, in line with \citet[(23), p. 4]{Taulbee_vanOsdol_1991a}, the inverse of the relative density $\bar\rho\rho^{-1}$ was used in \eqref{Eq_PDTEFCTPCF_s_B_ss_DVMF_001b}. 

\begin{figure}
\begin{center}
\begin{picture}(450,220)
\put(-45,-340){\includegraphics[angle=0,width=465pt]{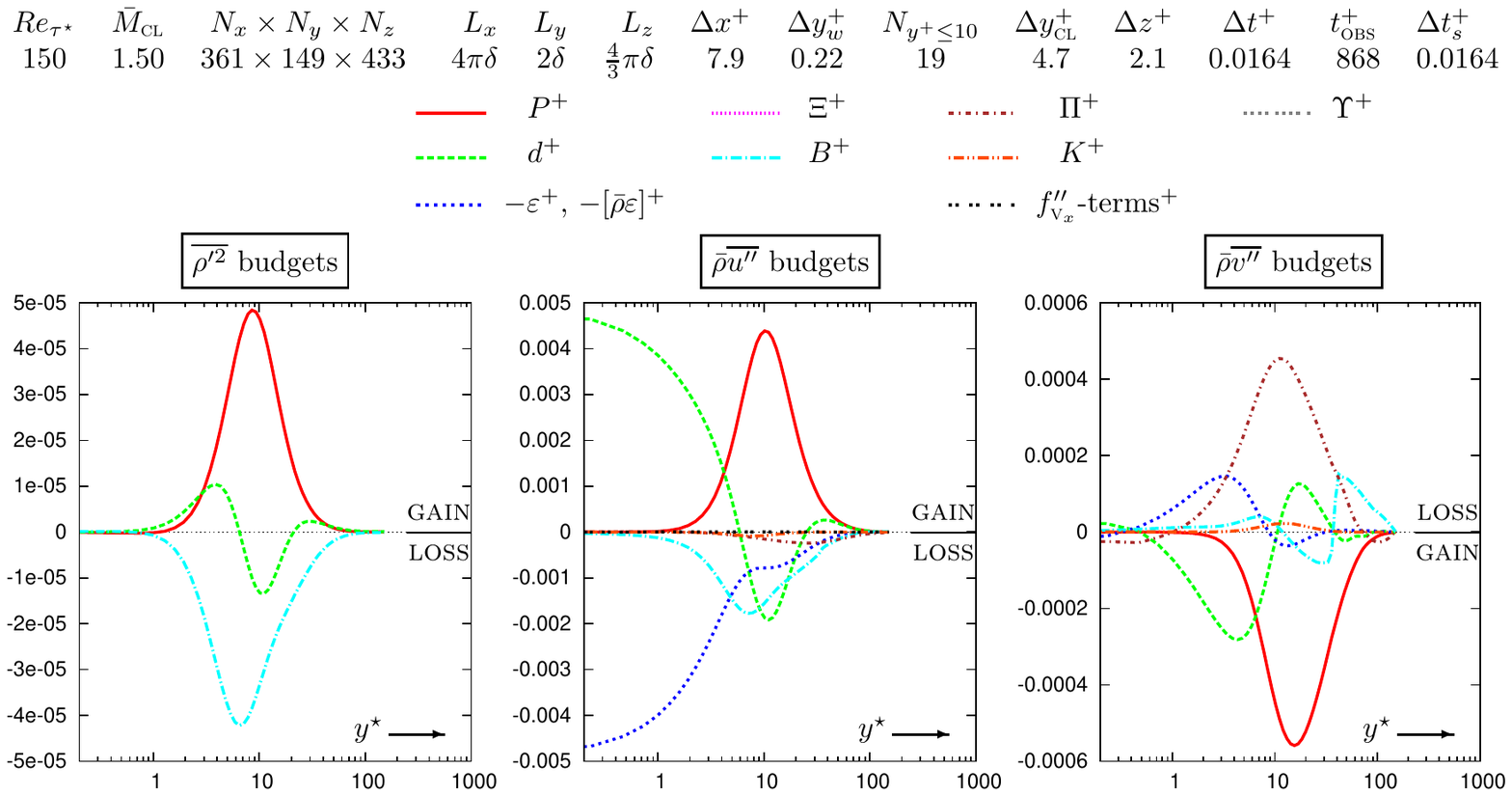}}
\end{picture}
\end{center}
\caption{Budgets, in wall-units \parref{PDTEFCTPCF_s_S_ss_WUs}, of the transport equations
\eqref{Eq_PDTEFCTPCF_s_B_ss_DVMF_001a} for the density variance $\overline{\rho'^2}$ ($d_{(\rho'^2)}$, $P_{(\rho'^2)}$, $B_{(\rho'^2)}$) and
\eqref{Eq_PDTEFCTPCF_s_B_ss_DVMF_001b} for the mass-fluxes $\overline{u_i''}$ ($d_{(u_i'')}$, $P_{(u_i'')}$, $B_{(u_i'')}$, $\Pi_{(u_i'')}$, $-\bar\rho\varepsilon_{(u_i'')}$, $K_{(u_i'')}$, $\bar\rho\overline{f_{\tsn{V}_x}''}$),
obtained from \tsn{DNS} computations ($Re_{\tau^\star}\approxeq150$, $\bar M_\tsn{CL}\approxeq1.50$; \tabrefnp{Tab_PDTEFCTPCF_s_BEqsDNSC_ss_DNSCs_001}), plotted against $y^\star$ \eqref{Eq_PDTEFCTPCF_s_S_ss_CCF_001c} .}
\label{Fig_PDTEFCTPCF_s_B_ss_DVMF_001}
\end{figure}

As already stated, \eqref{Eq_PDTEFCTPCF_s_B_001} implies that for streamwise invariant compressible turbulent plane channel flow
$C_{(\rho'^2)}\stackrel{\eqrefsab{Eq_PDTEFCTPCF_s_B_ss_DVMF_001a}
                                 {Eq_PDTEFCTPCF_s_B_001}}{=}0$ and
$C_{(u_i'')}\stackrel{\eqrefsab{Eq_PDTEFCTPCF_s_B_ss_DVMF_001b}
                               {Eq_PDTEFCTPCF_s_B_001}}{=}0$.
Furthermore, $\eqref{Eq_PDTEFCTPCF_s_B_001}\Longrightarrow\breve\Theta=0$, so that, for the present flow $\overline{\rho'^2}$ is produced by density stratification only,
\ie $P_{(\rho'^2)}\stackrel{\eqrefsab{Eq_PDTEFCTPCF_s_B_ss_DVMF_001a}
                                     {Eq_PDTEFCTPCF_s_B_001}}{=}-2\overline{\rho'v'}d_y\bar\rho\stackrel{\eqref{Eq_PDTEFCTPCF_s_BEqsDNSC_ss_SA_sss_RFD_001c}}{=}2\bar\rho\overline{v''}d_y\bar\rho\geq0$
because $\overline{v''}\leq0$ \figref{Fig_PDTEFCTPCF_s_B_001} and $d_y\bar\rho\leq0$ \figref{Fig_PDTEFCTPCF_s_S_ss_CCF_002}.
There is no molecular mechanism (related to viscosity or heat-conductivity) in \eqref{Eq_PDTEFCTPCF_s_B_ss_DVMF_001a}.
Examination of the $y^\star$-distributions of the terms in \eqref{Eq_PDTEFCTPCF_s_B_ss_DVMF_001a} indicates \figref{Fig_PDTEFCTPCF_s_B_ss_DVMF_001} that the principal mechanism governing $\overline{\rho'^2}$
is a balance between production $P_{(\rho'^2)}\stackrel{\eqrefsabc{Eq_PDTEFCTPCF_s_B_ss_DVMF_001a}{Eq_PDTEFCTPCF_s_B_001}{Eq_PDTEFCTPCF_s_BEqsDNSC_ss_SA_sss_RFD_001c}}{=}2\bar\rho\overline{v''}d_y\bar\rho\geq0\;\forall y^\star$ (gain)
and the dilatational term $B_{(\rho'^2)}\leq0\;\forall y^\star$ (loss).
The peak of production is located \figref{Fig_PDTEFCTPCF_s_B_001} at $y^\star\approxeq9$, very near the $\overline{\rho'^2}$-peak \figref{Fig_PDTEFCTPCF_s_B_001},
while the peak of $B_{(\rho'^2)}$ is located \figref{Fig_PDTEFCTPCF_s_B_001} at $y^\star\approxeq6$.
Notice several equivalent expressions
\begin{alignat}{6}
-B_{({\rho'}^2)}\stackrel{\eqref{Eq_PDTEFCTPCF_s_B_ss_DVMF_001a}}{=}\overline{\left(2\bar\rho \rho'+{\rho'}^2\right)\Theta^{\backprime\backprime}}
                \stackrel{\eqref{Eq_PDTEFCTPCF_s_BEqsDNSC_ss_SA_sss_RFD_001}}{=}\overline{\left(\rho^2-\bar\rho^2\right)\Theta^{\backprime\backprime}}
                \stackrel{\eqref{Eq_PDTEFCTPCF_s_BEqsDNSC_ss_SA_sss_RFD_001}}{=}\overline{\left(2\rho\rho'-\rho'^2\right)\Theta^{\backprime\backprime}}
                                                                                                                                    \label{Eq_PDTEFCTPCF_s_B_ss_DVMF_002}
\end{alignat}
for the dilatational term in \eqref{Eq_PDTEFCTPCF_s_B_ss_DVMF_001a}.
Expectedly, diffusion \smash{$d_{({\rho'}^2)}\stackrel{\eqrefsab{Eq_PDTEFCTPCF_s_B_ss_DVMF_001a}
                                                                {Eq_PDTEFCTPCF_s_B_001}}{=}-d_y(\overline{\rho'^2v''})$}
contributes positively (gain) to $\overline{\rho'^2}$ in the region $y^\star\lessapprox7$, where $\abs{B_{({\rho'}^2)}}>P_{(\rho'^2)}$, by transporting high-$\overline{\rho'^2}$ fluid towards the wall \figref{Fig_PDTEFCTPCF_s_B_001}.

The limiting values at the wall of all of the mechanisms ($d_{(\rho'^2)}$, $P_{(\rho'^2)}$, $B_{(\rho'^2)}$) in \eqref{Eq_PDTEFCTPCF_s_B_ss_DVMF_001a} are $0$ \figref{Fig_PDTEFCTPCF_s_B_001}. This can be shown analytically,
by combining the no-slip condition at the wall
\begin{subequations}
                                                                                                                                    \label{Eq_PDTEFCTPCF_s_B_ss_DVMF_003}
\begin{alignat}{6}
u_w=v_w=w_w\qquad\forall\;x,z,t
                                                                                                                                    \label{Eq_PDTEFCTPCF_s_B_ss_DVMF_003a}
\end{alignat}
with the flow symmetries \eqref{Eq_PDTEFCTPCF_s_B_001} and with the continuity equation \eqrefsab{Eq_PDTEFCTPCF_s_BEqsDNSC_ss_FM_001a}
                                                                                                 {Eq_PDTEFCTPCF_s_BEqsDNSC_ss_TTV_003a}.
Notice first that
\begin{alignat}{6}
\Theta_w^{\backprime\backprime}\stackrel{\eqrefsab{Eq_PDTEFCTPCF_s_BEqsDNSC_ss_SA_sss_NAN_002b}
                                                  {Eq_PDTEFCTPCF_s_B_001}                     }{=}
\Theta_w \stackrel{\eqrefsab{Eq_PDTEFCTPCF_s_BEqsDNSC_ss_FM_001e}
                            {Eq_PDTEFCTPCF_s_B_ss_DVMF_003a}     }{=}
\left.\dfrac{\partial v}{\partial y}\right|_w\stackrel{\eqref{Eq_PDTEFCTPCF_s_B_001}}{=}
\left.\dfrac{\partial v''}{\partial y}\right|_w
                                                                                                                                    \label{Eq_PDTEFCTPCF_s_B_ss_DVMF_003b}
\end{alignat}
since by \eqref{Eq_PDTEFCTPCF_s_B_ss_DVMF_003a} $(\partial_x u)_w=(\partial_x u'')_w=(\partial_z w)_w=(\partial_z w'')_w=0$.
By \eqrefsab{Eq_PDTEFCTPCF_s_B_ss_DVMF_003a}
            {Eq_PDTEFCTPCF_s_B_ss_DVMF_003b}
the limiting form of the continuity equation \eqrefsab{Eq_PDTEFCTPCF_s_BEqsDNSC_ss_FM_001a}
                                                      {Eq_PDTEFCTPCF_s_BEqsDNSC_ss_TTV_003a}
at the wall reads
\begin{alignat}{6}
\!\!\!\!\!\!\!\!
\eqrefsab{Eq_PDTEFCTPCF_s_BEqsDNSC_ss_FM_001a}
         {Eq_PDTEFCTPCF_s_BEqsDNSC_ss_TTV_003a}\!\stackrel{\eqrefsabc{Eq_PDTEFCTPCF_s_B_001}
                                                                     {Eq_PDTEFCTPCF_s_B_ss_DVMF_003a}
                                                                     {Eq_PDTEFCTPCF_s_B_ss_DVMF_003b}}{\Longrightarrow}\!
\dfrac{\partial\rho_w}{\partial t}+\rho_w\Theta_w^{\backprime\backprime}=0\Longrightarrow
\left.\dfrac{\partial\overline{v''}}{\partial y}\right|_w\!\!\!\stackrel{\eqref{Eq_PDTEFCTPCF_s_B_ss_DVMF_003b}}{=}\!\overline{\Theta_w^{\backprime\backprime}}
                                                                                                           =\dfrac{\partial}{\partial t}\!\!\left(\overline{\ln\dfrac{\bar\rho_w}{\rho_w}}\right)\!\!\stackrel{\eqref{Eq_PDTEFCTPCF_s_B_001}}{=}\!\!0
                                                                                                                                    \label{Eq_PDTEFCTPCF_s_B_ss_DVMF_003c}
\end{alignat}
where we used the fact that Reynolds-averaging (nonweighted ensemble averaging) commutes with differentiation and that the flow is steady in the mean \eqref{Eq_PDTEFCTPCF_s_B_001}.
Hence the continuity equation \eqrefsab{Eq_PDTEFCTPCF_s_BEqsDNSC_ss_FM_001a}
                                       {Eq_PDTEFCTPCF_s_BEqsDNSC_ss_TTV_003a} implies that,
although by \eqref{Eq_PDTEFCTPCF_s_B_ss_DVMF_003b} the instantaneous velocity near the wall $[v''(x,y,z,t)]^+\sim-[\partial_t(\ln(\bar\rho_w^{-1}\rho_w))]y^++O({y^+}^2)$ contains a linear with $y^+$ term,
on the average $[\overline{v''}(y)]^+\sim O({y^+}^2)$, a behaviour verified by the \tsn{DNS} data \figref{Fig_PDTEFCTPCF_s_B_001}, implying also that $\overline{\Theta_w^{\backprime\backprime}}=0$ \eqref{Eq_PDTEFCTPCF_s_B_ss_DVMF_003c}.
Multiplying the limiting form \eqref{Eq_PDTEFCTPCF_s_B_ss_DVMF_003c} of the continuity equation at the wall by $\rho_w'$ and $\rho_w$, respectively, yields upon averaging
\begin{alignat}{6}
\bar\rho_w\overline{\rho_w'\Theta_w^{\backprime\backprime}}+\overline{\rho_w'^2\Theta_w^{\backprime\backprime}}\stackrel{\eqref{Eq_PDTEFCTPCF_s_BEqsDNSC_ss_SA_sss_RFD_001a}}{=}
\overline{\rho_w\rho_w'\Theta_w^{\backprime\backprime}}\stackrel{\eqref{Eq_PDTEFCTPCF_s_B_ss_DVMF_003c}}{=}-\overline{\rho_w'\dfrac{\partial \rho_w}{\partial t}}
\stackrel{\eqrefsab{Eq_PDTEFCTPCF_s_BEqsDNSC_ss_SA_sss_RFD_001a}
                   {Eq_PDTEFCTPCF_s_BEqsDNSC_ss_SA_sss_RFD_001b}}{=}-\tfrac{1}{2}\dfrac{\partial\overline{\rho_w'^2}}{\partial t}\stackrel{\eqref{Eq_PDTEFCTPCF_s_B_001}}{=}0
                                                                                                                                    \label{Eq_PDTEFCTPCF_s_B_ss_DVMF_003d}\\
\bar\rho_w^2\overline{\Theta_w^{\backprime\backprime}}+2\bar\rho_w\overline{\rho_w'\Theta_w^{\backprime\backprime}}+\overline{\rho_w'^2\Theta_w^{\backprime\backprime}}\stackrel{\eqref{Eq_PDTEFCTPCF_s_BEqsDNSC_ss_SA_sss_RFD_001a}}{=}
\overline{\rho^2_w\Theta_w^{\backprime\backprime}}\stackrel{\eqref{Eq_PDTEFCTPCF_s_B_ss_DVMF_003c}}{=}-\tfrac{1}{2}\dfrac{\partial\overline{\rho_w^2}}{\partial t}\stackrel{\eqref{Eq_PDTEFCTPCF_s_B_001}}{=}0
                                                                                                                                    \label{Eq_PDTEFCTPCF_s_B_ss_DVMF_003e}
\end{alignat}
which can be solved for $\overline{\rho_w'\Theta_w^{\backprime\backprime}}$ and $\overline{\rho_w'^2\Theta_w^{\backprime\backprime}}$, yielding using also \eqref{Eq_PDTEFCTPCF_s_B_ss_DVMF_003c}
\begin{alignat}{6}
\eqrefsabc{Eq_PDTEFCTPCF_s_B_ss_DVMF_003c}
          {Eq_PDTEFCTPCF_s_B_ss_DVMF_003d}
          {Eq_PDTEFCTPCF_s_B_ss_DVMF_003e}\Longrightarrow
\bar\rho_w\overline{\rho_w'\Theta_w^{\backprime\backprime}}=\overline{\rho_w'^2\Theta_w^{\backprime\backprime}}=0
                                                                                                                                    \label{Eq_PDTEFCTPCF_s_B_ss_DVMF_003f}
\end{alignat}
\end{subequations}
Therefore,
\begin{subequations}
                                                                                                                                    \label{Eq_PDTEFCTPCF_s_B_ss_DVMF_004}
\begin{alignat}{6}
&[P_{(\rho'^2)}]_w\stackrel{\eqrefsab{Eq_PDTEFCTPCF_s_B_ss_DVMF_001a}
                                     {Eq_PDTEFCTPCF_s_B_001}         }{=}
-\left[2\overline{\rho'v'}\dfrac{d\bar\rho}{dy}\right]_w\stackrel{\eqrefsabc{Eq_PDTEFCTPCF_s_BEqsDNSC_ss_SA_sss_RFD_001a}
                                                                            {Eq_PDTEFCTPCF_s_BEqsDNSC_ss_SA_sss_RFD_001b}
                                                                            {Eq_PDTEFCTPCF_s_B_ss_DVMF_003a}}{=}0
                                                                                                                                    \label{Eq_PDTEFCTPCF_s_B_ss_DVMF_004a}\\
&[d_{(\rho'^2)}]_w\stackrel{\eqrefsab{Eq_PDTEFCTPCF_s_B_ss_DVMF_001a}
                                     {Eq_PDTEFCTPCF_s_B_001}         }{=}
-\left[\dfrac{d}{dy}\overline{\rho'^2v''}\right]_w\stackrel{\eqref{Eq_PDTEFCTPCF_s_B_ss_DVMF_003a}}{=}
-\left[\overline{\rho'^2\dfrac{dv''}{dy}}\right]_w\stackrel{\eqref{Eq_PDTEFCTPCF_s_B_ss_DVMF_003b}}{=}
-\overline{\rho_w'^2\Theta_w^{\backprime\backprime}}\stackrel{\eqref{Eq_PDTEFCTPCF_s_B_ss_DVMF_003f}}{=}0
                                                                                                                                    \label{Eq_PDTEFCTPCF_s_B_ss_DVMF_004b}\\
&[B_{(\rho'^2)}]_w\stackrel{\eqrefsab{Eq_PDTEFCTPCF_s_B_ss_DVMF_001a}
                                     {Eq_PDTEFCTPCF_s_B_ss_DVMF_003f}}{=}0
                                                                                                                                    \label{Eq_PDTEFCTPCF_s_B_ss_DVMF_004c}
\end{alignat}
\end{subequations}
Regarding the streamwise mass-flux $\bar\rho\overline{u''}\stackrel{\eqref{Eq_PDTEFCTPCF_s_BEqsDNSC_ss_SA_sss_RFD_001c}}{=}-\overline{\rho'u'}$, production is related both with mean shear and with density  stratification,
$P_{(u'')}\stackrel{\eqrefsab{Eq_PDTEFCTPCF_s_B_ss_DVMF_001b}{Eq_PDTEFCTPCF_s_B_001}}{=}-\bar\rho\overline{v''}d_y\tilde{u}+\overline{u''v''}d_y\bar\rho\geq 0$, because $\overline{v''}\leq 0$ \figref{Fig_PDTEFCTPCF_s_B_001},
$d_y\tilde{u}\geq 0$ \figref{Fig_PDTEFCTPCF_s_S_ss_CCF_002}, $\overline{u''v''}\leq 0$ is of the same sign as $\overline{\rho u''v''}$ \figref{Fig_PDTEFCTPCF_s_S_ss_HCBMCL_001} and $c_{u'v'}$ \figref{Fig_PDTEFCTPCF_s_S_ss_CCs_006},
and $d_y\bar\rho\leq 0$ \figref{Fig_PDTEFCTPCF_s_S_ss_CCF_002}.
The production $P_{(u'')}$ peak is located \figref{Fig_PDTEFCTPCF_s_B_ss_DVMF_001} at $y^\star\approxeq10$ very near the  $\overline{u''}$-peak \figref{Fig_PDTEFCTPCF_s_B_001}. The direct compressibility term 
in \eqref{Eq_PDTEFCTPCF_s_B_ss_DVMF_001b} \smash{$K_{(u''^2)}\stackrel{\eqrefsab{Eq_PDTEFCTPCF_s_B_ss_DVMF_001b}{Eq_PDTEFCTPCF_s_B_001}}{=}0$} exactly, because of the streamwise invariance of the mean-flow \eqref{Eq_PDTEFCTPCF_s_B_001}.
The body-acceleration term \smash{$\bar\rho\overline{f''_{\tsn{V}_x}}\stackrel{\eqref{Eq_PDTEFCTPCF_s_BEqsDNSC_ss_SA_sss_RFD_001c}}{=}-\overline{\rho'f'_{\tsn{V}_x}}$} in \eqref{Eq_PDTEFCTPCF_s_B_ss_DVMF_001b}
is negligibly small everywhere ($\forall y^\star$) in the $\overline{u''}$-budgets \figref{Fig_PDTEFCTPCF_s_B_ss_DVMF_001}.
The term $\Pi_{(u'')}$ related with the fluctuating pressure gradient in \eqref{Eq_PDTEFCTPCF_s_B_ss_DVMF_001b} is weak everywhere \figref{Fig_PDTEFCTPCF_s_B_ss_DVMF_001}, 
and negligibly small both near the wall ($y^\star\lessapprox5$) and in the outer part of the flow \figref{Fig_PDTEFCTPCF_s_B_ss_DVMF_001}. Near the wall ($y^\star\lessapprox4$)
the budgets of the streamwise mass-flux are dominated \figref{Fig_PDTEFCTPCF_s_B_ss_DVMF_001} by the competition between destruction $-\bar\rho\varepsilon_{(u'')}$ (loss) and diffusion $d_{(u'')}$ (gain), both of which
reach their maxima at the wall where \smash{$[P_{(u'')}]_w\stackrel{\eqrefsab{Eq_PDTEFCTPCF_s_B_ss_DVMF_001b}{Eq_PDTEFCTPCF_s_B_ss_DVMF_003a}}{=}0$}.
The dilatational term $B_{(u'')}\leq0\;\forall y^\star$ (loss) is of the same magnitude as $-\bar\rho\varepsilon_{(u'')}$ for $y^\star\gtrapprox5$ \figref{Fig_PDTEFCTPCF_s_B_ss_DVMF_001}.

The relative importance of various mechanism in \eqref{Eq_PDTEFCTPCF_s_B_ss_DVMF_001b} governing the wall-normal mass-flux \smash{$\bar\rho\overline{v''}\stackrel{\eqref{Eq_PDTEFCTPCF_s_BEqsDNSC_ss_SA_sss_RFD_001c}}{=}-\overline{\rho'v'}\leq0$}
\figref{Fig_PDTEFCTPCF_s_B_001} differs \figref{Fig_PDTEFCTPCF_s_B_ss_DVMF_001} from the streamwise mass-flux case. Production, because of the flow symmetries \eqref{Eq_PDTEFCTPCF_s_B_001}, is related to mean-density stratification only,
\smash{$P_{(v'')}\stackrel{\eqrefsab{Eq_PDTEFCTPCF_s_B_ss_DVMF_001b}{Eq_PDTEFCTPCF_s_B_001}}{=}\overline{v''^2}d_y\bar\rho\leq 0$}, because $\overline{v''^2}\geq 0$ \figref{Fig_PDTEFCTPCF_s_B_001} and $d_y\bar\rho\leq 0$ \figref{Fig_PDTEFCTPCF_s_S_ss_CCF_002},
with a peak located at $y^\star\approxeq16$ \figref{Fig_PDTEFCTPCF_s_B_ss_DVMF_001}, very near the $\overline{v''}$ peak \figref{Fig_PDTEFCTPCF_s_B_001}. Notice that since $\overline{v''}\leq 0$ \figref{Fig_PDTEFCTPCF_s_B_001},
$P_{(v'')}\leq 0$ implies gain in the $\overline{v''}$-budgets. Contrary to the $\overline{u''}$-budgets, the principal mechanism opposing production $P_{(v'')}$ in the $\overline{v''}$-budgets \figref{Fig_PDTEFCTPCF_s_B_ss_DVMF_001}
is the fluctuating pressure gradient term $\Pi_{(v'')}\geq 0$ for $y^\star\gtrapprox 2$. Diffusion $d_{(v'')}$ and destruction $-\bar\rho\varepsilon_{(v'')}$ change sign 2 or 3 times across the channel
generally opposing one another \figref{Fig_PDTEFCTPCF_s_B_ss_DVMF_001}. Near the wall $y^\star\in[\tfrac{1}{2},3]$ destruction $-\bar\rho\varepsilon_{(v'')}\geq \Pi_{(v'')}$ becomes the main loss-mechanism 
in the $\overline{v''}$-budgets \figref{Fig_PDTEFCTPCF_s_B_ss_DVMF_001}, balanced by diffusion $d_{(v'')}$ (gain), while in the sublayer ($y^\star\lessapprox0.5$) $\Pi_{(v'')}\leq 0$ (gain)
opposing $d_{(v'')}\geq 0$ (loss) become the dominant mechanisms at the wall.
Finally, the dilatational term $B_{(v'')}$ is active mainly in the outer part of the flow ($y^\star\gtrapprox40$; \figrefnp{Fig_PDTEFCTPCF_s_B_ss_DVMF_001}).

%
%
%
%
%
\subsection{Entropy variance and fluxes}\label{PDTEFCTPCF_s_B_ss_EVF}
%
%
%
%
%

Entropy production \eqref{Eq_PDTEFCTPCF_s_BEqsDNSC_ss_TTV_003b} and the continuity equation \eqref{Eq_PDTEFCTPCF_s_BEqsDNSC_ss_TTV_003a} are the two generating relations in the development of transport equations for the thermodynamic variables 
\eqrefsatob{Eq_PDTEFCTPCF_s_BEqsDNSC_ss_TTV_003c}{Eq_PDTEFCTPCF_s_BEqsDNSC_ss_TTV_003g}. In analogy with the continuity equation \eqref{Eq_PDTEFCTPCF_s_BEqsDNSC_ss_TTV_003a} which expresses the substantial derivative of density $\bar\rho^{-1}D_t\rho$ in 
terms of the dilatation  $\Theta$ \eqrefsab{Eq_PDTEFCTPCF_s_BEqsDNSC_ss_FM_001e}{Eq_PDTEFCTPCF_s_BEqsDNSC_ss_SA_sss_NAN_002b}, the entropy transport equation expresses the substantial derivative of entropy $\rho T D_t s$ in terms of the viscous
stresses and heat-fluxes, representing the influence of viscosity and heat conductivity on the flow. Most of the studies  involving thermodynamic fluctuations in compressible turbulence have focussed  on flows away from solid walls and have used simplified 
assumptions for the entropy fluctuations \citep{Ristorcelli_1997a,
                                                Rubinstein_Erlebacher_1997a,
                                                Hamba_1999a}.

The evolution of the entropy variance and fluxes across the channel is very similar with those of density and temperature \figref{Fig_PDTEFCTPCF_s_B_001}, suggesting that the viscosity effects represented by $\rho T D_t s$ \eqref{Eq_PDTEFCTPCF_s_BEqsDNSC_ss_TTV_003b},
and hence entropy fluctuations, are important in wall-bounded flows. The exact equations for the mass-weighted entropy variance $\overline{\rho s''^2}$ and fluxes $\overline{\rho s'' u_i''}$ can be obtained from the fluctuating parts of the
entropy \eqref{Eq_PDTEFCTPCF_s_BEqsDNSC_ss_TTV_003b} and momentum \eqref{Eq_PDTEFCTPCF_s_BEqsDNSC_ss_FM_001b} equations and read 
\begin{subequations}
                                                                                                                                    \label{Eq_PDTEFCTPCF_s_B_ss_EVF_001}
\begin{alignat}{6}
&\underbrace{
 \dfrac{\partial\overline{\rho s''^2}             }
       {\partial                                 t}
+\dfrac{\partial\overline{\rho s''^2}\tilde u_\ell}
       {\partial                            x_\ell}}_{\displaystyle C_{(\rho s''^2)}}
                                                   = \underbrace{-\dfrac{\partial\overline{\rho s''^2u_\ell''}}
                                                                        {\partial                       x_\ell}
                                                                 -\dfrac{\partial       }
                                                                        {\partial x_\ell}\overline{\left(2\dfrac{s'q_\ell'}{T}\right)}}_{\displaystyle d_{(\rho s''^2)}}
                                                                                                                                  \notag\\
                                                    &\underbrace{-2\overline{\rho u_\ell''s''}\dfrac{\partial\tilde s}
                                                                                                    {\partial  x_\ell}
                                                                 +2\overline{\dfrac{s'}{T}\tau_{m\ell}'}\bar   S_{m\ell}
                                                                 +2\overline{\dfrac{s'}{T}   S_{m\ell}'}\bar\tau_{m\ell}
                                                                 +2\overline{\left(\dfrac{s'}{T}\right)}\bar\tau_{m\ell}\bar S_{m\ell}
                                                                 -2\overline{\left(\dfrac{s'}{T}\right)}\dfrac{\partial\bar q_\ell}
                                                                                                              {\partial     x_\ell}
                                                                 +2\overline{s'q_\ell'}\dfrac{\partial}{\partial x_\ell}\overline{\left(\dfrac{1}{T}\right)}}_{\displaystyle P_{(\rho s''^2)}}
                                                                                                                                  \notag\\
                                                    &-\underbrace{\left(-2\overline{\dfrac{q_\ell'}{T}\dfrac{\partial s'}
                                                                                                            {\partial  x_\ell}}\right)}_{\displaystyle \bar\rho\varepsilon_{(\rho s''^2)}}\;
                                                     \underbrace{+2\overline{s''}\left(\overline{\dfrac{\tau_{m\ell} S_{m\ell}}{T}}
                                                                                      -\overline{\dfrac{1}{T}\dfrac{\partial q_\ell}
                                                                                                                   {\partial x_\ell}}\right)}_{\displaystyle K_{(\rho s''^2)}}\;
                                                     \underbrace{+2\overline{s'q_\ell'\dfrac{\partial}{\partial x_\ell}\left(\dfrac{1}{T}\right)'}
                                                                 +2\overline{\dfrac{s'}{T}\tau_{m\ell}'S_{m\ell}'}}_{\displaystyle \Xi_{(\rho s''^2)}}\;
                                                                                                                                    \label{Eq_PDTEFCTPCF_s_B_ss_EVF_001a}
\end{alignat}
\begin{figure}
\begin{center}
\begin{picture}(450,210)
\put(-45,-340){\includegraphics[angle=0,width=465pt]{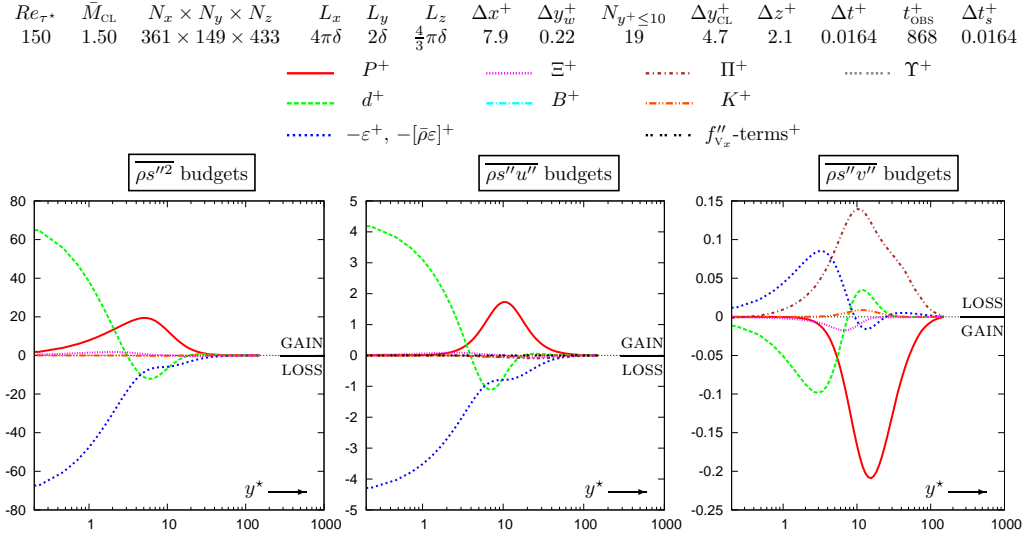}}
\end{picture}
\end{center}
\caption{Budgets, in wall-units \parref{PDTEFCTPCF_s_S_ss_WUs}, of the transport equations
\eqref{Eq_PDTEFCTPCF_s_B_ss_EVF_001a} for the entropy variance $\overline{\rho s''^2}$ ($d_{(\rho s''^2)}$, $P_{(\rho s''^2)}$, $-\bar\rho\varepsilon_{(\rho s''^2)}$, $K_{(\rho s''^2)}$, $\Xi_{(\rho s''^2)}$) and
\eqref{Eq_PDTEFCTPCF_s_B_ss_EVF_001b} for the entropy fluxes $\overline{\rho s''u_i''}$
($d_{(\rho s''u_i'')}$, $P_{(\rho s''u_i'')}$, $K_{(\rho s''u_i'')}$, $\Xi_{(\rho s''u_i'')}$, $-\bar\rho\varepsilon_{(\rho s''u_i'')}$, $\Pi_{(\rho s''u_i'')}$, $\overline{\rho s'' f''_{\tsn{V}_x}}$),
obtained from \tsn{DNS} computations ($Re_{\tau^\star}\approxeq150$, $\bar M_\tsn{CL}\approxeq1.50$; \tabrefnp{Tab_PDTEFCTPCF_s_BEqsDNSC_ss_DNSCs_001}), plotted against $y^\star$ \eqref{Eq_PDTEFCTPCF_s_S_ss_CCF_001c}.}
\label{Fig_PDTEFCTPCF_s_B_ss_EVF_001}
\end{figure}
\begin{alignat}{6}
&\underbrace{
 \dfrac{\partial\overline{\rho s''u_i''}             }
       {\partial                                    t}
+\dfrac{\partial\overline{\rho s''u_i''}\tilde u_\ell}
       {\partial                               x_\ell}}_{\displaystyle C_{(\rho s''u_i'')}}
                                                      = \underbrace{-\dfrac{\partial\overline{\rho s''u_i''u_\ell''}}
                                                                           {\partial                          x_\ell}
                                                                    +\dfrac{\partial\overline{s'\tau_{i\ell}'}}
                                                                           {\partial                    x_\ell}
                                                                    -\dfrac{\partial       }
                                                                           {\partial x_\ell}\overline{\left(\dfrac{u_i'q_\ell'}{T}\right)}}_{\displaystyle d_{(\rho s''u_i'')}}
                                                                                                                                  \notag\\
                                                    &\underbrace{-\overline{\rho u_\ell''s''}\dfrac{\partial\tilde u_i}
                                                                                                   {\partial    x_\ell}
                                                                 -\overline{\rho u_i''u_\ell''}\dfrac{\partial\tilde s}
                                                                                                     {\partial  x_\ell}
                                                                 +\overline{u_i'q_\ell'}\dfrac{\partial}{\partial x_\ell}\overline{\left(\dfrac{1}{T}\right)}
                                                                 -\overline{\left(\dfrac{u_i'}{T}\right)}\dfrac{\partial\bar q_\ell}
                                                                                                               {\partial     x_\ell}
                                                                 +\overline{\dfrac{u_i'}{T}\tau_{m\ell}'}\bar   S_{m\ell}
                                                                 +\overline{\dfrac{u_i'}{T}   S_{m\ell}'}\bar\tau_{m\ell}
                                                                 +\overline{\left(\dfrac{u_i'}{T}\right)}\bar\tau_{m\ell}\bar S_{m\ell}}_{\displaystyle P_{(\rho s''u_i'')}}
                                                                                                                                  \notag\\
                                                     &\underbrace{+\overline{u_i''}\left(\overline{\dfrac{\tau_{m\ell} S_{m\ell}}{T}}
                                                                                        -\overline{\dfrac{1}{T}\dfrac{\partial q_\ell}
                                                                                                                     {\partial x_\ell}}\right)
                                                                  +\overline{s''}\left(-\dfrac{\partial\bar p}
                                                                                             {\partial   x_i}
                                                                                      +\dfrac{\partial\bar\tau_{i\ell}}
                                                                                             {\partial          x_\ell}\right)}_{\displaystyle K_{(\rho s''u_i'')}}\;
                                                     \underbrace{+\overline{u_i'q_\ell'\dfrac{\partial}{\partial x_\ell}\left(\dfrac{1}{T}\right)'}
                                                                 +\overline{\dfrac{u_i'}{T}\tau_{m\ell}'S_{m\ell}'}}_{\displaystyle \Xi_{(\rho s''u_i'')}}\;
                                                                                                                                  \notag\\
                                                    -&\underbrace{\left(-\overline{\dfrac{q_\ell'}{T}\dfrac{\partial u_i'   }
                                                                                                           {\partial  x_\ell}}
                                                                        +\overline{\tau_{i\ell}'\dfrac{\partial s'     }
                                                                                                      {\partial x_\ell}}
                                                                 \right)}_{\displaystyle \bar\rho\varepsilon_{(\rho s''u_i'')}}\;
                                                      \underbrace{-\overline{ s'\dfrac{\partial p'     }
                                                                                      {\partial x_i   }}}_{\displaystyle \Pi_{(\rho s''u_i'')}}\;
                                                      +\overline{\rho s'' f''_{\tsn{V}_i}}
                                                                                                                                    \label{Eq_PDTEFCTPCF_s_B_ss_EVF_001b}
\end{alignat}
\end{subequations}
Because of \eqref{Eq_PDTEFCTPCF_s_BEqsDNSC_ss_TTV_003b}, the factor $T^{-1}$ appears in most of the terms of the transport equations for
$\overline{\rho s''^2}$ \eqref{Eq_PDTEFCTPCF_s_B_ss_EVF_001a} and for $\overline{\rho s''u_i''}$ \eqref{Eq_PDTEFCTPCF_s_B_ss_EVF_001b}.
In the retained form of the equations \eqref{Eq_PDTEFCTPCF_s_B_ss_EVF_001}, $T^{-1}$ was treated as a weight,
like $\rho$ is treated in Favre-averaging \eqref{Eq_PDTEFCTPCF_s_BEqsDNSC_ss_SA_sss_RFD_001c},
in line with the general presentation of weighted averaging by \citet{Favre_1965a}.

The production of entropy fluctuations $P_{(\rho s''^2)}$ involves several mechanisms. Taking into account the flow symmetries \eqref{Eq_PDTEFCTPCF_s_B_001}, these mechanisms include \eqrefsab{Eq_PDTEFCTPCF_s_B_ss_EVF_001a}{Eq_PDTEFCTPCF_s_B_001}
production by mean entropy stratification $-2\overline{\rho v''s''}d_y\tilde{s}$, by mean temperature stratification $2\overline{s'q_y'}d_y \overline{T^{-1}}$, by the gradient of the mean heat-flux $-2\overline{T^{-1}s'}d_y\bar{q}_y$ (which involves
because of Fourier's law \eqref{Eq_PDTEFCTPCF_s_BEqsDNSC_ss_FM_001f} the 2-derivative of mean temperature $d_{yy}^2\bar{T}$, and 3 viscous work terms coming from the expansion of $\tau_{m\ell}S_{m\ell}$ in \eqref{Eq_PDTEFCTPCF_s_BEqsDNSC_ss_TTV_003b}.
\tsn{DNS} data show that $P_{(\rho s''^2)}\geq0$ (gain) everywhere \figref{Fig_PDTEFCTPCF_s_B_ss_EVF_001} with a peak located at $y^\star\approxeq5$, and decays very slowly as $y^\star\to 0$.
Destruction $-\bar\rho\varepsilon_{(\rho s''^2)}\leq0$ (loss) everywhere \figref{Fig_PDTEFCTPCF_s_B_ss_EVF_001}, while diffusion $d_{(\rho s''^2)}\leq0$ (loss) for $y^\star\gtrapprox3$ and 
$d_{(\rho s''^2)}\geq0$ (gain) for $y^\star\lessapprox3$ \figref{Fig_PDTEFCTPCF_s_B_ss_EVF_001}. Very near the wall ($y^\star\lessapprox2$) diffusion $d_{(\rho s''^2)}\geq0$ (gain)
mainly balances \figref{Fig_PDTEFCTPCF_s_B_ss_EVF_001} destruction $-\bar\rho\varepsilon_{(\rho s''^2)}\leq0$ (loss).
Finally, direct compressibility effects $K_{(\rho s''^2)}$ and the triple correlations term $\Xi_{(\rho s''^2)}$ are negligibly small everywhere \figref{Fig_PDTEFCTPCF_s_B_ss_EVF_001}.

Regarding the streamwise entropy flux $\overline{\rho s'' u''}\geq0$ \figref{Fig_PDTEFCTPCF_s_B_001}, production  $P_{(\rho s'' u'')}$ \eqref{Eq_PDTEFCTPCF_s_B_ss_EVF_001b} involves all the mechanisms present in $P_{(\rho s''^2)}$ \eqref{Eq_PDTEFCTPCF_s_B_ss_EVF_001a},
and an extra-mechanism, which for the present flow \eqref{Eq_PDTEFCTPCF_s_B_001} is associated with mean shear, $-\overline{\rho s'' v''}d_y\tilde{u}$ \eqrefsab{Eq_PDTEFCTPCF_s_B_ss_EVF_001b}{Eq_PDTEFCTPCF_s_B_001}.
The \tsn{DNS} data indicate that $P_{(\rho s'' u'')}\geq 0$ everywhere \figref{Fig_PDTEFCTPCF_s_B_ss_EVF_001}, with a peak located at $y^\star\approxeq10$, and diminishes as $y^+\to 0$, at
a much faster rate compared with  $P_{(\rho s''^2)}$. Notice that all of the terms in $P_{(\rho s'' u''_i)}$ are proportional to correlations of the fluctuating velocities \eqref{Eq_PDTEFCTPCF_s_B_ss_EVF_001b}, implying by \eqref{Eq_PDTEFCTPCF_s_B_ss_DVMF_003a}
that $[P_{(\rho s'' u'')}]_w=[P_{(\rho s'' v'')}]_w=0$, whereas $[P_{(\rho s''^2)}]_w\neq 0$ \eqref{Eq_PDTEFCTPCF_s_B_ss_EVF_001a}.
The direct compressibility effects $K_{(\rho s'' u'')}$, the triple correlations term $\Xi_{(\rho s'' u'')}$, the term related with the fluctuating pressure gradient $\Pi_{(\rho s'' u'')}$ and the body-acceleration term $\overline{\rho s'' f''_{\tsn{V}_x}}$
are negligible everywhere \figref{Fig_PDTEFCTPCF_s_B_ss_EVF_001} in the $\overline{\rho s'' u''}$-budgets \eqref{Eq_PDTEFCTPCF_s_B_ss_EVF_001b}. Diffusion $d_{(\rho s'' u'')}$ and destruction $-\bar\rho\varepsilon_{(\rho s'' u'')}\leq0$ (loss) 
behave \figref{Fig_PDTEFCTPCF_s_B_ss_EVF_001} in a way similar with their counterparts in the $\overline{\rho s''^2}$-budgets, and approximately balance one another at the wall \figref{Fig_PDTEFCTPCF_s_B_ss_EVF_001}.

Finally, the budgets of the wall-normal flux $\overline{\rho s'' v''}\leq0$ \figref{Fig_PDTEFCTPCF_s_B_001} differ \figref{Fig_PDTEFCTPCF_s_B_ss_EVF_001} from those of the streamwise flux $\overline{\rho s'' u''}$ because of the relative importance of the term 
$\Pi_{(\rho s'' v'')}\geq0$ (loss) related with the fluctuating pressure-gradient in \eqref{Eq_PDTEFCTPCF_s_B_ss_EVF_001b}, which \figref{Fig_PDTEFCTPCF_s_B_ss_EVF_001}
is the main mechanism opposing production $P_{(\rho s'' v'')}\leq 0$ (gain). The peak of production $P_{(\rho s'' v'')}$ is located at $y^\star\lessapprox15$,
near the peak of $\overline{\rho s'' v''}$ \figref{Fig_PDTEFCTPCF_s_B_001}, while the peak of $\Pi_{(\rho s'' v'')}$ is located at $y^\star\lessapprox10$ \figref{Fig_PDTEFCTPCF_s_B_ss_EVF_001}.
Both the triple correlations $\Xi_{(\rho s'' v'')}\leq 0$ (gain) and the direct compressibility effects $K_{(\rho s'' v'')}\geq 0$ (loss) have weak contribution to the $\overline{\rho s''v''}$-budgets \figref{Fig_PDTEFCTPCF_s_B_ss_EVF_001}.
Finally, destruction $-\bar\rho\varepsilon_{(\rho s'' v'')}\geq0$ and diffusion $d_{(\rho s'' v'')}\leq0$ oppose each other everywhere, and are the dominant mechanisms in the near-wall region ($y^\star\lessapprox 2$) where
all of the other terms ($P_{(\rho s'' v'')}$, $\Pi_{(\rho s'' v'')}$, $\Xi_{(\rho s'' v'')}$ and $K_{(\rho s'' v'')}$) are negligibly small \figref{Fig_PDTEFCTPCF_s_B_ss_EVF_001}.

%
%
%
%
%
\subsection{Temperature variance and heat-fluxes}\label{PDTEFCTPCF_s_B_ss_TVHF}
%
%
%
%
%

Turbulent heat-fluxes $\overline{\rho h''u_i''}\stackrel{\eqref{Eq_PDTEFCTPCF_s_BEqsDNSC_ss_FM_001d}}{=}-c_p\;\overline{\rho T''u_i''}$ \citep{Bowersox_2009a} are important in compressible flows because they appear in the mean energy equation
\citep[(3.6), p.~196]{Huang_Coleman_Bradshaw_1995a} and affect the mean-flow via the mean static temperature field.
\citet{Tamano_Morinishi_2006a} studied the \tsn{DNS} budgets of the transport equation for the temperature variance $\overline{\rho  T''^2}$ in compressible turbulent plane channel flow, and \citet{Shahab_Lehnasch_Gatski_Comte_2011a} those of
the equation for temperature transport $\overline{\rho T''u_i''}$ in \tsn{ZPG} compressible turbulent boundary-layer flow ($\bar{M}_e\approxeq 2.25$) over both isothermal ($\bar{T}_e^{-1}\bar{T}_w=1.35$) and adiabatic walls.
Both these studies were based on the $c_v$-form \eqref{Eq_PDTEFCTPCF_s_BEqsDNSC_ss_TTV_003d} of the temperature equation, and contain dilatational ($[\cdot]\breve\Theta$ or $\overline{[\cdot]\Theta^{\backprime\backprime}}$) terms. 
Alternatively \citep{LeRibault_Friedrich_1997a} it is possible to work with the $c_p$-form, containing instead terms stemming from the correlation $c_p^{-1}\overline{T''(D_t p)'}$ in the equations for the temperature variance $\overline{\rho T''^2}$
or $c_p^{-1}\overline{u_i''(D_t p)'}$ in the equations for the temperature transport $\overline{\rho T''u_i''}$. 

\begin{figure}
\begin{center}
\begin{picture}(450,210)
\put(-45,-340){\includegraphics[angle=0,width=465pt]{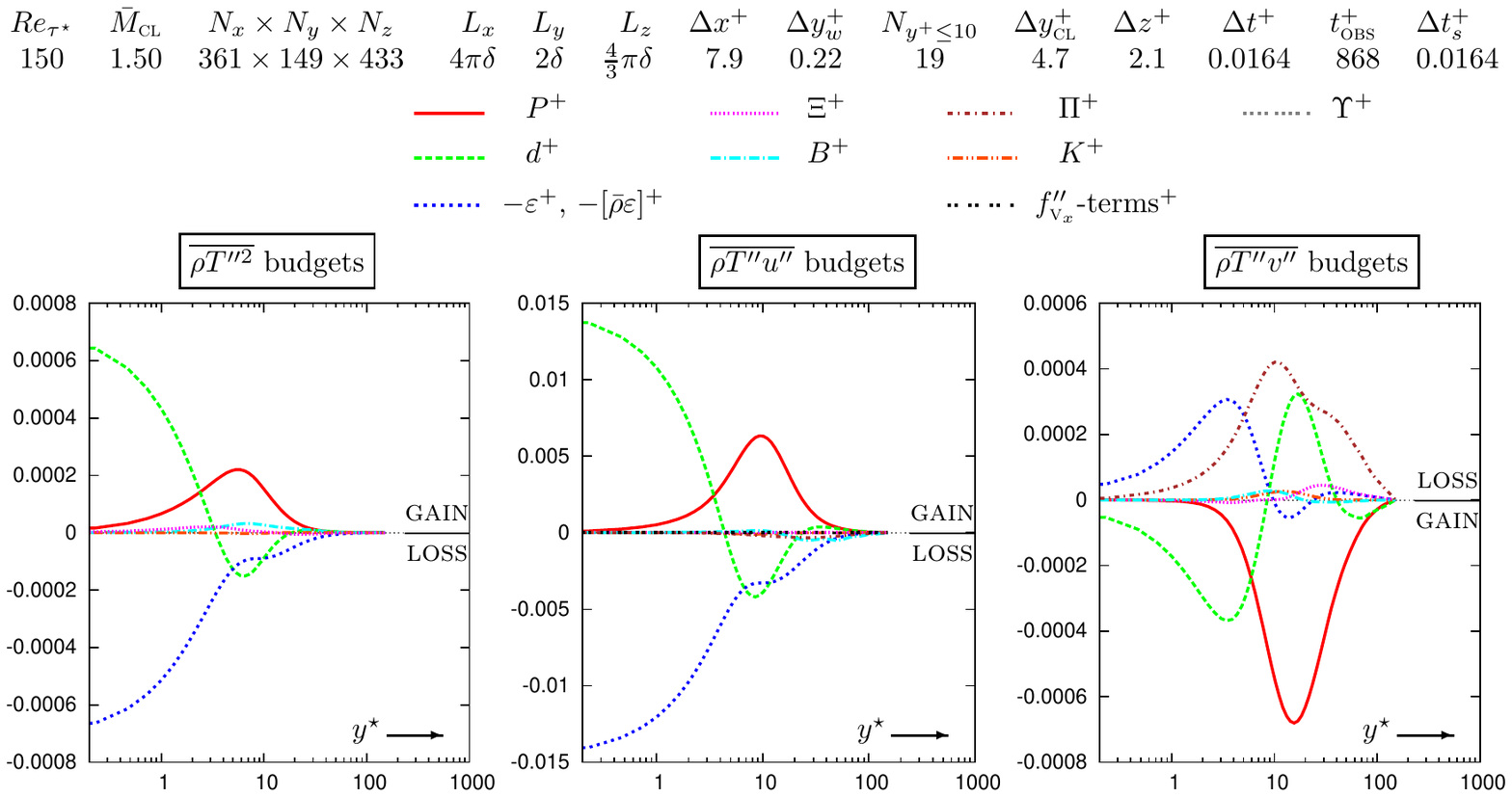}}
\end{picture}
\end{center}
\caption{Budgets, in wall-units \parref{PDTEFCTPCF_s_S_ss_WUs}, of the $c_v$-form of the transport equations
\eqref{Eq_PDTEFCTPCF_s_B_ss_TVHF_001a} for the temperature variance $\overline{\rho T''^2}$
($d_{(\rho T''^2; c_v)}$, $P_{(\rho T''^2; c_v)}$, $-\rho\varepsilon_{(\rho T''^2; c_v)}$, $K_{(\rho T''^2; c_v)}$, $\Xi_{(\rho T''^2; c_v)}$, $B_{(\rho T''^2; c_v)}$) and
\eqref{Eq_PDTEFCTPCF_s_B_ss_TVHF_001b} for the temperature fluxes $\overline{\rho T''u_i''}$
($d_{(\rho T''u_i''; c_v)}$, $P_{(\rho T''u_i''; c_v)}$, $K_{(\rho T''u_i''; c_v)}$, $\Xi_{(\rho T''u_i''; c_v)}$, $B_{(\rho T''u_i''; c_v)}$, $-\bar\rho\varepsilon_{(\rho T''u_i''; c_v)}$, $\Pi_{(\rho T''u_i'')}$, $\overline{\rho T'' f''_{\tsn{V}_x}}$),
obtained from \tsn{DNS} computations ($Re_{\tau^\star}\approxeq150$, $\bar M_\tsn{CL}\approxeq1.50$; \tabrefnp{Tab_PDTEFCTPCF_s_BEqsDNSC_ss_DNSCs_001}), plotted against $y^\star$ \eqref{Eq_PDTEFCTPCF_s_S_ss_CCF_001c}.}
\label{Fig_PDTEFCTPCF_s_B_ss_TVHF_001}
%
\begin{center}
\begin{picture}(450,210)
\put(-45,-340){\includegraphics[angle=0,width=465pt]{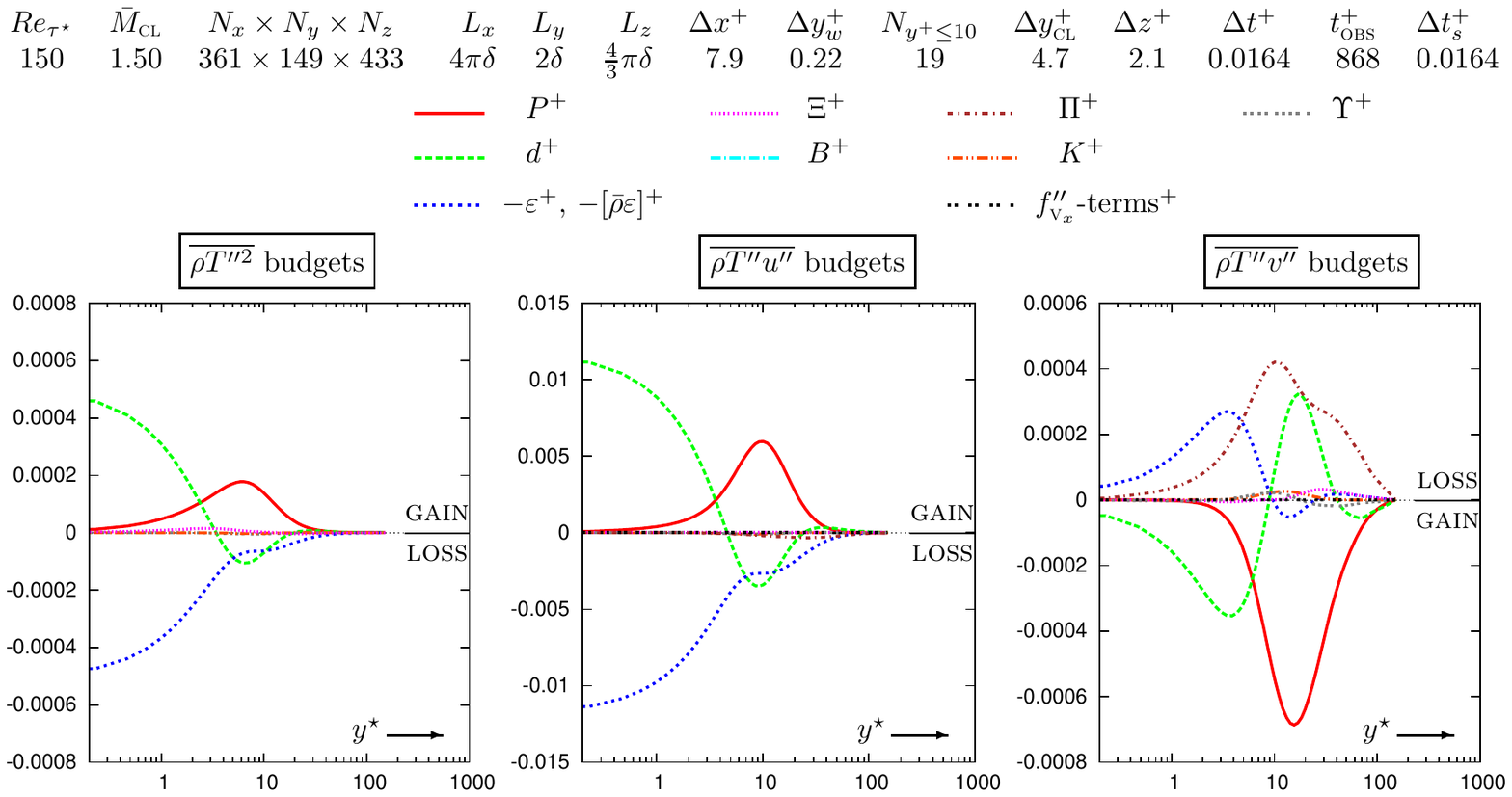}}
\end{picture}
\end{center}
\caption{Budgets, in wall-units \parref{PDTEFCTPCF_s_S_ss_WUs}, of the $c_p$-form of the transport equations
\eqref{Eq_PDTEFCTPCF_s_B_ss_TVHF_004a} for the temperature variance $\overline{\rho T''^2}$
($d_{(\rho T''^2; c_p)}$, $P_{(\rho T''^2; c_p)}$, $-\rho\varepsilon_{(\rho T''^2; c_p)}$, $K_{(\rho T''^2; c_p)}$, $\Xi_{(\rho T''^2; c_p)}$, $\Upsilon_{(\rho T''^2; c_p)}$) and
\eqref{Eq_PDTEFCTPCF_s_B_ss_TVHF_004b} for the temperature fluxes $\overline{\rho T''u_i''}$
($d_{(\rho T''u_i''; c_p)}$, $P_{(\rho T''u_i''; c_p)}$, $K_{(\rho T''u_i''; c_p)}$, $\Xi_{(\rho T''u_i''; c_p)}$, $\Upsilon_{(\rho T''u_i''; c_p)}$, $-\bar\rho\varepsilon_{(\rho T''u_i''; c_p)}$, $\Pi_{(\rho T''u_i'')}$, $\overline{\rho T'' f''_{\tsn{V}_x}}$),
obtained from \tsn{DNS} computations ($Re_{\tau^\star}\approxeq150$, $\bar M_\tsn{CL}\approxeq1.50$; \tabrefnp{Tab_PDTEFCTPCF_s_BEqsDNSC_ss_DNSCs_001}), plotted against $y^\star$ \eqref{Eq_PDTEFCTPCF_s_S_ss_CCF_001c}.}
\label{Fig_PDTEFCTPCF_s_B_ss_TVHF_002}
\end{figure}

The $c_v$-form of the transport equations for the mass-weighted temperature variance $\overline{\rho T''^2}$ and fluxes $\overline{\rho T''u_i''}$ is obtained from the fluctuating parts of the temperature \eqref{Eq_PDTEFCTPCF_s_BEqsDNSC_ss_TTV_003d}
and momentum \eqref{Eq_PDTEFCTPCF_s_BEqsDNSC_ss_FM_001b} equations, and reads
\begin{subequations}
                                                                                                                                    \label{Eq_PDTEFCTPCF_s_B_ss_TVHF_001}
\begin{alignat}{6}
 \underbrace{
 \dfrac{\partial\overline{\rho T''^2}             }
       {\partial                                 t}
+\dfrac{\partial\overline{\rho T''^2}\tilde u_\ell}
       {\partial                            x_\ell}}_{\displaystyle C_{(\rho T''^2)}}
                                                   =&\underbrace{-\dfrac{\partial\overline{\rho T''^2u_\ell''}}
                                                                        {\partial                       x_\ell}
                                                                 -\dfrac{\partial       }
                                                                        {\partial x_\ell}\left(\dfrac{2}{c_v}\overline{T'q_\ell'}\right)}_{\displaystyle d_{(\rho T''^2; c_v)}}\;
                                                                                                                                  \notag\\
                                                    &\underbrace{-2\overline{\rho u_\ell''T''}\dfrac{\partial\tilde T}
                                                                                                    {\partial  x_\ell}
                                                                 +\dfrac{2}{c_v}\overline{T'\tau_{m\ell}'}\bar   S_{m\ell}
                                                                 +\dfrac{2}{c_v}\overline{T'   S_{m\ell}'}\bar\tau_{m\ell}
                                                                 -\dfrac{2}{c_v}\overline{T'p'}\breve\Theta}_{\displaystyle P_{(\rho T''^2; c_v)}}\;
                                                                                                                                  \notag\\
                                                    &-\underbrace{\left(-\dfrac{2}{c_v}\overline{q_\ell'\dfrac{\partial T'}
                                                                                                 {\partial  x_\ell}}\right)}_{\displaystyle \bar\rho\varepsilon_{(\rho T''^2; c_v)}}\;
                                                     \underbrace{+\dfrac{2}{c_v}\overline{T''}\left(-\overline{p\Theta}
                                                                                                    +\overline{\tau_{m\ell} S_{m\ell}}
                                                                                                    -          \dfrac{\partial\bar q_\ell}
                                                                                                                     {\partial     x_\ell}\right)}_{\displaystyle K_{(\rho T''^2; c_v)}}\;
                                                                                                                                  \notag\\
                                                    &\underbrace{+\dfrac{2}{c_v}\overline{T'\tau_{m\ell}'S_{m\ell}'}}_{\displaystyle \Xi_{(\rho T''^2; c_v)}}\;
                                                     \underbrace{-\dfrac{2}{c_v}\overline{pT'\Theta^{\backprime\backprime}}}_{\displaystyle B_{(\rho T''^2; c_v)}}\;
                                                                                                                                    \label{Eq_PDTEFCTPCF_s_B_ss_TVHF_001a}
\end{alignat}
\begin{alignat}{6}
&\underbrace{
 \dfrac{\partial\overline{\rho T''u_i''}             }
       {\partial                                    t}
+\dfrac{\partial\overline{\rho T''u_i''}\tilde u_\ell}
       {\partial                               x_\ell}}_{\displaystyle C_{(\rho T''u_i'')}}
                                                      = \underbrace{-\dfrac{\partial\overline{\rho T''u_i''u_\ell''}}
                                                                           {\partial                          x_\ell}
                                                                    +\dfrac{\partial\overline{T'\tau_{i\ell}'}}
                                                                           {\partial                    x_\ell}
                                                                    -\dfrac{\partial       }
                                                                           {\partial x_\ell}\left(\dfrac{1}{c_v}\overline{u_i'q_\ell'}\right)}_{\displaystyle d_{(\rho T''u_i''; c_v)}}\;
                                                                                                                                  \notag\\
                                                    &\underbrace{-\overline{\rho u_\ell''T''}\dfrac{\partial\tilde u_i}
                                                                                                   {\partial    x_\ell}
                                                                 -\overline{\rho u_i''u_\ell''}\dfrac{\partial\tilde T}
                                                                                                     {\partial  x_\ell}
                                                                 +\dfrac{1}{c_v}\left(\overline{u_i'\tau_{m\ell}'}\bar   S_{m\ell}
                                                                                     +\overline{u_i'   S_{m\ell}'}\bar\tau_{m\ell}\right)
                                                                 -\dfrac{1}{c_v}\overline{p'u_i'}\breve\Theta}_{\displaystyle P_{(\rho T''u_i''; c_v)}}\;
                                                                                                                                  \notag\\
                                                     &\underbrace{+\dfrac{1}{c_v}\overline{u_i''}\left(\overline{\tau_{m\ell} S_{m\ell}}
                                                                                                      -        \dfrac{\partial\bar q_\ell}
                                                                                                                     {\partial     x_\ell}
                                                                                                      -\overline{p\Theta}\right)
                                                                  +\overline{T''}\left(-\dfrac{\partial\bar p}
                                                                                             {\partial   x_i}
                                                                                      +\dfrac{\partial\bar\tau_{i\ell}}
                                                                                             {\partial          x_\ell}\right)}_{\displaystyle K_{(\rho T''u_i''; c_v)}}\;
                                                     \underbrace{+\dfrac{1}{c_v}\overline{u_i'\tau_{m\ell}'S_{m\ell}'}}_{\displaystyle \Xi_{(\rho T''u_i''; c_v)}}\;
                                                     \underbrace{-\dfrac{1}{c_v}\overline{u_i'p\Theta^{\backprime\backprime}}}_{\displaystyle B_{(\rho T''u_i''; c_v)}}\;
                                                                                                                                  \notag\\
                                                    -&\underbrace{\left(-\dfrac{1}{c_v}\overline{q_\ell'\dfrac{\partial u_i'   }
                                                                                                              {\partial  x_\ell}}
                                                                        +\overline{\tau_{i\ell}'\dfrac{\partial T'     }
                                                                                                      {\partial x_\ell}}
                                                                 \right)}_{\displaystyle \bar\rho\varepsilon_{(\rho T''u_i''; c_v)}}\;
                                                      \underbrace{-\overline{ T'\dfrac{\partial p'     }
                                                                                      {\partial x_i    }}}_{\displaystyle \Pi_{(\rho T''u_i'')}}\;
                                                      +\overline{\rho T'' f''_{\tsn{V}_i}}
                                                                                                                                    \label{Eq_PDTEFCTPCF_s_B_ss_TVHF_001b}
\end{alignat}
\end{subequations}
The terms in \eqref{Eq_PDTEFCTPCF_s_B_ss_TVHF_001} which correspond to production by mean dilatation $\breve\Theta$, in $P_{(\rho T''^2; c_v)}$ \eqref{Eq_PDTEFCTPCF_s_B_ss_TVHF_001a} and $P_{(\rho T''u_i''; c_v)}$ \eqref{Eq_PDTEFCTPCF_s_B_ss_TVHF_001b},
and those containing  the fluctuating dilatation $\Theta^{\backprime\backprime}$, {\ie}$B_{(\rho T''^2; c_v)}$ \eqref{Eq_PDTEFCTPCF_s_B_ss_TVHF_001a} and  $B_{(\rho T''u_i''; c_v)}$ \eqref{Eq_PDTEFCTPCF_s_B_ss_TVHF_001b},
are generated by the term $-c_v^{-1}(p\Theta)'$ in the fluctuating part of \eqref{Eq_PDTEFCTPCF_s_BEqsDNSC_ss_TTV_003d}, and can be grouped together as 
\begin{subequations}
                                                                                                                                    \label{Eq_PDTEFCTPCF_s_B_ss_TVHF_002}
\begin{align}
-\dfrac{2}{c_v}\overline{T'p'}\tilde\Theta+B_{(\rho T''^2; c_v)}\stackrel{\eqrefsab{Eq_PDTEFCTPCF_s_B_ss_TVHF_001a}{Eq_PDTEFCTPCF_s_BEqsDNSC_ss_SA_sss_RFD_001}}{=}
-\dfrac{2}{c_v}\overline{T''(p\Theta)'}\stackrel{\eqref{Eq_PDTEFCTPCF_s_BEqsDNSC_ss_SA_sss_RFD_001d}}{=}
-\dfrac{2}{c_v}\overline{T'(p\Theta)'}
                                                                                                                                    \label{Eq_PDTEFCTPCF_s_B_ss_TVHF_002a}\\
-\dfrac{2}{c_v}\overline{p'u_i'}\tilde\Theta+B_{(\rho T''u_i''; c_v)}\stackrel{\eqrefsab{Eq_PDTEFCTPCF_s_B_ss_TVHF_001b}{Eq_PDTEFCTPCF_s_BEqsDNSC_ss_SA_sss_RFD_001}}{=}
-\dfrac{2}{c_v}\overline{u_i''(p\Theta)'}\stackrel{\eqref{Eq_PDTEFCTPCF_s_BEqsDNSC_ss_SA_sss_RFD_001d}}{=}
-\dfrac{2}{c_v}\overline{u_i'(p\Theta)'}
                                                                                                                                    \label{Eq_PDTEFCTPCF_s_B_ss_TVHF_002b}
\end{align}
\end{subequations}
Examination of the budgets of \eqref{Eq_PDTEFCTPCF_s_B_ss_TVHF_001} reveals a striking similarity between the $y^\star$-distributions across the channel of the budgets of the mass-weighted variance and fluxes of entropy
($\overline{\rho s''^2}$, $\overline{\rho s''u_i''}$; \figrefnp{Fig_PDTEFCTPCF_s_B_ss_EVF_001}) with those of temperature ($\overline{\rho T''^2}$, $\overline{\rho T''u_i''}$; \figrefnp{Fig_PDTEFCTPCF_s_B_ss_TVHF_001}), as could be expected because of the
strong correlation coefficient $c_{s'T'}\approxeq 1\;\forall y^\star\in[1,50]$ \figref{Fig_PDTEFCTPCF_s_S_ss_CCs_001}, and generaly $c_{s'T'}\geq\tfrac{6}{10}\;\forall y$ \figref{Fig_PDTEFCTPCF_s_S_ss_CCs_002},
in line with the statement on "{\em temperature (entropy) spots}" of \citet[p.~661]{Kovasznay_1953a} when referring to the
entropy mode of compressible turbulent fluctuations.

Alternatively, the fluctuating part of \eqref{Eq_PDTEFCTPCF_s_BEqsDNSC_ss_TTV_003e} can be used in lieu of \eqref{Eq_PDTEFCTPCF_s_BEqsDNSC_ss_TTV_003d}, and yields the $c_p$-form of the transport equations for the temperature variance 
$\overline{\rho T''^2}$ and fluxes $\overline{\rho T''u_i''}$
\begin{subequations}
                                                                                                                                    \label{Eq_PDTEFCTPCF_s_B_ss_TVHF_003}
\begin{alignat}{6}
&\underbrace{
 \dfrac{\partial\overline{\rho T''^2}             }
       {\partial                                 t}
+\dfrac{\partial\overline{\rho T''^2}\tilde u_\ell}
       {\partial                            x_\ell}}_{\displaystyle C_{(\rho T''^2)}}
                                                   = \underbrace{-\dfrac{\partial}{\partial x_\ell}\left (\overline{\rho T''^2u_\ell''}+\dfrac{2}{c_p}\overline{T'q_\ell'}\right)}_{\displaystyle d_{(\rho T''^2; c_p)}}\;
                                                     \underbrace{-2\overline{\rho u_\ell''T''}\dfrac{\partial\tilde T}
                                                                                                    {\partial  x_\ell}
                                                                 +\dfrac{2}{c_p}\left (\overline{T'\tau_{m\ell}'}\bar   S_{m\ell}+\overline{T'S_{m\ell}'}\bar\tau_{m\ell}\right)}_{\displaystyle P_{(\rho T''^2; c_p)}}\;
                                                                                                                                  \notag\\
                      &\!\!-\underbrace{\left(-\dfrac{2}{c_p}\overline{q_\ell'\dfrac{\partial T'}
                                                                                                 {\partial  x_\ell}}\right)}_{\displaystyle \bar\rho\varepsilon_{(\rho T''^2; c_p)}}
                                                    +\underbrace{\dfrac{2}{c_p}\overline{T''}\left( \overline{\dfrac{Dp}{Dt}}
                                                                                                    +\overline{\tau_{m\ell} S_{m\ell}}
                                                                                                    -          \dfrac{\partial\bar q_\ell}
                                                                                                                     {\partial     x_\ell}\right)}_{\displaystyle K_{(\rho T''^2; c_p)}}
                                                    +\underbrace{\dfrac{2}{c_p}\overline{T'\tau_{m\ell}'S_{m\ell}'}}_{\displaystyle \Xi_{(\rho T''^2; c_p)}}
                                                    +\underbrace{\dfrac{2}{c_p}\overline{T'\left (\dfrac{D p}{D t}\right )'}}_{\displaystyle \Upsilon_{(\rho T''^2; c_p)}}
                                                                                                                                    \label{Eq_PDTEFCTPCF_s_B_ss_TVHF_003a}
\end{alignat}
\begin{alignat}{6}
&\underbrace{
 \dfrac{\partial\overline{\rho T''u_i''}             }
       {\partial                                    t}
+\dfrac{\partial\overline{\rho T''u_i''}\tilde u_\ell}
       {\partial                               x_\ell}}_{\displaystyle C_{(\rho T''u_i'')}}
                                                      = \underbrace{-\dfrac{\partial\overline{\rho T''u_i''u_\ell''}}
                                                                           {\partial                          x_\ell}
                                                                    +\dfrac{\partial\overline{T'\tau_{i\ell}'}}
                                                                           {\partial                    x_\ell}
                                                                    -\dfrac{\partial       }
                                                                           {\partial x_\ell}\left(\dfrac{1}{c_p}\overline{u_i'q_\ell'}\right)}_{\displaystyle d_{(\rho T''u_i''; c_p)}}\;
                                                                                                                                  \notag\\
                                                    &\underbrace{-\overline{\rho u_\ell''T''}\dfrac{\partial\tilde u_i}
                                                                                                   {\partial    x_\ell}
                                                                 -\overline{\rho u_i''u_\ell''}\dfrac{\partial\tilde T}
                                                                                                     {\partial  x_\ell}
                                                                 +\dfrac{1}{c_p}\left(\overline{u_i'\tau_{m\ell}'}\bar   S_{m\ell}
                                                                                     +\overline{u_i'   S_{m\ell}'}\bar\tau_{m\ell}\right)}_{\displaystyle P_{(\rho T''u_i''; c_p)}}\;
                                                     -\underbrace{\left(-\dfrac{1}{c_p}\overline{q_\ell'\dfrac{\partial u_i'   }{\partial  x_\ell}}
                                                                        +\overline{\tau_{i\ell}'\dfrac{\partial T'     }{\partial x_\ell}}\right)}_{\displaystyle \bar\rho\varepsilon_{(\rho T''u_i''; c_p)}}\;
                                                                                                                                  \notag\\
                                                     &+\underbrace{\dfrac{1}{c_p}\overline{u_i''}\left(\overline{\dfrac{Dp}{Dt}}
                                                                                                      +\overline{\tau_{m\ell} S_{m\ell}}
                                                                                                      -\dfrac{\partial\bar q_\ell}{\partial x_\ell}\right)
                                                                  +\overline{T''}\left(-\dfrac{\partial\bar p}
                                                                                             {\partial   x_i}
                                                                                      +\dfrac{\partial\bar\tau_{i\ell}}
                                                                                             {\partial          x_\ell}\right)}_{\displaystyle K_{(\rho T''u_i''; c_p)}}\;
                                                     +\underbrace{\dfrac{1}{c_p}\overline{u_i'\tau_{m\ell}'S_{m\ell}'}}_{\displaystyle \Xi_{(\rho T''u_i''; c_p)}}\;
                                                                                                                                  \notag\\
                                                      &\underbrace{-\overline{ T'\dfrac{\partial p'     }
                                                                                      {\partial x_i    }}}_{\displaystyle \Pi_{(\rho T''u_i'')}}\;
                                                      +\overline{\rho T'' f''_{\tsn{V}_i}}
                                                      +\underbrace{\dfrac{1}{c_p}\overline{u_i'\left(\dfrac{D p}{D t}\right )'}}_{\displaystyle \Upsilon_{(\rho T''u_i''; c_p)}}
                                                                                                                                    \label{Eq_PDTEFCTPCF_s_B_ss_TVHF_003b}
\end{alignat}
\end{subequations}
where
\begin{subequations}
                                                                                                                                    \label{Eq_PDTEFCTPCF_s_B_ss_TVHF_004}
\begin{align}
\overline{\dfrac{D(\cdot)}{Dt}}&:=\overline{\dfrac{\partial(\cdot)}{\partial t}+u_j\dfrac{\partial(\cdot)}{\partial x_j}}
                                 =\dfrac{\partial\overline{(\cdot)}}{\partial t}+\bar u_j\dfrac{\partial\overline{(\cdot)}}{\partial x_j}
                                                                                    +\overline{u_j'\dfrac{\partial(\cdot)'}{\partial x_j}}
                                                                                                                                    \label{Eq_PDTEFCTPCF_s_B_ss_TVHF_004a}\\
\left[\dfrac{D(\cdot)}{Dt}\right]'&:=\dfrac{D(\cdot)}{Dt}-\overline{\dfrac{D(\cdot)}{Dt}}
                                     =\dfrac{\partial(\cdot)'}{\partial t}+\bar u_j \dfrac{\partial          (\cdot)'}{\partial x_j}
                                                                          +     u_j'\dfrac{\partial\overline{(\cdot)}}{\partial x_j}
                                                                          +     u_j'\dfrac{\partial          (\cdot)'}{\partial x_j}
                                                                          -\overline{u_j'\frac{\partial      (\cdot)'}{\partial x_j}}
                                                                                                                                    \label{Eq_PDTEFCTPCF_s_B_ss_TVHF_004b}
\end{align}
\end{subequations}
The general shape and relative importance of the dominant mechanisms in the budgets of the transport equations
for $\overline{\rho T''^2}$ \eqrefsab{Eq_PDTEFCTPCF_s_B_ss_TVHF_001a}{Eq_PDTEFCTPCF_s_B_ss_TVHF_003a} and for $\overline{\rho T''u_i''}$ \eqrefsab{Eq_PDTEFCTPCF_s_B_ss_TVHF_001b}{Eq_PDTEFCTPCF_s_B_ss_TVHF_003b}
are quite similar between the $c_v$-form \figref{Fig_PDTEFCTPCF_s_B_ss_TVHF_001} and the $c_p$-form \figref{Fig_PDTEFCTPCF_s_B_ss_TVHF_002},
because some of the terms are identical, \eg the diffusion mechanisms $-\partial_{x_\ell}\overline{\rho T''u_\ell''}$ \eqrefsab{Eq_PDTEFCTPCF_s_B_ss_TVHF_001a}{Eq_PDTEFCTPCF_s_B_ss_TVHF_003a}
and $\partial_{x_\ell}(-\overline{\rho T''u_i''u_\ell''}+\overline{T'\tau_{i\ell}'})$ \eqrefsab{Eq_PDTEFCTPCF_s_B_ss_TVHF_001b}{Eq_PDTEFCTPCF_s_B_ss_TVHF_003b}, the pressure scrambling term
$\Pi_{(\rho T''u_i'')}$ \eqrefsab{Eq_PDTEFCTPCF_s_B_ss_TVHF_001b}{Eq_PDTEFCTPCF_s_B_ss_TVHF_003b}, and the production term $-2\overline{\rho u_\ell''T''}\partial_{x_\ell}\tilde{T}$ \eqrefsab{Eq_PDTEFCTPCF_s_B_ss_TVHF_001a}{Eq_PDTEFCTPCF_s_B_ss_TVHF_003a}
and $-\overline{\rho u_\ell''T''}\partial_{x_\ell}\tilde{u}_i-\overline{\rho u_i''u_\ell''}\partial_{x_\ell}\tilde{T}$ \eqrefsab{Eq_PDTEFCTPCF_s_B_ss_TVHF_001b}{Eq_PDTEFCTPCF_s_B_ss_TVHF_003b}, while many others have a constant ratio
 $\gamma=c_v^{-1}c_p=1.4$ in the present \tsn{DNS} computations.
Therefore, the main differences between the 2 forms \eqrefsab{Eq_PDTEFCTPCF_s_B_ss_TVHF_001}{Eq_PDTEFCTPCF_s_B_ss_TVHF_003} are in the compressible terms $B_{(.)}$ \eqref{Eq_PDTEFCTPCF_s_B_ss_TVHF_001}
and $\Upsilon_{(.)}$ \eqref{Eq_PDTEFCTPCF_s_B_ss_TVHF_003}. Notice that the flow symmetries $\eqref{Eq_PDTEFCTPCF_s_B_001}\Longrightarrow\breve\Theta=0$, so that only $B_{(.)}$ remains in \eqref{Eq_PDTEFCTPCF_s_B_ss_TVHF_002}.
All compressible terms in \eqrefsab{Eq_PDTEFCTPCF_s_B_ss_TVHF_001}{Eq_PDTEFCTPCF_s_B_ss_TVHF_003}, \viz $K_{(.)}$, $B_{(.)}$ or $\Upsilon_{(.)}$, are generally weak \figrefsab{Fig_PDTEFCTPCF_s_B_ss_TVHF_001}{Fig_PDTEFCTPCF_s_B_ss_TVHF_002}. The term
\smash{$\Upsilon_{(\rho T''^2; c_p)}\stackrel{\eqref{Eq_PDTEFCTPCF_s_B_ss_TVHF_003a}}{=}2 c_p^{-1}\overline{T'(D_tp)'}$} in \eqref{Eq_PDTEFCTPCF_s_B_ss_TVHF_003a} is negligibly small \figref{Fig_PDTEFCTPCF_s_B_ss_TVHF_002} and much smaller
than the weak term $B_{(\rho T''^2; c_v)}$ \figref{Fig_PDTEFCTPCF_s_B_ss_TVHF_001} in \eqref{Eq_PDTEFCTPCF_s_B_ss_TVHF_001a}. In both forms \figrefsab{Fig_PDTEFCTPCF_s_B_ss_TVHF_001}{Fig_PDTEFCTPCF_s_B_ss_TVHF_002}, the main mechanisms in the
budgets of $\overline{\rho T''^2}$ are similar one with another and with the budgets of $\overline{\rho s''^2}$ \figref{Fig_PDTEFCTPCF_s_B_ss_EVF_001}.
Notice that \smash{$\varepsilon_{(\rho T''^2; c_p)}^{-1}\varepsilon_{(\rho T''^2; c_v)}\stackrel{\eqrefsab{Eq_PDTEFCTPCF_s_B_ss_TVHF_001a}{Eq_PDTEFCTPCF_s_B_ss_TVHF_001b}}{=}c_v^{-1}{c_p}=\gamma$} exactly, and that the same applies
to the every weak triple correlations \smash{$\Xi_{(\rho T''^2; c_p)}^{-1}\Xi_{(\rho T''^2; c_v)}\stackrel{\eqrefsab{Eq_PDTEFCTPCF_s_B_ss_TVHF_001a}{Eq_PDTEFCTPCF_s_B_ss_TVHF_001b}}{=}\gamma$} \figrefsab{Fig_PDTEFCTPCF_s_B_ss_TVHF_001}{Fig_PDTEFCTPCF_s_B_ss_TVHF_002}.
Diffusion $d_{(\rho T''u''; .)}$, destruction $-\bar\rho\varepsilon_{(\rho T''u''; .)}$
and production $P_{(\rho T''u''; .)}$ are the main mechanisms genering the $\overline{\rho T''u''}$-budgets \figrefsab{Fig_PDTEFCTPCF_s_B_ss_TVHF_001}{Fig_PDTEFCTPCF_s_B_ss_TVHF_002},
while $\Upsilon_{(\rho T''u''; c_p)}$ \figref{Fig_PDTEFCTPCF_s_B_ss_TVHF_002} is negligibly small and $B_{(\rho T''u''; c_v)}$ \figref{Fig_PDTEFCTPCF_s_B_ss_TVHF_001} is very weak but larger than $\Upsilon_{(\rho T''u''; c_p)}$.
Finally, like for $s''$ \figref{Fig_PDTEFCTPCF_s_B_ss_EVF_001}, the pressure scrambling term $\Pi_{(\rho T''v''; .)}\geq0\;\forall y^\star$, which is common to both forms \eqrefsab{Eq_PDTEFCTPCF_s_B_ss_TVHF_001b}{Eq_PDTEFCTPCF_s_B_ss_TVHF_003b}, is a major loss
mechanism in the budgets of $\overline{\rho T''v''}\leq 0$ \figref{Fig_PDTEFCTPCF_s_B_001}.
\begin{figure}
\begin{center}
\begin{picture}(450,210)
\put(-45,-340){\includegraphics[angle=0,width=465pt]{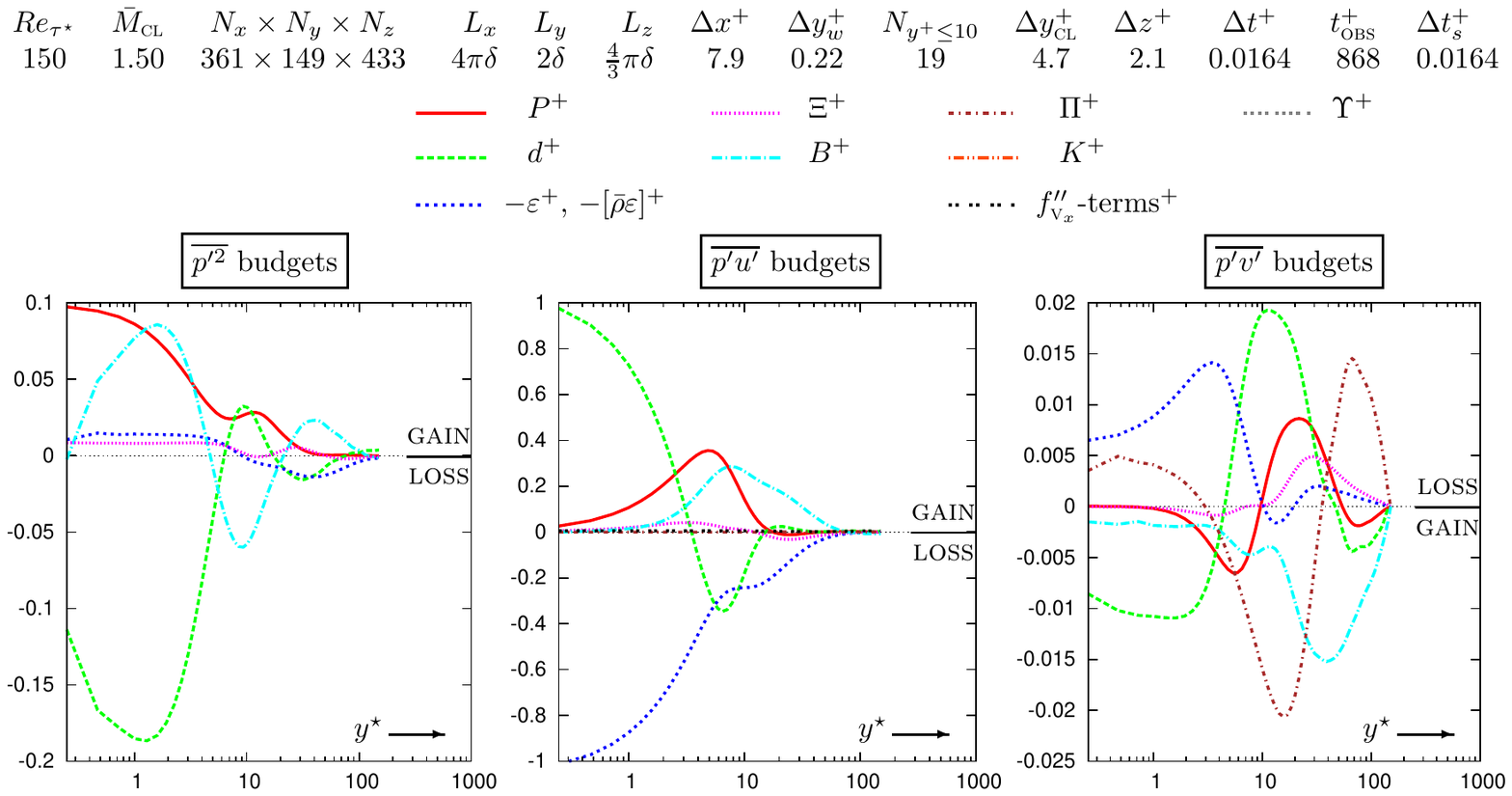}}
\end{picture}
\end{center}
\caption{Budgets, in wall-units \parref{PDTEFCTPCF_s_S_ss_WUs}, of the transport equations
\eqref{Eq_PDTEFCTPCF_s_B_ss_PVT_001a} for the pressure variance $\overline{p'^2}$ ($d_{(p'^2)}$, $P_{(p'^2)}$, $-\varepsilon_{(p'^2)}$, $\Xi_{(p'^2)}$, $B_{(p'^2)}$) and
\eqref{Eq_PDTEFCTPCF_s_B_ss_PVT_001b} for the pressure transport $\overline{p'u_i'}$
($d_{(p'u_i')}$, $P_{(p'u_i')}$, $-\varepsilon_{(p'u_i')}$, $\Xi_{(p'u_i')}$, $\Pi_{(p'u_i')}$, $B_{(p'u_i')}$, $\overline{p'f'_{\tsn{V}_x}}$),
obtained from \tsn{DNS} computations ($Re_{\tau^\star}\approxeq150$, $\bar M_\tsn{CL}\approxeq1.50$; \tabrefnp{Tab_PDTEFCTPCF_s_BEqsDNSC_ss_DNSCs_001}), plotted against $y^\star$ \eqref{Eq_PDTEFCTPCF_s_S_ss_CCF_001c}.}
\label{Fig_PDTEFCTPCF_s_B_ss_PVT_001}
\end{figure}

The terms related with compressibility, apart from the direct compressibility effetcs $K_{(.)}$, are $B_{(.)}$ in \eqref{Eq_PDTEFCTPCF_s_B_ss_TVHF_001}, recalling that \eqref{Eq_PDTEFCTPCF_s_B_001} $\Longrightarrow \breve\Theta=0$ in
\eqref{Eq_PDTEFCTPCF_s_B_ss_TVHF_002}, and $\Upsilon_{(.)}$ in \eqref{Eq_PDTEFCTPCF_s_B_ss_TVHF_003}, and are generally weak. 
In particular, the very small contribution of $\Upsilon_{(.)}$ in the budgets of \eqref{Eq_PDTEFCTPCF_s_B_ss_TVHF_003} \figref{Fig_PDTEFCTPCF_s_B_ss_TVHF_001}, implies that the corresponding term can probably be neglected, or modelled as a single term without applying
further decomposition as in \eqref{Eq_PDTEFCTPCF_s_B_ss_TVHF_004}.

%
%
%
%
%
\subsection{Pressure variance and transport}\label{PDTEFCTPCF_s_B_ss_PVT}
%
%
%
%
%
It is generally accepted that pressure fluctuations $p'$ behave differently than the fluctuations of density $\rho'$ or temperature $T'$, because of their strong dependence on the dynamic field (velocity fluctuations $u_i'$). In incompressible
flow $p'$ are simply a consequence of the velocity-field through the Poisson equation for $p'$ \citep{Chou_1945a}, while studies of the compressible analog of this equation \citep{Gerolymos_Senechal_Vallet_2013a} suggest that, in wall
bounded compressible flows, the extra compressible terms do not alter substantially this subordination to the dynamic field \citep{Foysi_Sarkar_Friedrich_2004a}. Related to this conceptual view of pressure fluctuations is the non-locality
of $p'$ \citep{Kim_1989a,
               Chang_Piomelli_Blake_1999a}.
In this context, \citet{Hamba_1999a} applied a Helmholtz decomposition splitting the fluctuating velocity field, into a solenoidal and a compressible part,
and defined solenoidal pressure fluctuations obtained by the solution of a Poisson equation driven by the solenoidal velocity field \citep[(16), p. 1625]{Hamba_1999a}
as a part of the actual thermodynamic pressure fluctuations. \citet{Yoshizawa_Matsuo_Mizobuchi_2013a} applied a similar distinction between the actual thermodynamic pressure fluctuations and those subordinated to the fluctuating velocity field,
to ensure smooth limiting behaviour (incompressible flow limit) of their model as $[\bar\rho^{-1}\rho']_\mathrm{rsm}\to0$ \citep[Appendix A, pp. 45--51]{Gerolymos_Senechal_Vallet_2013a}.
For these reason, all models for the pressure terms in Reynolds-stress transport closures, even in compressible flow with shock-waves \citep{Gerolymos_Lo_Vallet_Younis_2012a},
are based on tensorial representations constructed from the strain-rate, vorticity-rate and anisotropy tensors \citep{Gerolymos_Lo_Vallet_2012a},
eventually augmented by gradients of turbulent quantities \citep{Gerolymos_Lo_Vallet_Younis_2012a}.

These observations not withstanding, it is possible, formally, to derive, using the fluctuating part of \eqref{Eq_PDTEFCTPCF_s_BEqsDNSC_ss_TTV_003c}, 
$(D_tp)'\stackrel{\eqref{Eq_PDTEFCTPCF_s_BEqsDNSC_ss_TTV_003c}}{=}-\gamma(p\Theta)'+(\gamma-1)(\tau_{m\ell}S_{m\ell}-\partial_{x_\ell}q_\ell)'$, transport equations for the variance $\overline{p'^2}$
\begin{subequations}
                                                                                                                                  \label{Eq_PDTEFCTPCF_s_B_ss_PVT_001}
\begin{alignat}{6}
&\underbrace{\dfrac{\partial\overline{p'^2}}{\partial t}+\tilde u_\ell\dfrac{\partial\overline{p'^2}}{\partial x_\ell}}_{\displaystyle C_{(p'^2)}}\; = \underbrace{-\dfrac{\partial\overline{p'^2u_\ell''}}{\partial x_\ell}-\dfrac{\partial }
                                                                        {\partial x_\ell}\left(2(\gamma-1)\overline{p'q_\ell'}\right)}_{\displaystyle d_{(p'^2)}}\;  \notag\\
                                                     & \underbrace{-2\overline{p'u_\ell'}\dfrac{\partial \bar p}{\partial x_\ell}-2\gamma\overline{p'^2}\breve\Theta
                                                                 +2(\gamma-1)\left(\overline{p'\tau_{m\ell}'}\bar S_{m\ell}
                                                                                  +\overline{p'   S_{m\ell}'}\bar\tau_{m\ell}\right)}_{\displaystyle P_{(p'^2)}}\;\notag\\
                                                     &-\underbrace{\left(-2(\gamma-1)\overline{q_\ell'\dfrac{\partial p'}{\partial x_\ell}}\right)}_{\displaystyle \varepsilon_{(p'^2)}}\;
                                                     \underbrace{+2(\gamma-1)\overline{p'\tau_{m\ell}'S_{m\ell}'}}_{\displaystyle \Xi_{(p'^2)}}\;
                                                     \underbrace{-\overline{\biggl((2\gamma-1)p'^2+2\gamma\bar p p'\biggr)\Theta^{\backprime\backprime}}}_{\displaystyle B_{(p'^2)}}\;
                                                                                                                                  \label{Eq_PDTEFCTPCF_s_B_ss_PVT_001a}
\end{alignat}
and, combining with the fluctuating part of the momentum equation \eqref{Eq_PDTEFCTPCF_s_BEqsDNSC_ss_FM_001b}, for the transport by the fluctuating velocity field $\overline{p'u_i'}$
\begin{alignat}{6}
&\underbrace{
               \dfrac{\partial \overline{p'u_i'}}
                     {\partial                 t}
+\tilde u_\ell \dfrac{\partial \overline{p'u_i'}}
                     {\partial            x_\ell}}_{\displaystyle C_{(p'u_i')}}=
                                                 \underbrace{-\dfrac{\partial       }
                                                                      {\partial x_\ell}\left(          \overline{p'u_i'u_\ell''}\right)
                                                               -\dfrac{\partial       }
                                                                      {\partial x_\ell}\left((\gamma-1)\overline{u_i'q_\ell'}\right)
                                                               +\dfrac{\partial       }
                                                                      {\partial x_\ell}\overline{\left(\dfrac{p'}{\rho}\tau_{i\ell}'\right)}}_{\displaystyle d_{(p'u_i')}}
                                                                                                                                  \notag\\
                                                  &\underbrace{-                 \overline{p'u_\ell'}\dfrac{\partial\bar u_i   }
                                                                                                           {\partial     x_\ell}
                                                               -           \gamma\overline{p'u_i'}\breve\Theta
                                                               -\overline{u_i'u_\ell'}\dfrac{\partial \bar p}{\partial x_\ell}
                                                               +\overline{\left(\dfrac{p'}{\rho}\right)}\left (\dfrac{\partial \bar \tau_{i\ell}}{\partial x_\ell}-\dfrac{\partial \bar p}{\partial x_i}\right )
                                                               +(\gamma-1)(\overline{u_i'S_{m\ell}'} \bar \tau_{m\ell}+\overline{u_i'\tau_{m\ell}'}\bar S_{m\ell})
                                                               -\overline{p'\tau_{i\ell}'}
                                                                \dfrac{\partial}{\partial x_\ell}\overline{\left(\dfrac{1}{\rho}\right)}}_{\displaystyle P_{(p'u_i')}}\;
                                                                                                                                  \notag\\
                                                  -&\underbrace{\left(-(\gamma-1)\overline{q_\ell'\dfrac{\partial u_i'}{\partial x_\ell}}\;
                                                                     +          \overline{\dfrac{\tau_{i\ell}'}{\rho}\dfrac{\partial p'}{\partial x_\ell}}\right)}_{\displaystyle \varepsilon_{(p'u_i')}}
                                                   \underbrace{+(\gamma-1)\overline{u_i'\tau_{m\ell}'S_{m\ell}'}
                                                               -\overline{p'\tau_{i\ell}'\dfrac{\partial}{\partial x_\ell}\left(\dfrac{1}{\rho}\right)'}}_{\displaystyle \Xi_{(p'u_i')}}\;
                                                   \underbrace{-\overline{\dfrac{p'}{\rho}\dfrac{\partial p'}{\partial x_i}}}_{\displaystyle \Pi_{(p'u_i')}}\;\notag\\
                                                   &\underbrace{-\overline{\bigl(\gamma\bar p+(\gamma-1)p'\bigr)u_i'\Theta^{\backprime\backprime}}}_{\displaystyle B_{(p'u_i')}}\;
                                                   +\overline{p'f'_{\tsn{V}_i}}
                                                                                                                                  \label{Eq_PDTEFCTPCF_s_B_ss_PVT_001b}
\end{alignat}
\end{subequations}
The $y^\star$-distributions \figref{Fig_PDTEFCTPCF_s_B_ss_PVT_001} of the budgets of \eqref{Eq_PDTEFCTPCF_s_B_ss_PVT_001} are fundamentally different from those of the other thermodynamic variables, $\rho'$ \figref{Fig_PDTEFCTPCF_s_B_ss_DVMF_001},
$s''$ \figref{Fig_PDTEFCTPCF_s_B_ss_EVF_001} and $T''$ \figrefsab{Fig_PDTEFCTPCF_s_B_ss_TVHF_001}{Fig_PDTEFCTPCF_s_B_ss_TVHF_002}.
The obvious importance of the dilatational terms $B_{(.)}$, containing $\Theta^{\backprime\backprime}$ \eqref{Eq_PDTEFCTPCF_s_BEqsDNSC_ss_SA_sss_NAN_002b} in \eqref{Eq_PDTEFCTPCF_s_B_ss_PVT_001}, which are invariably comparable
in magnitude with the corresponding production terms $P_{(.)}$ \figref{Fig_PDTEFCTPCF_s_B_ss_PVT_001} contrasts with the weak level of these mechanisms in the budgets of $\{\rho',s'',T''\}$ \figrefsatob{Fig_PDTEFCTPCF_s_B_ss_DVMF_001}{Fig_PDTEFCTPCF_s_B_ss_TVHF_002}.

The production of pressure fluctuations $P_{(p'^2)}\geq 0\geq0$ (gain) increases from centerline to the wall \figref{Fig_PDTEFCTPCF_s_B_ss_PVT_001}, where, contrary to $\{P_{(\rho'^2)},P_{(\rho s''^2)},P_{(\rho T''^2)}\}$ 
\figrefsatob{Fig_PDTEFCTPCF_s_B_ss_DVMF_001}
            {Fig_PDTEFCTPCF_s_B_ss_TVHF_002},
it reaches its maximum value $[P_{(p'^2)}]_w$, $3.3$ times higher than the local maximum at $y^\star\approxeq12$ \figref{Fig_PDTEFCTPCF_s_B_ss_PVT_001}. This explains the relatively constant shape
of $\overline{p'^2}$ in the region $0\leq y^\star\lessapprox 80$ where $[\overline{p'^2}]^\frac{1}{2}$ varies by only 20\% \figref{Fig_PDTEFCTPCF_s_B_001} in contrast to the rms of the other thermodynamic fluctuations
$\{[\overline{\rho'^2}]^\frac{1}{2},[\overline{\rho s''^2}]^\frac{1}{2},[\overline{\rho  T''^2}]^\frac{1}{2}\}$ which exhibit much larger variations.
In the near-wall region ($y^\star\lessapprox 3$; \figrefnp{Fig_PDTEFCTPCF_s_B_ss_PVT_001}), production $P_{(p'^2)}$, diffusion $d_{(p'^2)}$ and the dilatational terms $B_{(p'^2)}$ are the main mechanisms in the $\overline{p'^2}$-budgets,
while destruction $-\varepsilon_{(p'^2)}\geq0\;\forall y^\star\lessapprox 10$ and the triple correlation term $\Xi_{(p'^2)}\geq0\;\forall y^\star\lessapprox 10$ both contribute as a weak gain
in the near-wall $\overline{p'^2}$-budgets (\figrefnp{Fig_PDTEFCTPCF_s_B_ss_PVT_001}). Further away from the
wall ($y^\star\gtrapprox 20$; \figrefnp{Fig_PDTEFCTPCF_s_B_ss_PVT_001}), all of these mechanisms ($P_{(p'^2)}$, $d_{(p'^2)}$, $B_{(p'^2)}$, $\Xi_{(p'^2)}$, $-\varepsilon_{(p'^2)}$) are of equal importance with 
destruction $-\varepsilon_{(p'^2)}\leq0\;\forall y^\star\lessapprox 20$ (loss). Notice also the wave-like shape of $B_{(p'^2)}$ across the channel, generally opposing diffusion \figref{Fig_PDTEFCTPCF_s_B_ss_PVT_001}.

The main mechanisms contributing to the budgets of the streamwise transport $\overline{p'u'}\geq0$ \figref{Fig_PDTEFCTPCF_s_B_001}
are production $P_{(p'u')}\geq0\;\forall y^\star$ (gain), diffusion $d_{(p'u')}$, destruction $-\varepsilon_{(p'u')}\leq0\;\forall y^\star$ (loss) and the dilatational term $B_{(p'u')}\geq0\;\forall y^\star$ \figref{Fig_PDTEFCTPCF_s_B_ss_PVT_001},
with a weak but discernible contribution of the triple correlations term $\Xi_{(p'u')}$. As for all of the other streamwise fluxes 
\smash{$\{\overline{\rho'u'}\stackrel{\eqref{Eq_PDTEFCTPCF_s_BEqsDNSC_ss_SA_sss_RFD_001c}}{=}-\bar\rho\overline{u''},\overline{\rho s'' u''}, \overline{\rho T'' u''}\}$} \figrefsatob{Fig_PDTEFCTPCF_s_B_ss_DVMF_001}{Fig_PDTEFCTPCF_s_B_ss_TVHF_002},
the body-acceleration term $\overline{p'f'_{\tsn{V}_x}}$ \eqref{Eq_PDTEFCTPCF_s_B_ss_PVT_001b} is negligible\figref{Fig_PDTEFCTPCF_s_B_ss_PVT_001}.
The dilatational term $B_{(p'u')}\geq 0\;\forall y^\star$ is practically the unique gain mechanism in the major part of the channel ($y^\star\gtrapprox 13$; \figrefnp{Fig_PDTEFCTPCF_s_B_ss_PVT_001}),
where it is principally balanced by destruction $-\varepsilon_{(p'u')}\leq0\;\forall y^\star$ (loss). In the near-wall buffer-layer ($2\lessapprox y^\star\lessapprox 7$; \figrefnp{Fig_PDTEFCTPCF_s_B_ss_PVT_001}),
$B_{(p'u')}$ decreases rapidly and is replaced by the production term $P_{(p'u')}\geq 0$ as the main gain production mechanism,
balanced by diffusion $d_{(p'u')}$ and destruction $-\varepsilon_{(p'u')}$. Finally, in the very-near-wall region ($y^\star\lessapprox 2$), diffusion $d_{(p'u')}\geq0\;\forall y^\star\leq2$, which changes sign 
at $y^\star\approxeq3.5$, becomes \figref{Fig_PDTEFCTPCF_s_B_ss_PVT_001} the major gain mechanism at the wall, and is principally balanced by destruction $-\varepsilon_{(p'u')}\leq0\;\forall y^\star$ (loss).

Finally, the budgets of the wall-normal transport $\overline{p'v'}\leq0$ \figref{Fig_PDTEFCTPCF_s_B_001} are unlike all the others, in that all of the mechanism in \eqref{Eq_PDTEFCTPCF_s_B_ss_PVT_001b} 
($P_{(p'v')}$, $d_{(p'v')}$, $B_{(p'v')}$, $\Pi_{(p'v')}$ and $-\varepsilon_{(p'v')}$) are of equal importance in the major part of the flow \figref{Fig_PDTEFCTPCF_s_B_ss_PVT_001}.
The only similarity of the $\overline{p'v'}$-budgets \figref{Fig_PDTEFCTPCF_s_B_ss_PVT_001} with the other wall-normal fluxes 
($\smash{\overline{\rho'v'}\stackrel{\eqref{Eq_PDTEFCTPCF_s_BEqsDNSC_ss_SA_sss_RFD_001c}}{=}-\bar\rho\overline{v''}},\overline{\rho s'' v''}, \overline{\rho T'' v''}$; \figrefsatobnp{Fig_PDTEFCTPCF_s_B_ss_DVMF_001}{Fig_PDTEFCTPCF_s_B_ss_TVHF_002})
is the importance of the pressure-gradient term $\Pi_{(p'v')}$, compared to its negligible influence in the streamwise transport budgets
($\smash{\overline{\rho'u'}\stackrel{\eqref{Eq_PDTEFCTPCF_s_BEqsDNSC_ss_SA_sss_RFD_001c}}{=}-\bar\rho\overline{u''}},\overline{\rho s'' u''}, \overline{\rho T'' u''}, \overline{p'u_i'}$; \figrefsatobnp{Fig_PDTEFCTPCF_s_B_ss_DVMF_001}{Fig_PDTEFCTPCF_s_B_ss_PVT_001}).
On the other hand the particular form of the $\overline{p'v'}$-budgets \figref{Fig_PDTEFCTPCF_s_B_ss_PVT_001}, where most of the terms change sign across the channel,
highlights the specific nonlocal behaviour of $p'$ \citep{Chang_Piomelli_Blake_1999a} and explains the extreme difficulty of single-point closure modelling of pressure diffusion $\Pi_{ij}:=-\partial_{x_j}\overline{p'u_i'}-\partial_{x_i}\overline{p'u_j'}$
in Reynolds-stress transport closures \citep{Vallet_2007a,
                                             Sauret_Vallet_2007a,
                                             Gerolymos_Lo_Vallet_Younis_2012a}.
Notice, nonetheless, that the dilatational term $B_{(p'v')}\leq0\;\forall y^\star$ (gain), always contributes \figref{Fig_PDTEFCTPCF_s_B_ss_PVT_001} in generating $\overline{p'v'}$,
and so does $B_{(p'u')}\geq0\;\forall y^\star$ (gain) in the $\overline{p'u'}$-budgets \figref{Fig_PDTEFCTPCF_s_B_ss_PVT_001}.

%
%
%
%
%
%
%
%
%
\section{Conclusions}\label{PDTEFCTPCF_s_C}
%
%
%
%
%
%
%
%
%

Mixed semi-local \tsn{HCB}-scaling, $y^\star$ and $Re_{\tau^\star}$ \eqref{Eq_PDTEFCTPCF_s_S_ss_CCF_001c}, introduced by \citet{Huang_Coleman_Bradshaw_1995a},
represents reasonably well $Re$-effects in compressible turbulent plane channel flow, not only for the Reynolds-stresses, but also for statistics
involving thermodynamic variations. Regarding Mach-number effects, it is suggested in the present work, based on analysis of \tsn{DNS} results,
that the centerline Mach number, $\bar M_\tsn{CL}$ \eqref{Eq_PDTEFCTPCF_s_S_ss_CCF_001d}, is the appropriate parameter. In particular, the coefficients
of variation of thermodynamic fluctuations ($[\bar{\rho}^{-1}\rho']_{\rm rms}$, $[\bar{T}^{-1}T']_{\rm rms}$ and $[\bar{p}^{-1}p']_{\rm rms}$) scale reasonably well with $\bar M^2_\tsn{CL}$.
Notice that, because of the intense viscous heating which causes the ratio of centerline-to-wall temperature, $\bar T_w^{-1}\bar T_\tsn{CL}$, to increase with Mach number,
the relation between $\bar M_\tsn{CL}$ \eqref{Eq_PDTEFCTPCF_s_S_ss_CCF_001d} and the bulk Mach number $M_{\tsn{B}_w}$ \eqref{Eq_PDTEFCTPCF_s_S_ss_CCF_001a} is nonlinear,
the later increasing faster than the former. The \tsn{DNS} results analysed in the present work cover the range $Re_{\tau^\star}\in[64,344]$ and $\bar M_\tsn{CL}\in[0.34,2.47]$.

Outside of the viscous sublayer ($y^\star\gtrapprox3$), the ratio of the correlation coefficients of wall-normal transport of thermodynamic fluctuations
($c_{\rho'v'}$, $c_{T'v'}$, $c_{p'v'}$, $c_{s'v'}$) on the shear correlation coefficient $c_{u'v'}$ is a unique function of $y^\star$, practically independent of $(Re_{\tau^\star},\bar M_\tsn{CL})$,
implying a $Re_{\tau^\star}$-dependence of the centerline limiting values of these ratios. The same observation applies to the correlation coefficients between thermodynamic variables
($c_{\rho'T'}$, $c_{p'\rho'}$, $c_{p'T'}$, $c_{s'\rho'}$, $c_{s'p'}$, $c_{s'T'}$), but with a larger scatter between different values of $(Re_{\tau^\star},\bar M_\tsn{CL})$.
The correlation coefficient between entropy and pressure fluctuation, $c_{s'p'}\approxeq0\;\forall y^\star\gtrapprox5$, with very little sactter ($\abs{c_{s'p'}}<0.1\;\forall y^\star\gtrapprox5$),
implying statistical independence of $s'$ and $p'$ everywhere in the channel except in the viscous sublayer, presumably because of the localised influence of the
isothermal-wall boundary-condition. On the other hand, the correlation coefficients of wall-normal transport of thermodynamic fluctuations ($c_{\rho'v'}$, $c_{T'v'}$, $c_{p'v'}$, $c_{s'v'}$)
appear to vary as functions of the outer-scaled wall-distance $\delta^{-1}(y-y_w)\;\forall y-y_w\gtrapprox\tfrac{1}{10}\delta$,
with very little scatter for different values of $(Re_{\tau^\star},\bar M_\tsn{CL})$.

The coefficients of variation of the thermodynamic state-variables ($p$, $\rho$, $T$) are always ($\forall (Re_{\tau^\star},\bar M_\tsn{CL})\;\forall y^\star$) of the same order-of-magnitude,
\ie $O([\bar\rho^{-1}\rho']_\mathrm{rms})=O([\bar T^{-1}T']_\mathrm{rms})=O([\bar p^{-1}p']_\mathrm{rms})$, even at the low-$\bar M_\tsn{CL}$ limit, in agreement with sustained
compressible \tsn{HIT} results. Because of the equation-of-state $p=\rho R_g T$, the correlation coefficients between thermodynamic fluctuations ($c_{\rho'T'}$, $c_{p'\rho'}$, $c_{p'T'}$)
are related. To $O([\bar\rho^{-1}\rho']_\mathrm{rms})$, these relations can be solved to express the correlation coefficients ($c_{\rho'T'}$, $c_{p'\rho'}$, $c_{p'T'}$)
as functions of the coefficients of variation ($[\bar\rho^{-1}\rho']_\mathrm{rms}$, $[\bar T^{-1}T']_\mathrm{rms}$, $[\bar p^{-1}p']_\mathrm{rms}$).
The \tsn{DNS} results confirm these approximations. The approximation error for $c_{p'\rho'}$ and $c_{p'T'}$ increases with increasing coefficient of variation of density $[\bar\rho^{-1}\rho']_\mathrm{rms}$,
while the approximation of $c_{\rho'T'}$ appears to be particularly accurate, even at the highest values of $[\bar\rho^{-1}\rho']_\mathrm{rms}\approxeq0.14$ corresponding to the peak of the density variance
for the $(Re_{\tau^\star},\bar M_\tsn{CL})\approxeq(111,2.47)$ case. A detailed study of the approximation error is required to explain this behaviour.

The evolution equations for the thermodynamic variables ($\rho$, $s$, $T$, $p$) obtained from the Navier-Stokes equations using thermodynamic derivatives imply exact transport equations
for their variances ($\overline{\rho'^2}$, $\overline{\rho s''^2}$, $\overline{\rho T''^2}$ and $\overline{p'^2}$)
and fluxes ($\overline{\rho'u_i'}$, $\overline{\rho s''u_i''}$, $\overline{\rho T''u_i''}$ and $\overline{p'u_i'}$).
The point is made that the Favre-averaging operator does not commute with differentation,
\eg $\tilde S_{ij}\neq\tfrac{1}{2}(\partial_{x_j}\tilde u_i+\partial_{x_i}\tilde u_j)$ and $S_{ij}''\neq\tfrac{1}{2}(\partial_{x_j}u_i''+\partial_{x_i}u_j'')$,
and to avoid widespread notational misuse we introduce specific notation $\breve S_{ij}:=\tfrac{1}{2}(\partial_{x_j}\tilde u_i+\partial_{x_i}\tilde u_j)$ and $S_{ij}^{\backprime\backprime}:=\tfrac{1}{2}(\partial_{x_j}u_i''+\partial_{x_i}u_j'')$.
The budgets of the transport equations for the thermodynamic variables were studied using \tsn{DNS} data from a well-resolved simulation at $(Re_{\tau^\star},\bar M_\tsn{CL})\approxeq(150,1.5)$.
There is a strong similarity between the budgets of entropy correlation ($\overline{\rho s''^2}$ and $\overline{\rho s''u_i''}$) and the budgets of the corresponding correlation of temperature ($\overline{\rho T''^2}$ and $\overline{\rho T''u_i''}$),
associated with the fact that the dilatational terms in the $c_v$-form or $(D_t p')'$ terms in the $c_p$-form of the temperature-correlations budgets are very weak and often negligeable.
Pressure stands apart from the other thermodynamic variables, and in particular the complexity of the budgets of the wall-normal transport $\overline{p'v'}$  is probably the signature of the nonlocality of $p'$ and of its subordination to the velocity field.
The coefficient of variation of density $[\bar\rho^{-1}\rho']_\mathrm{rms}$ is undoubtedly the best measure of the effects of compressiblity, beyond Morkovin's hypothesis, on turbulence. Therefore, modelling of density variance (and of the massfluxes
$\overline{\rho'u_i'}$ which govern its production by mean density stratification) is the necessary step in that direction. The main difficulty is probably the dilatational term 
$B_{(\rho'^2)}=-\overline{\left(\rho^2-\bar\rho^2\right)\Theta^{\backprime\backprime}}$ \eqref{Eq_PDTEFCTPCF_s_B_ss_DVMF_002},
which does not contain viscous terms, and further studies on its dynamics and spectra are required.

It is hoped that the present investigation adds to our understanding of compressible turbulent plane channel flow. Nonetheless, to fully understand the influence of ($Re_{\tau^\star}$, $\bar M_\tsn{CL}$) on turbulent statistics and budgets, further
\tsn{DNS} data, with systematic variation of both these parameters is required and is the subject of ongoing work.

%
%
%
%
%
%
%
%
%
\bibliographystyle{jfm}\footnotesize\bibliography{Aerodynamics,GV,GV_news}\normalsize
%
%
%
%
%
%
%
%
%

\end{document}

%% file: JFluidMech_01_Tab_DNS_computations.tex
\scalebox{0.90}{
\begin{tabular}{ccccrcrcccccccccc}
$Re_{\tau^\star}$&$\bar M_\tsn{CL}$&$N_x\times N_y\times N_z$&  $L_x$      &$L_y$    & $L_z$                 &$\Delta x^+$&$\Delta y_w^+$&$N_{y^+\leq10}$&$\Delta y_\tsn{CL}^+$&$\Delta z^+$&$\Delta t^+$&$t_\tsn{OBS}^+$&$\Delta t_s^+$\\
 $~64$           & $1.62$          &$137\times113\times201$  &$~8\pi\delta$&$2\delta$&$           4\pi\delta$&  $18.4$    &  $0.17$      & $22$          &  $~3.1$             & $~6.2$     &$0.0114$    & $~988$       &$0.0115$       \\
 $111$           & $2.47$          &$401\times153\times601$  &$~8\pi\delta$&$2\delta$&$           4\pi\delta$&  $30.9$    &  $0.22$      & $18$          &  $10.2$             & $10.3$     &$0.0136$    & $~446$       &$0.0136$       \\
 $113$           & $1.51$          &$257\times129\times385$  &$~8\pi\delta$&$2\delta$&$           4\pi\delta$&  $16.6$    &  $0.21$      & $21$          &  $~4.7$             & $~5.5$     &$0.0150$    & $1373$       &$0.0150$       \\
 $152$           & $1.50$          &$345\times137\times529$  &$~8\pi\delta$&$2\delta$&$           4\pi\delta$&  $16.6$    &  $0.23$      & $19$          &  $~5.6$             & $~5.4$     &$0.0166$    & $1001$       &$0.0166$       \\
 $150$           & $1.50$          &$361\times149\times433$  &$~4\pi\delta$&$2\delta$&$\tfrac{4}{3}\pi\delta$&  $~7.9$    &  $0.22$      & $19$          &  $~4.7$             & $~2.1$     &$0.0164$    & $~868$       &$0.0164$       \\
 $176$           & $0.34$          &$257\times129\times385$  &$~8\pi\delta$&$2\delta$&$           4\pi\delta$&  $17.6$    &  $0.23$      & $20$          &  $~5.0$             & $~5.8$     &$0.0066$    & $~369$       &$0.0066$       \\
 $176$           & $0.34$          &$193\times129\times169$  &$~4\pi\delta$&$2\delta$&$\tfrac{4}{3}\pi\delta$&  $11.7$    &  $0.22$      & $20$          &  $~4.9$             & $~4.4$     &$0.0064$    & $2329$       &$0.0064$       \\
 $344$           & $1.51$          &$401\times153\times601$  &$~8\pi\delta$&$2\delta$&$           4\pi\delta$&  $32.9$    &  $0.23$      & $17$          &  $10.9$             & $10.9$     &$0.0209$    & $~813$       &$0.0209$       \\
\end{tabular}
}

%% file: JFluidMech_01_Tab_DNS_global_parameters.tex
\scalebox{0.90}{
\begin{tabular}{ccccrcrcccccccccc}
$Re_{\tau^\star}$&$\bar M_\tsn{CL}$&$M_{\tsn{B}_w}$&$M_{\tau_w}$&$Re_{\tau_w}$&$Re_{\tsn{B}_w}$&$Re_{\theta_w}$&$Re_{\theta_\tsn{CL}}$&$Re_{\theta_{\tsn{CL}w}}$&$\bar T_w$       &$\dfrac{\bar T_\tsn{CL}}{\bar T_w}$&$B_{q_w}$&$\displaystyle\max_y\left[\dfrac{\rho'_\mathrm{rms}}{\bar\rho}\right]$\\
                                                                                                                                                                                                                                                                                     \\
 $~64$           & $1.62$          &$1.50$         &$0.0848$    &$~99$        &$1300$          &$~238$         &$~129$                &$~168$                   &$298\;\mathrm{K}$&$1.417$                            &$-0.0508$&$0.04256$                                                             \\
 $111$           & $2.47$          &$3.83$         &$0.1293$    &$492$        &$4438$          &$1588$         &$~190$                &$~444$                   &$298\;\mathrm{K}$&$3.542$                            &$-0.1990$&$0.14274$                                                             \\
 $113$           & $1.51$          &$1.48$         &$0.0825$    &$169$        &$2259$          &$~326$         &$~185$                &$~235$                   &$298\;\mathrm{K}$&$1.379$                            &$-0.0484$&$0.04390$                                                             \\
 $152$           & $1.50$          &$1.50$         &$0.0815$    &$227$        &$3100$          &$~439$         &$~249$                &$~316$                   &$298\;\mathrm{K}$&$1.383$                            &$-0.0485$&$0.04600$                                                             \\
 $176$           & $0.34$          &$0.30$         &$0.0191$    &$180$        &$2785$          &$~291$         &$~283$                &$~286$                   &$298\;\mathrm{K}$&$1.015$                            &$-0.0022$&$0.00237$                                                             \\
 $344$           & $1.51$          &$1.56$         &$0.0754$    &$525$        &$7812$          &$1105$         &$~611$                &$~785$                   &$298\;\mathrm{K}$&$1.402$                            &$-0.0466$&$0.04948$                                                             \\
\end{tabular}
}